VLSI DESIGN OF ADVANCED DIGITAL FILTERS

Author: Asst. Prof. Dr. Rozita Teymourzadeh, CEng.
Formatted by: Kok Wai Chan

FACULTY OF ENGINEERING
NATIONAL UNIVERSITY OF MALAYSIA

**TO MY LOVELY MOTHER**

# ABSTRACT

The Cascaded Integrator Comb filters (CIC) find many applications in recent electronic devices such as frequency selection functions in a digital radio or modem and any filter structure that is required to efficiently process large sample rate factor. These filters are normally located after the sigma delta modulator and have regular structure. These types of filters do not require multipliers and the coefficient storage unlike in the normal digital FIR and IIR filters because of all filter coefficients are unity. Hence, it can be efficiently implemented to operate at high speed. Hence, this book describes the Very Large Scale Integration (VLSI) implementation of the CIC filters that are suitable for high performance audio applications. A total of five cascaded integrators and comb pairs were chosen to meet the design requirement, particularly the effect of signal aliasing to increase the performance. The CIC filter makes use of pipeline architecture that consists of fast adders. The proposed filter was designed and implemented and from FPGA implementation and Xilinx ISE synthesis result shows that the filter can perform at the maximum frequency of 189MHz, which is required for high performance audio applications. For area and power consumption, the CIC filter was optimized in ASIC under SilTerra $0.18\,\mu m$ and MIMOS $0.35\,\mu m$ technologies file. The layout was performed by the $0.18\,\mu m$ technology.

# CONTENTS

# LIST OF FIGURES

# LIST OF TABLES

# LIST OF SYMBOLS

| | | |
|---|---|
| $\|\ \|$ | Absolute value |
| $A_s$ | Stop band attenuation |
| $A_p$ | Pass band attenuation |
| $a(k)$ | Feedback filter coefficient |
| $B_{in}$ | Input data word length |
| $B_{max}$ | Maximum word length |
| $b$ | Discarded bit in truncation |
| $c$ | Carry |
| $CLK$ | Master clock |
| $D_{in}$ | Input data to the CIC chip |
| $D_{out}$ | Output data to the CIC chip |
| $D_j$ | Filter coefficients |
| $E(z)$ | Additive White noise modeling the $\sum\Delta$ quantization noise |
| $E_j$ | Truncation error in stage $j$ |
| $f_B$ | Bandwidth frequency |
| $f_s$ | Sampling frequency |
| $f_{pass}$ | Passband frequency |
| $f_{stop}$ | Stopband frequency |
| $f_T$ | Transition band |
| $F_j^2$ | Power of two filter coefficient |
| $G_{max}$ | Maximum register growth |
| $g_i$ | Generate bit in MCLA |
| $G_{out}$ | Group generate bit in MCLA |
| $H(z)$ | Transfer function in z domain |
| $H_C(z)$ | Transfer function of the comb stage |
| $H_I(z)$ | Transfer function of the integrator stage |
| $h(k)$ | Feedforward filter coefficient |
| $L$ | Sigma delta modulator order |
| $LD_{in}$ | Load in |
| $M$ | Differential delay |
| $\mu$ | Mean of the truncation error |

| | | |
|---|---|---|
| $\mu_{Tj}$ | | Total mean of the truncation error |
| $N$ | | CIC filter order |
| $ND$ | | New data |
| $p(f)$ | | Power frequency response |
| $p_A(f)$ | | Power frequency response approximation |
| $p_i$ | | Propagate bit in MCLA |
| $P_{out}$ | | Group Propagate bit in MCLA |
| $R$ | | Decimation ratio |
| $RDY$ | | Filter output sample ready pin |
| $RFD$ | | Ready for data pin |
| $s$ | | Sum |
| $\sum \Delta$ | | Sigma delta |
| $T_s$ | | Sampled clock period |
| $\partial_j^2$ | | Variance of the truncation error |
| $\partial_{Tj}^2$ | | Total variance of the truncation error |
| $WE$ | | Write enable |
| $\omega$ | | Frequency in rad/s |
| $X(z)$ | | Input data in z domain |
| $x[n]$ | | Input data in discrete-time |
| $Y(z)$ | | Output data in z domain |
| $y[m]$ | | Output data in discrete-time |

# LIST OF ABBREVIATIONS

| | |
|---|---|
| ADC | Analog to Digital Converter |
| ASIC | Application Specific Integrated Circuit |
| CD | Compact Disc |
| CIC | Cascaded Integrator Comb |
| CLA | Carry Look-ahead Adder |
| CPLD | Complex Programmable Logic Device |
| CSA | Carry Save Adder |
| DAC | Digital to Analog Converter |
| DAT | Digital Audio Tape |
| DDC | Digital Down Converter |
| DDF | Digital Decimation Filter |
| DSP | Digital Signal Processing |
| FA | Full Adder |
| FFT | Fast Fourier Transform |
| FIR | Finite Impulse Response |
| FPGA | Field Programmable Gate Array |
| FPLD | Field Programmable Logic Device |
| FSK | Frequency Shift Keying |
| HA | Half Adder |
| HDTV | High-Definition TeleVision |
| IC | Integrated Circuit |
| IF | Intermediate Frequency |
| IIR | Infinite Impulse Response |
| I/O | Input/Output |
| ISE | Integrated Synthesis Environment |
| ISDN | Integrated Services Digital Network |
| ISOP | Interpolated Second-Order Polynomial |
| LPTV | Linear Periodically Time-Variant system |
| LSB | Least Significant Bit |
| LTI | Linear Time Invariant |
| MCLA | Modified Carry look-ahead Adder |
| MCSA | Modified Carry Save Adder |
| MHBFs | Modified Half Band Filters |
| MOS | Metal-Oxide Silicon |
| MSB | Most Significant Bit |
| OSR | Over Sampling Ratio |
| PAR | Place And Route |
| PFA | Partial Full Adder |
| RCA | Ripple Carry Adder |
| RCAS | Ripple Carry Adder Subtractor |
| RF | Radio Frequency |
| RNS | Residue Number System |
| $\sum\Delta$ | Sigma Delta |
| SAR | Successive Approximation Register |
| SLD | System Level Design |
| SNR | Signal to Noise Ratio |
| SOC | System On Chip |
| SPFA | Sum of Partial Full Adder |
| SRC | Sample Rate Conversion |
| VHDL | Verilog Hardware Description Language |
| VLSI | Very Large Scale Integration |

# CHAPTER I

# INTRODUCTION

This chapter describes the research background, problem statement, motivation, methodology, objectives of the research, scope of work, significance and contribution of the research, chapter organization and finally the summary of the chapter.

The research work proposes the design and implementation of high-speed and low power of a class of filters known as Cascaded Integrator Comb filters (CIC). To achieve the above objectives, the pipelined architecture and fast adders were incorporated into the design. The filters were designed using Verilog hardware description language (Verilog HDL) and were implemented on the field programmable gate array (FPGA) to test the functionality of the design.

## 1.1 BACKGROUND

In many audio systems such as the HiFi compact disc (CD), digital audio tape (DAT) and speech processing make use of analog to digital converters (ADC) like sigma delta modulator ($\sum\Delta$), (Crochiere & Rabiner 1983; Vaidyanathan 1993; Rabii & Wooley 1997; Peluso & Vancoreland 1998; Dessouky & Kaiser 2000), successive approximation register (SAR), (Mortezapour & Lee 2000; Fayomi et al. 2001; Ginsburg & Chandrakasan 2005), pipeline ADC (Jipeng & Moon 2004; Yuh-Min et al. 1991; Waltari & Halonen 2000; Chiu et al. 2004) and re-circulating ADC (Song et al. 1990; Onodera et al. 1998) in their operations. Among the modulators mentioned above the sigma delta modulator is considered the best choice for audio process due to the high resolution obtained. $\sum\Delta$ modulator is an oversampled modulation technique which provides high resolution sample output in contrast to the standard Nyquist sampling technique.

To obtain the high resolution conversion, it is necessary to have high order sigma delta modulator. However this is not acceptable because of the high cost of implementation and the instability. To overcome this problem, the integration of high order decimation filters and the sigma delta modulator can be used. Thus the CIC filter is cascaded to the sigma delta modulator as the next stage so as to increase the resolution of the sigma delta modulator without sacrificing the area and low cost.

## 1.2    PROBLEM STATEMENT

In the standard design of oversampling and decimation system, two important digital signal processors are used: i) the sigma delta modulator and ii) the decimation low-pass filter (Crochiere & Rabiner 1983; Oppenheim & Schafer 1989). To obtain high resolution sigma delta ADC (more than 20-bit), the high order type of the modulator is required and this will produce disadvantages such as the instability. Additionally the use of digital decimation low-pass filter to cooperate with the modulator is complicated since the filter requires multiplications that result high power consumption and higher area die size. Thus the research work proposes the use of a special filter called the CIC filter which can produce high resolution and low power consumption in the sigma delta modulator due to the CIC filter does not require multiplier and the modulator can be relaxed of high computation.

In communication devices like mobile systems and wireless transceivers, low power and small device size are important issues. In transceivers, the programmable digital filter is required to select the desirable channel. This filter causes excessive of power consumption at also higher die area. Using sigma delta modulator in mobile devices is well suited because of the channel selection is carried out by decimation filter automatically. Additionally anti-aliasing low-pass filter requirements are also relaxed due to the CIC filter and oversampling. Furthermore, on FPGA chip or ASIC implementation of the CIC filters make it low cost for DSP and audio application (Cummings & Haruyama 1999; Dick & Harris 1999).

## 1.3    MOTIVATION

Assume that oversampling technique is required for audio applications. Hence it is necessary to achieve a minimum of 70 dB signal to noise ratio (SNR) in the conversion process. To achieve 70 dB SNR, a minimum of 12-bits converter resolution after oversampling is required. Thus the CIC filter is applied to increase resolution for enhancing SNR and makes it well suited for audio application. Furthermore, the digital decimation system has been shown to reduce the power consumption due to sampling rate conversion (Crochiere & Rabiner 1983). The filter that uses this technique is expected to drive down the power consumption and thus allows the filter to operate at low voltage. Additionally, the CIC decimation filters are multirate filters (operates at different sampling frequencies) but unlike other multirate filters, the CIC filters only allow fixed-point arithmetic (Bit stream), since these filters are inherently fixed-point filters (MATLAB7 software 1994-2006). The fixed point CIC filters are much more efficient that is less area and relatively faster to design which can lead to cheaper products in the market.

By eliminating the zero number of multipliers, the CIC filter has the advantage of very compact architecture compared to other low-pass filters. This filter also permits the high rate of changing signal decimation or interpolation factor.

## 1.4  RESEARCH METHODOLOGY

From the discussion in the previous sections, the research methodology presented in the following:

1) To present the basic architecture of the fast CIC filters. The research also continues to achieve high speed CIC filter algorithm by using fast adder and pipelined structure.

2) To perform simulation of the $3^{rd}$ order Sigma delta modulator and decimation process followed by CIC filter, associated half band filters and FIR filter to obtain desire frequency response.

3) To conduct Verilog implementation of CIC filter with synthesizing and defining constrained for the design and test the filter with Logic Analyzer.

4)         To implement the design and download the design to Xilinx Virtex II FPGA board and test the filter with logic analyzer. The FPGA board output is compared with MATLAB simulation result.

5)         To synthesize the CIC filter in ASIC in order to power consumption and area optimization.

## 1.5   OBJECTIVES

The following research objectives have been applied during the course of the research to ensure a complete and quality research is carried out.

1) To design the CIC filters that have the SNR of 145 dB that can be cascaded with sigma delta modulator to give desired frequency response.

2) To optimize the architecture of the CIC filters for high-speed and low power with minimum area die size.

3) To design and to test the CIC filter implemented on FPGA board.

## 1.6   SCOPE OF WORK

Based on the outlined objectives above, available hardware and software resources, and the time frame allocated, this research project is narrowed down to the following scope of work.

Building on prevalent the CIC decimation filters theory and practices a new architecture and its implementation is proposed. Besides, simulation of the analogue to digital converter and designing the whole multi-stage decimation filters including the half band FIR and droop correction filters have been carried out.

The third order sigma delta modulators are the architectures being explored in recent times for use in audio applications. This work focuses on the decimation digital filters that are needed at the output of the modulators.

The aims of the project are achieve the improved CIC filters taking into consideration the power-speed trade off. In the filter implementation, we make use of both Silterra 0.18 µm technology and Mimos 0.35 µm technology. The design is also implemented on the FPGA for quick prototyping.

For this research, three methods to increase the speed of the filter are proposed whilst trying to minimizing the area and power consumption. Firstly, new algorithm based on the carry look-ahead adder (CLA) has been implemented in the CIC filter which gives for addition. Modified Carry Look-ahead Adder (MCLA) was designed and implemented with the speed of 270 MHz for 8 bit adder. Secondly, the pipeline structure of CIC filter has been considered with the use of minimum number of registers and finally truncation lead us to have low power and decreases the complexity of calculation.

The CIC filter must be able to receive digital signal from $3^{rd}$ Sigma delta modulator with low resolution and a 5 bit quantizer and decimate the signal by factor of 16 to achieve the high resolution of 25 bit. While decimation process is going on, the noise should be removed to out of band quantization noise. The SNR of CIC filter is estimated to be higher than 140 dB. The filter also provides the necessary additional aliasing rejection for the input signal as opposed to the internally generation quantization noise.

To achieve clean audio signal from sigma delta modulator, its output needs to pass through multistage decimation filters consist of low pass CIC filter, two half-band filters and droop correction filter. The next chapters will explain the design and implementation of this filter in more detail. The overall frequency response of multistage decimation filters has the pass-band frequency of 20 kHz. Among these multistage filters, the CIC filter is used as a decimator with maximum decimation ratio of 16. It also has pass-band frequency, stop-band frequency and transition band of 7 kHz, 384 kHz and 377 kHz respectively.

Finally the non-recursive filter is introduced to operate as decimator in order to high speed and low power consumption. This filter is suitable decimation filter to replace with the CIC filter when the decimation ratio and filter order are high. It is described that why the non-recursive comb filter structure is easier to implement compare to the CIC filter.

## 1.7   RESEARCH CONTRIBUTION

A new architecture of CIC filters that can be used in the ADC was designed and implemented. These filters with incorporated a pipeline architecture and MCLA adder as summation block can provide the speed up to 189 MHz with SNR of more than 140 dB. The estimated chip die area has been measured. The result shows that the chip size is $0.308mm \times 0.308mm$ in Silterra $0.18\,\mu m$ technology library and $1.148mm \times 1.148mm$ in Mimos 0.35 $\mu m$ technology library.

The speed improvement in decimation system is most important point of the research. The summation block which is called MCLA lead us to have high speed performance of the CIC filter. This adder is considered fast adder (270 MHz in 8 bit) which can be used as pipeline adder too. The advantage of the MCLA adder is the simple structure compared to the CLA adder especially when it is used in high number of bit. It is easy to develop in high number of bit. In next sections, the property of this adder will be explained in detail. Additionally the pipeline structure of the CIC filter is presented and implemented to improve the speed in decimation system.

Saving the power consumption is one of the key issues for designing and implementing the decimation filter. It will be achieved by decreasing the number of calculation (achieved by truncation) and eliminating unnecessary elements (achieved by pipeline structure and using MCLA) in the filter structure. The CIC filter was synthesized in 0.18 $\mu m$ Silterra and 0.35 $\mu m$ Mimos technology libraries. The total estimated power consumption for the design after defining the constrained was less than 3.5 mW in 0.18 $\mu m$ Silterra technology and 6.03 mW in 0.35 $\mu m$ Mimos technology.

The CIC filter was tested by logic analyzer and a certain configuration circuit after its VLSI implementation. Testing by the certain configuration circuit takes the advantage of finding and selecting each part of the CIC filter which can not work properly.

## 1.8 CHAPTER ORGANIZATION

The work in this conveniently organized into seven chapters. Introduction is presented the first chapter which is consist of background, problem statement, motivation, research methodology, objective, scope of work, research contribution, organization and finally summary.

The second chapter provides the review of oversampling technique, sigma delta modulator and the necessity of using decimator after sigma delta modulator. This chapter also describes brief summaries of the literature review prior to engaging the mentioned scope of work. The several topics related to this research are reviewed to give an overall picture of the background knowledge involved. The summary of the literature review is given to clarify the research rationale.

Chapter three presents the design methodologies that are employed. Besides, the application of the decimation system which the CIC filters is a part of that is described. To clarify the decimation system, all decimator which are involved are introduced in brief. The system simulation is done by MATLAB software to verify the system properties.

Chapter four covers the principle definition of the CIC filter followed by introducing its structure in detail. The frequency analysis is done in this chapter. Additionally the comparison between the conventional decimation filter and the CIC filter as decimator is explained. The advantages of using The CIC filter in digital circuit are described. Besides, the basic introduction of the interpolation is presented.

Chapter five focuses on realization of the CIC filter stages and its implementation. After implementing the standard CIC filter structure on FPGA board, the optimization of the filter architecture to achieve high speed and low power is presented. The new architecture of pipeline high speed CIC filter is designed and programmed to the FPGA board. The implementation is done by using Xilinx software and Synopsis tools.

The implementation and simulation result is given in chapter six. In this chapter the simulation result of overall modulator and decimation system are given by using MATLAB, Modelsim and cad tools. Frequency component evaluation is presented in this chapter. The CIC filter implementation results are shown by using Xilinx software and Synopsys tools. This filter is tested by using logic analyzer and a certain configuration digital circuit.

Chapter seven extracts the conclusions and proposes future work.

## 1.9    SUMMARY

In this chapter, an introduction was given on the background and motivation and objectives of the project. The need to design low order sigma delta modulator, leads us to design and implement the low power decimation filter called CIC filters. The CIC filters were proposed to reduce the sampling frequency, avoid the presence of aliasing in base band and finally to push quantization noise to out of frequency bandwidths. Several objectives were presented and scope of project was set to achieve the desired objectives.

# CHAPTER II

# FILTER BACKGROUND AND LITERATURE REVIEW

This chapter reviews the important DSP concepts that bridge to the full understanding of the CIC filters which is the subject of this research. The designed CIC filter could be used in the high resolution sigma delta ADC. In the sigma delta modulators, the oversampling technique is used in order to provide such a high signal to noise ratio. This chapter starts with the principle definition of the oversampling technique to show the importance of the oversampling sigma delta modulator and continues with the necessity of the CIC filter in the system. Furthermore, the literature review of the CIC filter will be given, that leads to improving the performance of the CIC filter.

## 2.1  OVERSAMPLING TECHNIQUE

The audio bandwidth is the range of audio frequencies which directly influence the fidelity of a sound. The higher audio bandwidth creates the better sound fidelity. The highest practical frequency which the human ear can normally hear is typically 20 kHz or flat frequency response from 20 Hz to 20 kHz. Thus from the Nyquist criteria, in order to avoid aliasing to the original signal, the sampling rate must be at least more than 48 kHz. Every signal above Nyquist rate must be filtered by analog anti-aliasing filter. Since none of the filters are ideal, they have a finite slope as they begin attenuating frequencies which can corrupt the original audio signal. Additionally the implementation of analog anti-aliasing filters are very difficult due to sharp cut off necessary to maximize use of the available bandwidth without exceeding the Nyquist limit. Thus converting the signal into digital by using the analog to digital converter reduces the complexity of filtering significantly. It is due to using digital filter in comparison to implementation the same filter in analog.

Based on above explanation, it is good enough motivation to use oversampling technique (Baird & Fiez 1996; King et. al 1998; Jovanovic & Beferull-Lozano 2004; Yonghong & Tenhunen 1999) to overcome aliasing in bandwidth (Figure 2.1). Oversampling technique has immediate benefits for anti-aliasing filters. Because of making relaxed transition band requirements for anti-aliasing filters and reduced baseband quantization noise power. Oversampling means generating more samples from a waveform has already been digitally recorded. It is a technique which can be applied for signal processing or audio application. The sampling frequency significantly is much higher than twice the bandwidth. The signal is called oversampled signal if it oversampled by factor of R.

$$R = \frac{f_s}{2f_B} \qquad (2.1)$$

$$f_S = 2 \times R \times f_B \qquad (2.2)$$

Where $f_s$ is the sampling frequency, $f_B$ is the bandwidth or highest frequency of the signal.

The primary drawback is that oversampling needs the sigma delta modulator and the digital decimation filters to operate at the significantly higher clock rate (Galton & Jensen 1996). Thus the requirement of using sigma delta modulator comes from utilization over sampling technique.

Figure 2.1 shows how oversampling changes the frequency response which has effect on anti-aliasing filter.

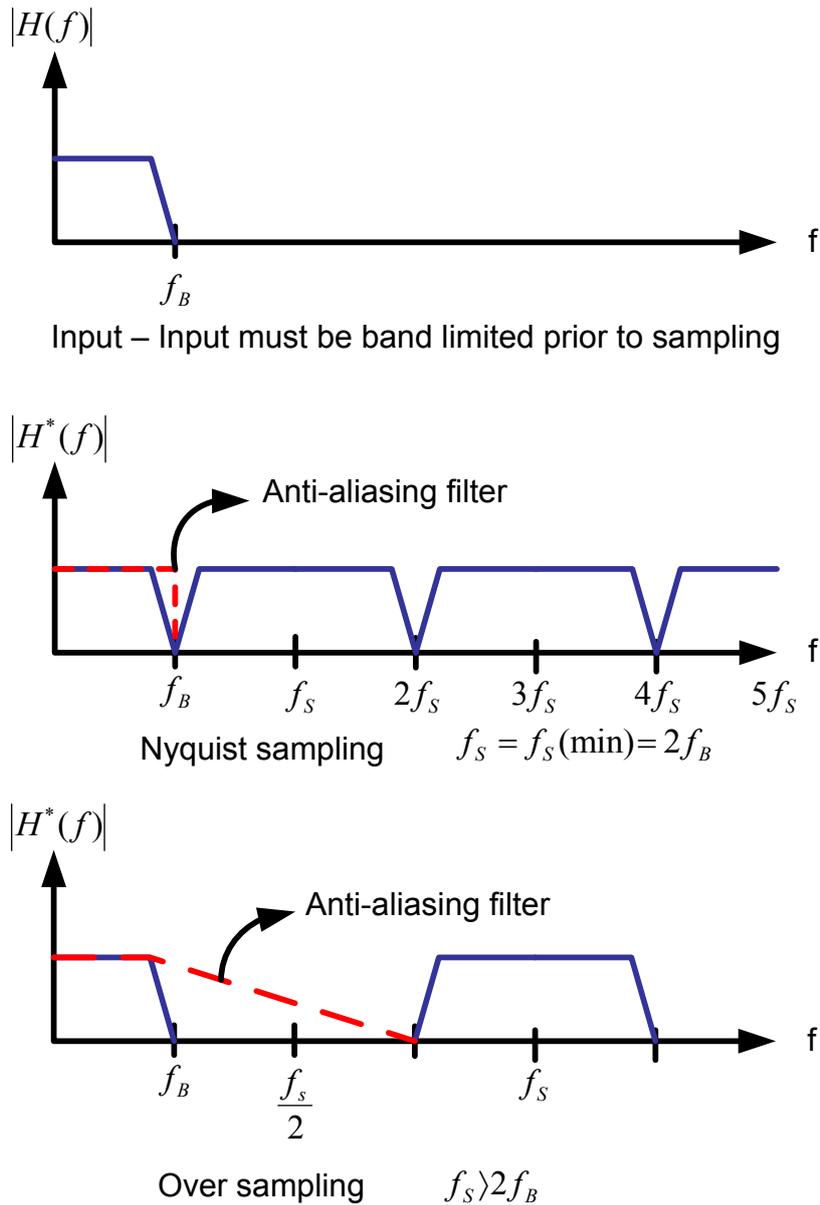

Figure 2.1 Oversampling techniques to relax anti-aliasing filter

Oversampling has become popular in recent years because it avoids many of the difficulties encountered with conventional methods for analogue-to-digital and digital–to-analogue conversion, especially for those applications that call for high-resolution representation of relatively low-frequency signals. Recently, oversampling sigma delta technique was introduced in signal processing and audio application to draw out the audio data from audio devices (Leung 1991; Xueshen et. al 2000).

## 2.2 OVERSAMPLING SIGMA DELTA MODULATOR (∑Δ)

An early description of sigma delta modulator concept was given in a patent by Cutler (1960) and also in 1946 (Steele 1975). The name of the sigma delta modulator comes from putting the integrator (sigma) in front of the delta in structure of the modulator (Park 1990). The Sigma-delta ADC (Rabii & Wooley 1997; Peluso & Vancoreland 1998; Hu et al. 2003; Baird & Fiez 1996; Dessouky & Kaiser 2000) is the analog-to-digital converter that uses oversampling technique to produce disproportionately more quantization noise in the upper portion of their output spectrum. In this technique, the signal is codified using a time-density called bit-stream. Sigma delta modulator is particularly an attractive idea and has been successfully used in analog to digital converters in audio applications over the last two decades.

Although sigma delta concepts existed since the middle of the century, only recent advantages in VLSI technologies have made possible the appropriate handling of bit stream generated by n-bit ADC. The scheme of sigma delta modulator is simple (Candy & Benjamin 1991). The last two terms are in fact evocative of two basic principles involved in the operation of sigma delta ADC, oversampling and noise shaping that achieve high SNR in low resolution.

The first order sigma delta modulator circuit is shown in Figure 2.2. It includes a difference amplifier, an integrator, and a comparator with feedback loop that contains a 1-bit DAC.

The analogue input is fed to an integrator, whose output is fed to a quantizer. The output of quantizer forms a subtracted feedback from the analogue input. This feedback forces the average value of the quantized signal to track the average input. The output of the modulator is a digital representation of the signal, which can be demodulated by smoothing the impulses in a low pass filter. The sigma delta is an example of pulse-density modulation (Paramesh & Jouanne 2001). The number of pulse per unit time increases with the magnitude of the input. The sigma delta ADCs are well suited for low-frequency high resolution applications (Figure 2.3) such as for digital audio and integrated services digital network (ISDN).

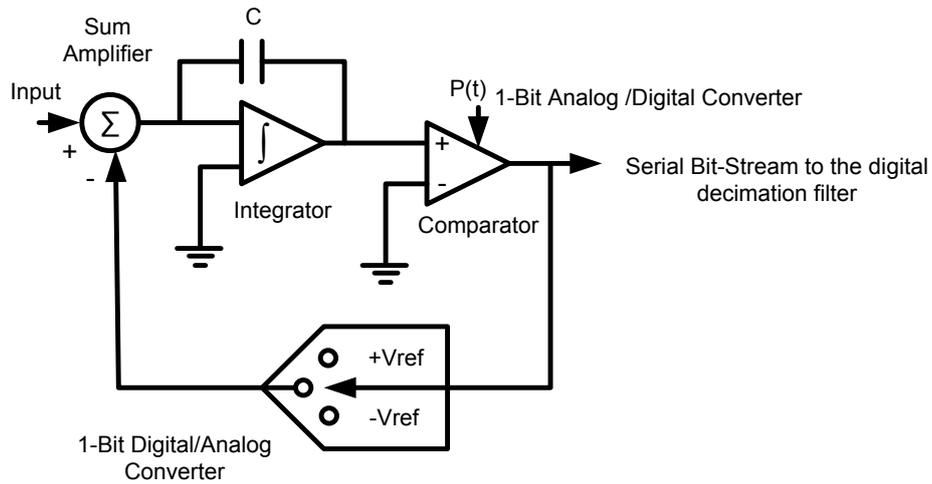

Figure 2.2  First order sigma delta modulator

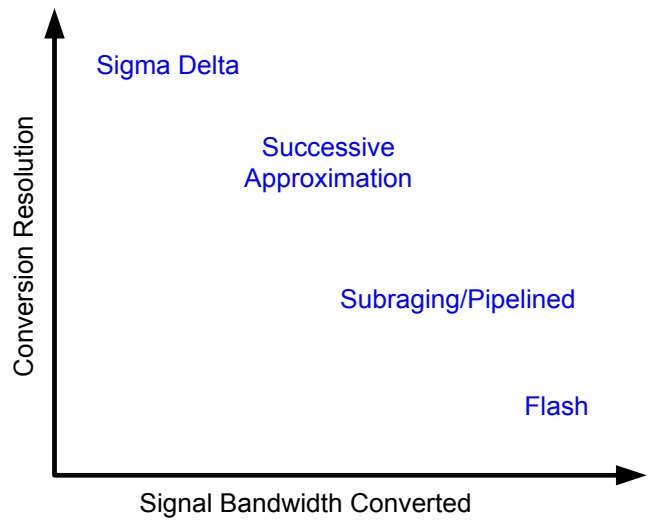

Figure 2.3  Bandwidth resolution tradeoffs

Extending this architecture to instrumentation and HDTV applications which require both high resolution and high bandwidth (Baird & Fiez 1996) necessitates in reducing the converters oversampling ratio (OSR) since the maximum sampling rate is limited by the IC process.

In z-domain, first order sigma delta modulator is shown in Figure 2.4. The noise $E(z)$ which is created by the quantizer, adds to the modulator structure.

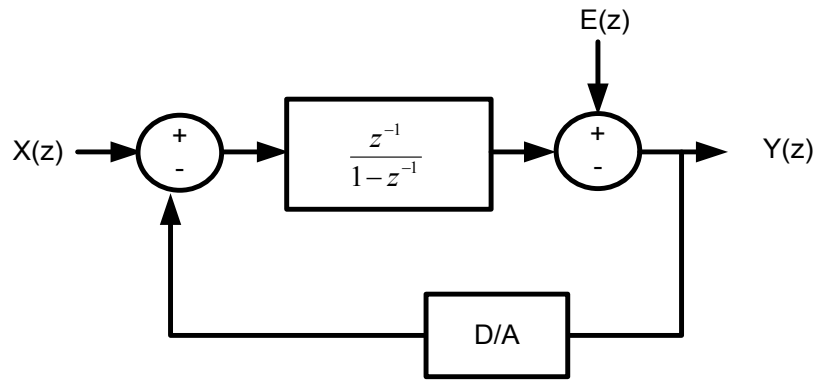

Figure 2.4  First order sigma delta modulator in z-domain

The 1-bit D/A converter is assumed to be ideal and does not introduce any additional noise. In the *z*-domain, the output of the first-order modulator is

$$Y(z) = z^{-1}X(z) + (1-z^{-1})E(z) \qquad (2.3)$$

which consists of the input, *X(z)*, delayed by one sample period, plus a first-order high pass shaping of the additive quantizer noise, *E(z)*.

Decreasing the over sampling ratio (OSR) of a particular sigma delta A/D converter causes a corresponding reduction in the dynamic range. This reduction in dynamic range may be compensated for by increasing the order of the modulator and/or using multibit quantization (Sarhang-Nejed & Temes 1993; Fattaruso et al. 1993; Brandt & Wooley 1991).

In a conventional ADC, the sine wave to the converter is sampled with the frequency of at least twice the bandwidth of the input signal in the frequency domain. The result of Fast Fourier Transform (FFT) diagram shows a single tone and lots of random noise extending from DC to Fs/2. SNR in this ADC is given by equation 2.4 below:

$$SNR = 6.02N + 1.76 dB \qquad (2.4)$$

Where N is number of bit in ADC.

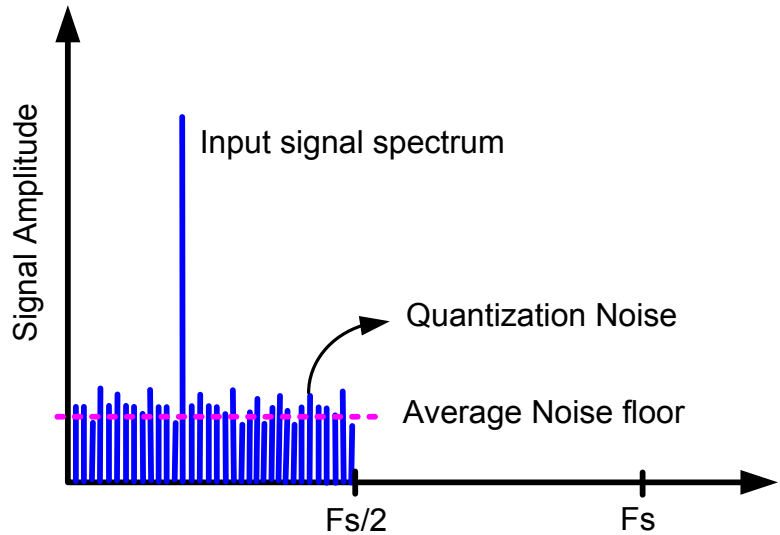

Figure 2.5 Conventional ADC spectrum with sampling frequency of $f_s$

The noise comes from converting the continuous signal to digital discrete signal which result in the information lost. Figure 2.5 shows the spectrum of conventional ADC when input sine wave is converted to digital signal.

To improve the SNR in conventional ADC, the number of bit should be increased. However in order to decrease the noise and to enhance the conversion process the oversampling method (Temes & Candy 1990; Joseph et. al 1999) is proposed. Thus according the oversampling rule, the sampling frequency is R times higher than sampling frequency:

$$F_{S_{oversampling}} = R \times F_{S_{conventional}} \tag{2.5}$$

Figure 2.6 shows the diagram of amplitude of signal in frequency domain after oversampling. An FFT analysis shows that the noise floor has decreased significantly. The signal tone is the same as before but the noise energy is spread out over the wider frequency range.

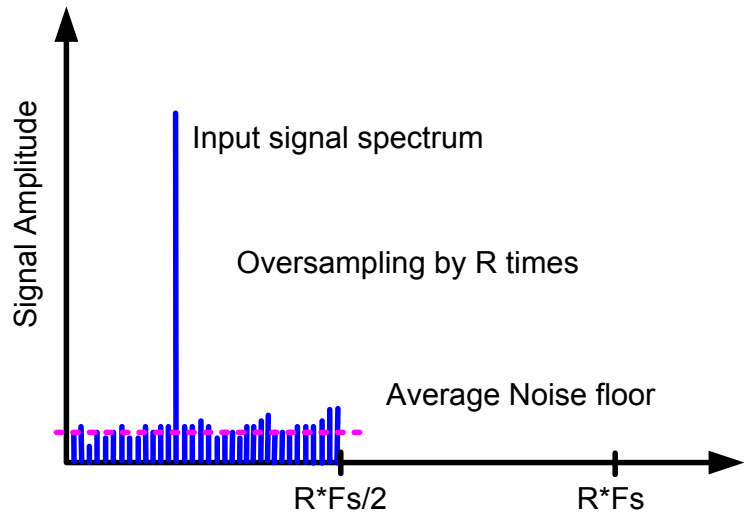

Figure 2.6 Conventional ADC spectrum with sampling frequency of $R \times f_s$

The structure of the sigma delta modulator (with oversampling) has the effect of pushing the broadband noise to upper at high frequency which is outside of the range of interest as shown in Figure 2.7. Feedback loop in sigma delta modulator creates error by subtracting output with input signal. By summing the error voltage, the integrator acts as low pass filter to the input signal and high pass filter to the quantization noise (Dunn & Sandler 1994), thus most of the quantization noise is pushed in to higher frequencies. Actually the oversampling has changed not the total noise power but it is distribution.

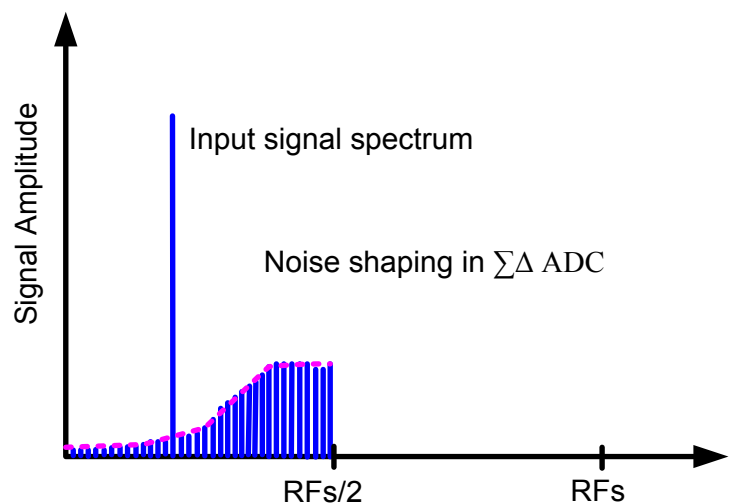

Figure 2.7 Sigma-delta ADC Spectrum with sampling frequency of $R \times f_s$

Further noise reduction can be obtained by using higher order modulators. The higher order sigma delta modulator (Chao-liang & Wilson 1997; Ribner 1991) requires adding more integrators (Figure 2.8) however the circuit is highly unstable. Thus in this research we propose to use low-pass filter in order to obtain a stable sigma delta circuit but with improved SNR.

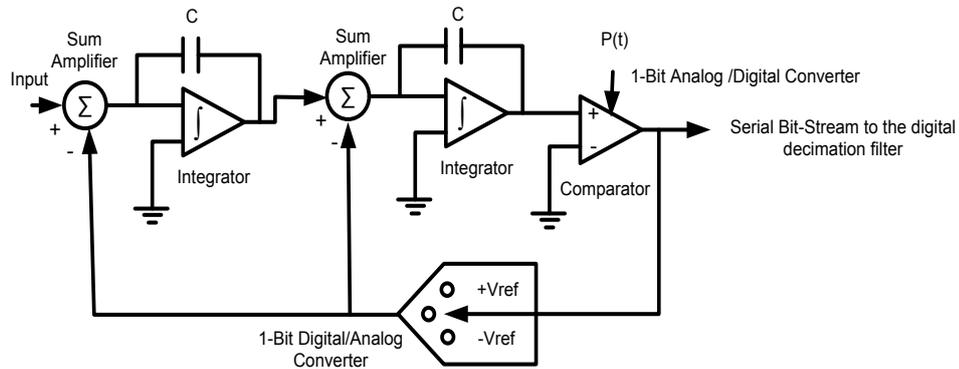

Figure 2.8  Second order sigma delta modulator

If the quantizer in a first-order ΣΔ modulator is itself replaced by a first-order ΣΔ modulator, and the first-stage integrator is implemented without a delay, then the second-order modulator of Figure 2.9 is obtained.

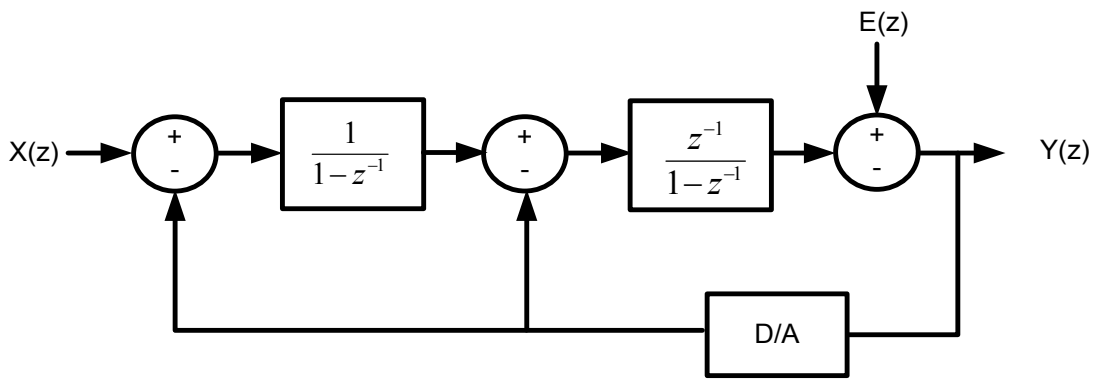

Figure 2.9  Second order sigma delta modulator in z-domain

The output of the second-order modulator is given by equation 2.6.

$$Y(z) = z^{-1}X(z) + (1-z^{-1})^2 E(z) \qquad (2.6)$$

Figure 2.10 shows the relationship between the order of sigma delta modulator and the amount of oversampling necessary to achieve a particular SNR.

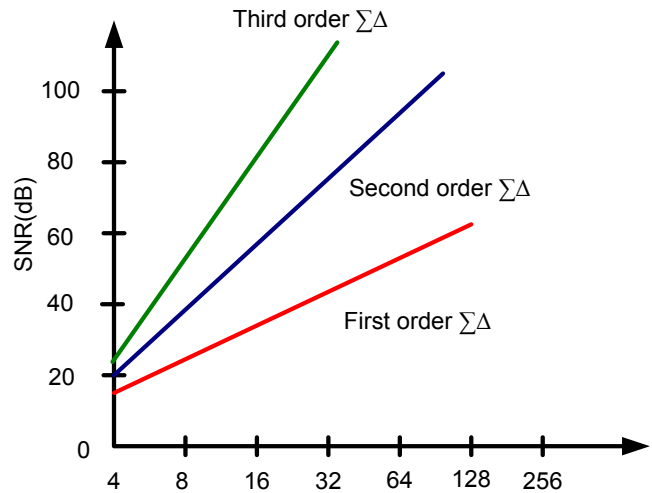

Figure 2.10  Oversampling ratios versus SNR in sigma delta modulator

**2.3  DIGITAL DECIMATION CASCADED INTEGRATOR COMB FILTER**

The decimation CIC filter - to be discussed - is low-pass filter. It is located after the sigma delta modulator in order to extract information from n-bit stream to increase the SNR by removing quantization noise to outside the band of interest. The CIC filter greatly attenuates out of band quantization noise and finally down sampling the signal to the nyquist rate. As a result the ADC resolution and the precision are increased.  The digital cascaded integrator comb (CIC) filter is strongly selected to carry out the decimation process. Figure 2.11 shows the effect of the low pass decimation CIC filter on signal spectrum and the necessity of using it in signal processing.

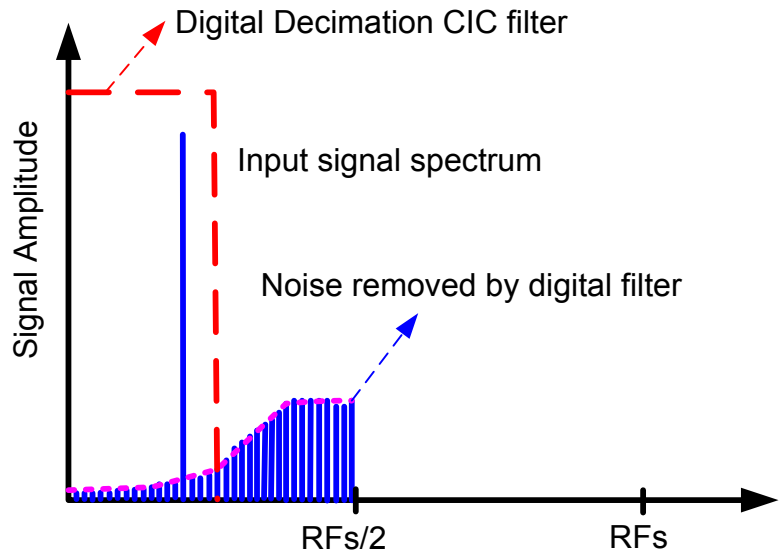

Figure 2.11 Noise Shape removing by digital decimation CIC filter

## 2.4 LITERATURE REVIEW

The CIC filter was introduced by Hogenauer (1981). He introduced a class of linear phase FIR filters for decimation and interpolation. Decimation filter is the simplest and most economical filter to reduce the input sampling rate. Additionally the CIC filter does not require storage for filter coefficients and multipliers as all coefficients are unity (Park 1990) thereby leading to more economical hardware implementations. He used cascaded integrator and cascaded comb section which are connected together by one down sampler in decimation or up sampler in interpolation. He implemented four order filter with sampling frequency, $f_s = 5MHz$, decimation factor $R = 64$ to 512 and the differential delay $M = 2$ resulting in stop band attenuation of 53 dB. He also described pass band attenuation table and aliasing/Imaging attenuation for different filter stages and different bandwidths.

Abu-Al-Saud (2003), constructed a simple modification to the CIC filter that enhanced its Sample Rate Conversion (SRC) performance at the expense of requiring a few extra computations per output sample. Modified CIC filter provides higher SNRs (+28 dB) and better image attenuation than the conventional CIC filter by adjusting the Zero of the filter to target high power image components. He used forth order CIC filter to perform SRC by a factor of 9/10.

Losada and Lyons (2006) have attempted to reduce the CIC filter complexity for interpolation and decimation. They described the new structure of decimation filter which is called non-recursive filters. These kinds of filters were used to avoid register bit-growth complexity. One of the difficulties in using the CIC filter is accommodating large data word growth, particularly when implementing integrators in multistage CIC filters. Thus non-recursive comb filter was introduced to overcome this problem. This method was also designed and implemented by Yonghong Gao et al. (1999). Non-recursive comb filter is attractive when decimation ratio and filter order are high.

Ritoniemi et al. (1994) presented a stereo audio sigma-delta A/D converter followed by decimation filters for audio application. They implemented the fourth order sigma delta modulator and linear phase decimation filters. The result implementation showed the stop band attenuation of 120 dB in decimation filters, the pass band ripple less than 0.0003 dB, signal-to-noise ratio of 95 dB, power consumption of 30 mW in digital part and the die are of 4.7 mm x 5.5 mm for overall system. The design was implemented using a 1.2 $\mu m$ double-poly BiCMOS process.

Djadi et al. (1994) implemented CIC filters as decimator and interpolator which have programmable capability in decimation ratio from 0 to 256. The number of filter was fixed in their design. They used MOSIS 1.2 micron technology CMOSEN standard cell libraries.

Mulpur (2003) designed and explained System-Level Design (SLD) of fixed point down converter filter by the method of faster embedded signal processing system. He implemented digital down converter (DDC) for simplified digital radio application. The system was designed for GSM type application and it was implemented on FPGA/PLD platforms. From an SLD point of view the fixed point scaling lends itself quite well for automation so he used fixed point down converter to ensure that final implementation would be free of saturation and overflow problems.

Moreover, interpolation is widely used in D/A converter. Karema et al. (1991) introduced an oversampled high resolution sigma-delta D/A converter which is primary digital and suits very well for audio applications. Interpolation filter also implements as digital part using two in-house IC-chips and is very flexible due to the filters are realized by programmable filter processor. The silicon area of the chips is small and can still be reduced considerably by using modern MOS technologies.

Yonghong Gao et al. (1999) presented a comparison design of comb decimators based on the non-recursive algorithm and the recursive algorithm. The implementation of non-recursive decimation filter showed that the filter can work with maximum clock frequency of 110 MHz by using power consumption of 44 mW (at 25 MHz). He also implemented high speed CIC decimation filter based on the partial-polyphase non recursive decomposition and parallel processing techniques. Non-recursive CIC filters also presented by using carry save adder (CSA) for addition. He divided the chip to several blocks due to high speed operation. Gao also recommended using separate power supply pads for internal circuitry and I/O pads.

Residue number system (RNS) for pipelined Hogenauer CIC filter was introduced by Garcia et al. (1998). They designed high decimation rate digital filter which demonstrated the RNS-FPL synergy. The Implementation adders and accumulators with CPLDs have been shown in the paper. Compared to the two's complement design, the RNS based Hogenauer filter enjoys an improved speed advantage by approximately 54%. Similar structure by Meyer-Baese et al. (2005) has been implemented to reduce cost in Hogenauer CIC filter. In that design Modified carry save adder (MCSA) was used for summation. The result shows that the CIC filter can work in maximum clock frequency of 96.19 MHz with CSA and 164.1 with MCSA on Altera FPLD and 69.4 MHz with CSA and 82.64 with MCSA on Synopsys cell-based IC design.

Grati et al. (2002) presented an optimized technique design for direct conversation wireless transceivers. They decimated the digital signal by multimode filtering structure compose of comb filters followed by a half-band and a FIR filter for both GSM and DECT standards. They tried to find the way to reduce complexity architecture. To decrease computational complexity and meet standard specifications a practical method introduced to compute filter stages relaxed specifications.

Henker et al. (1999), presented time-variant implementation of CIC-filters which circumvents the high intermediate sample rate. This time-variant implementation results in a linear periodically time-variant system (LPTV) which is completely equivalent to its original linear time invariant system (LTI) consisting of the interpolator and decimator be used by analysing the LTI system, while implementation the system as an LPTV system, avoiding the high intermediate sample rates of the LTI system. Anti-alias and anti image filtering were principally realized in that paper and was implemented in LPTV systems.

In 2005, Dolecek and Carmona changed the structure of the CIC filter and modified it by multistage structure. This structure consists of cascaded modified CIC and cascaded cosine prefilters. The first section is realized in non recursive form and other stages were designed recursively. This structure was defined to improve magnitude response of this filter due to cosine prefilters in higher rate. This prefilter also can be shifted in lower rate.

Kei-yang Khoo et al. (1998) proposed an efficient architecture for the first integrator stage in a high-speed CIC filter implemented using carry save architecture. The design is based on exploiting the carry propagation properties in a carry save accumulator. It will reduce the number of register to save power consumption. In this design a large number of full adder (FA) is replaced with half adder (HA). The result shows that the design reduces the number of registers in the integrator stages by 6.3 % to 13.5 % and replaces 18% to 42% of the FA.

Hyuk Jun Oh et al. (1999) used a simple Interpolated Second-Order Polynomial filter (ISOP) cascaded with a CIC decimation filter to effectively reduce the pass band distortion caused by CIC filtering with little degradation in aliasing attenuation. Their proposed decimation filter for a programmable down converter cascaded a CIC filter, an ISOP, modified half band filters (MHBFs) and programmable FIR filter. An optimization method was also presented to simultaneously design the ISOP and programmable FIR (Oh & Lee 1997). A design example demonstrated that this leads to a more efficient decimation filter implementation.

In 2003, Dolecek and Mitra proposed an efficient sharpening of a CIC decimation filter for an even decimation factor. The design structure consists of two main sections, a cascaded of the first-order moving average filters, and sharpening filter sections. The sharpening is moved to lower section which operates at the half of the high input rate and has only one half of the original sharpened CIC filter length. The result shows the gain responses of proposed CIC filter in the frequency range (0, 0.3) and dB scale (-200, 0). This New computationally efficient structure is well suited for sharpened CIC decimation filter.

Three leading semiconductor companies fabricated ICs with CIC decimation filters, namely Harris Semiconductor (now called Intersil) (Harris Semiconductor 1998; Harris Semiconductor 1990; Riley et al. 1991), Graychip Inc. (Texas Instrument 1996) and Analog Devices Inc. (Analog Devices 1998). Their functionality and performance are compared below.

In Harris' HSP50214 programmable DDC (Harris Semiconductor 1998), a five-stage CIC decimation filter is used, and the decimation ratio can be programmed from 4 to 32. The throughput can be as high as 52 MHz.

Harris' HSP43220 (Harris Semiconductor 1990; Riley et al. 1991) is a pure digital decimation filter. It composed of two modules, namely, a five-stage CIC decimation filter followed by a FIR decimation filter. The input word width is fixed at 16 bits, and the output at 24 bits. It accepts sample in 2's complement format with a sample rate of up to 35 MHz. The decimation ratio ranges from 10 to 1024.

In Gray chip's GC4016 Multi-Standard Quad DDC chip (Texas Instrument 1996), a five-stage CIC decimation filter is also used. The input data sampling rate can be up to 80 MHz, and the decimation ratio is programmable as any integer between 8 and 4K. Note: in this work, for throughput or data sampling rate, the unit is normally referred to as MHz.

Analog Device's AD6620 digital receiver (Analog Devices 1998) also uses a five-stage CIC decimation filter. The decimation ratio range is from 1 to 32 and the highest data input rate for the CIC decimator is 32.5 MHz.

Laddomada and Mondin (2006) presented the design of a novel decimation filter, especially suitable for sigma delta converter, which showed better performances in terms of both pass-band drop and quantisation noise rejection with respect to conventional CIC decimation filters. In particular, the combination of the sharpened filters and the modified CIC filters with the goal of increasing the rejection of the sigma delta quantisation noise around the folding bands and reducing the pass-band drop of the designed decimation filters has been designed with respect to CIC structures.

Hong- Kui Yang and W. Martin Snelgrove (1996) took another approach to implementing a CIC decimation filter. They decomposed the decimation ratio R into two factors, namely, $R_1 \times R_2$ (the parameters are integer). Instead of using one CIC filter to decimate the high speed digital signal, they used two CIC decimation filters: one CIC filter with a decimation ratio of $R_1$, and one with a decimation ratio of $R_2$. The implementation of the first decimator was based on polyphase decomposition. It has been shown that the polyphase components can be realized with only a small number of adders and shifters.

Kewentus et al. (1997) described a new architecture for the implementation of high order decimation filters. Proposed design combined the CIC multirate filter structure with filter sharpening technique to improve filter pass band response. They implemented programmable high-order decimation filter in 1.0 $\mu m$ CMOS technology. The chip can accommodate throughput rates up to 100 MHz for high performance.

In 2003, Uusikartano and Takala presented a power-efficient CIC decimator architecture for $f_s/4$ down converting digital receivers. The power saving was based on running all the integrator at half the sampling frequency, regardless of the decimation ratio. The implementation result shows that the system can work with maximum throughput of 124 MHz with the power consumption of 4.5 mW. (Note that reference CIC filters utilize 5.8 mW power supply) The proposed architecture can utilize any decimation ratio even a prime or fractional number.

In 1991, Riley et al. presented the architecture of an efficient high- decimation rate filter. A high decimation rate can extract narrow band signal intelligence from a wide band signal. The proposed system was multirate design consisting of a CIC section and a finite impulse response (FIR) filter. The function of the CIC section is to perform efficient high decimation rate filtering over the entire frequency band. The paper investigated the implementation of a 33 MHz high decimation rate filter and its hard ware implementation and performance. Riley reported the architecture and characteristics of an optimized decimating digital filter (DDF). The design can support decimation rates up to $2^{14}$ over a 16-bit 2's complement data field with 20-bit coefficient precision in single 33 MHz 84 pin low power CMOSIC.

## 2.5    SUMMARY

This chapter describes audio signal bandwidth and the necessity of using the oversampling technique to push quantization noise created by sigma delta modulator to higher frequency bandwidth. Utilizing the oversampling technique, the stringent requiring for anti-aliasing filter can be relaxed.

The chapter explains the reason of using the decimation filter to reduce sampling frequency rate and to get the audio signal from bandwidth. The CIC filter also rejects the quantization noise to achieve higher SNR.

The literature study presented above gives outline of the relevant fields that are required and used in the duration of this work. In next chapter, the study will focus on research methodology and the detailed explanation of the decimation filters.

# CHAPTER III

# RESEARCH METHODOLOGY AND DECIMATION PROCESS

This chapter explains the research methodology of the project followed by implementation workflow in detail. After understanding of overall system workflow, the applications of the CIC filter are discussed completely. It is continued by the system verifications with the express of the advantages of using multi-stage decimation filters. The oversampling and multi-stage decimation filter is briefly proposed and the system simulation is given by using MATLAB software.

## 3.1 PROJECT WORKFLOW

With the scope of work being successfully narrowed down to realistic goals, the design and development of the CIC filter is embarked up on. This overall research work involves the effort of simulation and VLSI implementation of high speed CIC filter to meet the required system functionalities. The design flow of the research work will be explained in the following section. Once the algorithm for the CIC filter is identified, a comprehensive MATLAB simulation was carried out to determine the functionality of the algorithm. The hardware coding (using Verilog) is written then followed by implementation using in ASIC or FPGA.

The literature reviews and documentation of project are done throughout the research process. Figure 3.1 illustrates the overview of the research in this work.

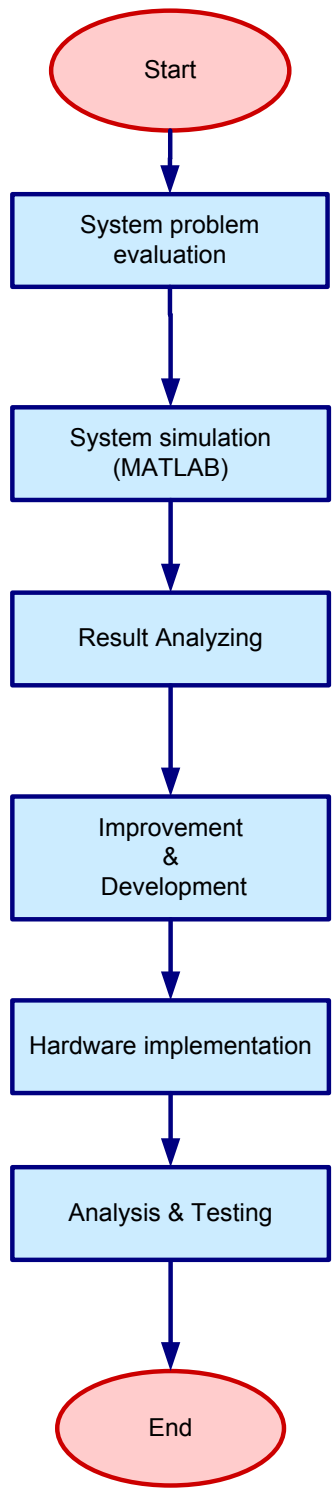

Figure 3.1 Workflow of the project

Figure 3.1 shows the overall strategy adopted in our research work. It consists of two components that are the system level simulation to determine the functionality of the CIC algorithm. The other component is the hardware implementation of the algorithm developed as Figure 3.2. In this figure, the detailed hardware implementation of the CIC chip is shown. The architecture of the CIC is implemented using Verilog code. The implementation started by research study base on efficient hardware design. The flowchart is followed by defining the algorithm, modification and the Verilog HDL implementation. The project is contributed to achieve the goals to enhance the previous decimation research work. The hardware code is transferred to UNIX machine to check the filter properties in gate level by CAD tools (Synopsys and Cadence). The workflow will go ahead if the design meets the desirable specifications and goes through the next stage which is the FPGA prototype and testing. In the same time, the design also is synthesized, timing verified and finally simulated at gate level on the 0.18 SilTerra and 0.35 MIMOS process technology. We also made a comparison between these two technologies in term of their performance (Speed, power consumption, area,…). Based on the filter architecture and design specifications, the hardware design flowchart of the CIC filter is proposed as shown in Figure 3.2.

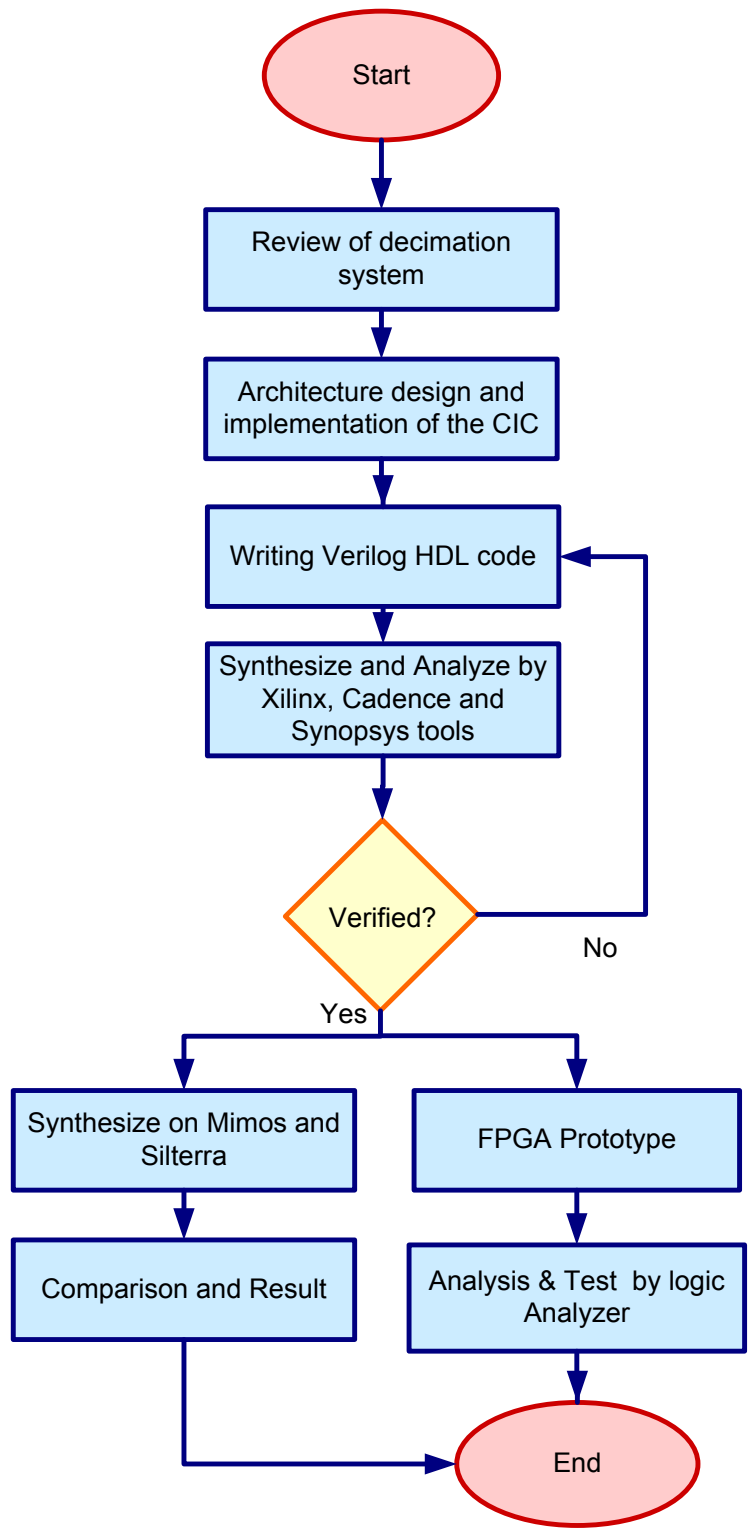

Figure 3.2 Hardware design methodology flowchart

## 3.2 APPLICATIONS OF THE CIC FILTER

The digital CIC filter is part of the decimation system in the DSP block. This block is located after the digital-to-analog converter. In the converter creates sampled data by sampling analog signal and the DSP block optimizes the sampled data. A block diagram of a typical sampled data DSP system for the Radio Frequency (RF) transceiver is shown in Figure 3.3.

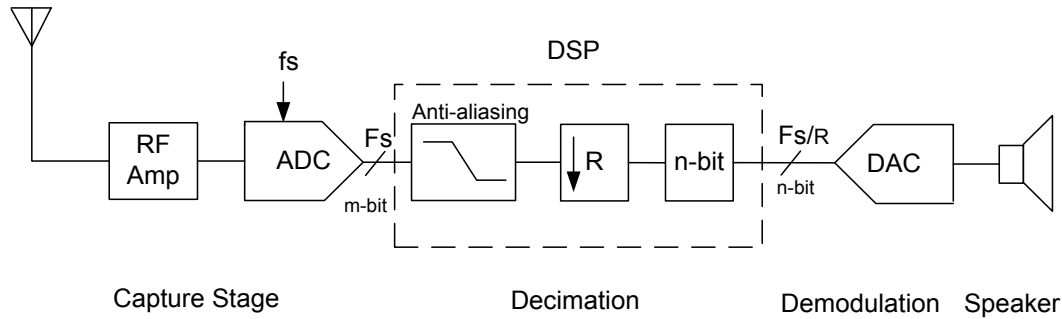

Figure 3.3 RF transceiver block diagram

In the RF transceiver, the analogue signal in coming (at the antenna) usually passes through a signal conditioning circuitry which performs functions such as amplification, the attenuation and filtering. After amplifying RF signal, it is digitized by the ADC converter. The signal to the ADC is continuously sampled at rate equal to $f_s$, and thus the ADC presents a new sample to DSP at this rate. Normally the sigma delta modulator ADC with will sample RF signal with extremely high sampling rate. This property makes the converter very costly in terms of processing requirements. Although sigma delta modulation to some extend is an analogue to digital converter, it does not produce a standard base representation for the analogue data (Fowler 1995). Therefore some computation on sigma delta modulator must be performed in order to convert it into the actual digital data which represents the analogue signal. This digital signal needs to be down-converted, filtered and decimated by the decimation digital filter before being used in audio and radio signal. Decimation converts m-bit one samples into n-bit one sample and filtering is used to reject quantization noise (Figure 3.3). The decimation filter eliminates frequency components beyond baseband and reduces the sampling rate down to $\frac{f_s}{R}$. The selection of the bandwidth and all other processing would be handled in the digital domain. The low pass filter is required to remove unwanted signals outside the bandwidth of interest and prevent aliasing. In order to maintain real-time operation, the DSP must perform all its required computation within the sampling interval and present an output sample to the DAC before the arrival of the next sample from the ADC. Finally the DAC is required so that the analog signal (voice) can be obtained (Hsu Kuan Chun Issac 1999; Felman 1997, 1998).

Besides being used in digital transceivers, decimation filters are widely used in many systems involving multirate processing and conversion. Those applications include frequency shift keying (FSK) modems, speech/audio coding and image/video coding (Vaidyanathan 1993). The comb filter which is part of the CIC filter is used in audio engineering to achieve special sound effects. It is also used in the construction of digital reverberation processors, where they represent the effects of multiple reflections of a sound wave off the walls of a listening space (Orfanidis 1996). In digital TV, comb filters are used in the case when both signals are periodic and must be separated from each other. The comb filters are used to separate the luminance (black & white) and chrominance (colour) signals from composite video signal and also to reduce video noise (Orfanidis 1996).

## 3.3    MULTI-STAGE DECIMATION FILTER

Oversampling technique limits the possibility of the aliasing present in the base-band. Thus the input signal could be sampled at higher rate to prevent aliasing. Later on, the sampled signal is decimated and filtered by using a digital decimation low-pass filter. The combination of the filtering and decimation prior to performing another filtering operation on the data is called multi-stage digital decimation filtering. It takes the advantages of having enabled the processing of oversampled signals, and continues to drive down cost and power consumption by decimation. By decimating in multiple stages, the complexity of the whole filter is reduced and the subsequent filters operate at lower sampling rate, further reducing the power consumption. Whereas single stage decimation filter use large number of filter coefficients which is not suitable for power-efficient implementation.

The multi-stage digital decimation filter also achieves computational savings by performing sampling rate conversion and filtering over several stages while utilizing efficient implementation structures such as polyphase filters (Crochiere 1983). The benefit of the multi-stage approach is that economical due to lower order filters operate at the higher sampling rate and higher order filters operate in lower sampling rate. The architecture of the filter is designed to minimize arithmetic and hardware complexity. By implementing the filter as a cascade of several linear-phase stages, the ratio of the input rate to the width of the transition band is dramatically reduced for each of the individual stages (Brandt & Wooley 1994; Fujimoni et al. 1997). This multi-stage decimation filter can achieve both greatly reduced filter lengths and computational rates as compared to standard single rate filter design and thereby provide a practical solution to an otherwise difficult problem.

In most existing implementations for audio applications, the filters have been implemented using several FIR filter stages where the first stage is normally a CIC filter. Two major design challenges of finding the proper number of stages as well as the appropriate sampling rate reduction per stage of decimation filter have so far been the topic of discussions (Mortazavi et al. 2005). The design architecture for audio application is based on achieving a proper frequency response, a low power design and a high SNR as well. Figure 3.4 shows the third order sigma delta modulator followed by the fifth order decimation filter. For this research, focus will be the VLSI implementation of the CIC filter. The other blocks in this figure are incorporated so that the design of the CIC can be carried out. Considering the multi-stage decimation filters operate at sampling frequency of 6.144 MHz with the decimation ratio of 128. When multi-stage decimation filters operate in the 48 kHz sample rate, the modulator output is decimated by $128 f_s$. To achieve this decimation ratio, the CIC filter provides 16x down-sampling. The two half band filters and the droop correction filter provide the remaining 8x down sampling factor. Figure 3.4 shows four cascade stages of decimation filter for audio application.

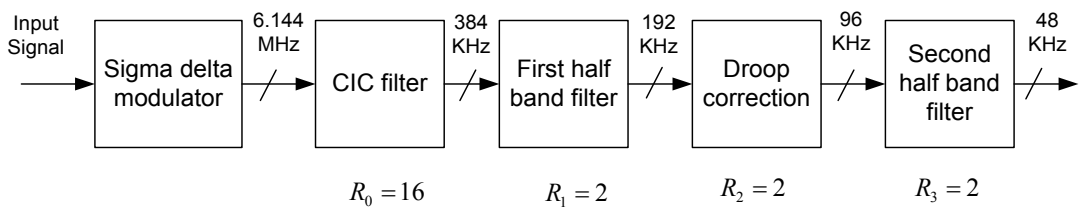

Figure 3.4    Third-order sigma delta modulator followed by the multi-stage decimation filters

### 3.3.1 THE SIGMA DELTA MODULATOR

According the above explanation, the analogue input signal is oversampled by a factor of 128. It passes through the third order sigma delta modulator. This modulator transforms a band limited input analogue signal into 5-bit digital output signal. The clock frequency of sigma delta modulator is much higher than the highest frequency of the input signal. To get a sufficiently high SNR ratio from the digital output of the audio signal (at least 16 bit) when truncation is applied, or 20 bit without truncation, is required. Among oversampled and decimation system above, the sigma delta modulator with 5-bit quantizer in its architecture provides the 5- bit resolution.

Signal conversion in the sigma delta modulator, is based on quantizing the change in the signal from sample to sample rather than absolute value of the signal at each sample. There is digital-to-analogue converter (DAC) in the sigma delta modulator structure called zero-order hold block in simulation which converts the digital signal back to analog form. It is then subtracted from the analogue input. The difference signal, which is the "delta" part, is accumulated in to the integrator which is the "sigma" part and provide a local average of the input the feedback loop causes the quantization noise generated by the ADC to be high pass filtered pushing its energy towards the higher frequencies and a way from the relevant signal band .

The multi-stage decimation stage reduces the sampling rate to 48 kHz. During this process, it removes the high frequency quantization noise that was introduced by the feedback loop and removes any undesired frequency components that were not removed by the simple analog prefilter.

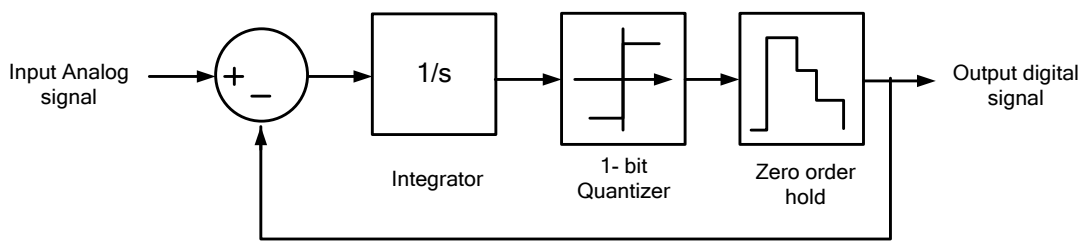

Figure 3.5  Matlab simulation of the first-order sigma delta modulator

First-order sigma delta modulator simulation is shown in Figure 3.5. The quantization noise is one limiting factor for the dynamic range of an ADC. This error is actually round-off error that occurs when an analog signal is quantized. The mean square value of the noise, $e_{rms}^2$ produced, by quantization in the sigma delta modulator (assuming that the signal value is in the range of one step of the quantized value with an equal distribution) is given by:

$$e_{rms}^2 = \frac{1}{\Delta} \int_{-\frac{1}{\Delta}}^{+\frac{1}{\Delta}} e^2 de = \frac{\Delta^2}{12} \tag{3.1}$$

where $\Delta$ is the quantization interval or one LSB size.

The equation above assumes that the quantization noise is white. The assumption is not true in general but it gives a good approximation when the input signal is busy.

When a quantized signal is sampled at frequency $f_s = \frac{1}{T}$, all of its power folds into the frequency band $0 \leq f < f_s$. Then, if the quantization noise is white, the spectral density of the sampled noise is given by:

$$E(f) = \sqrt{\frac{e^2}{\frac{f_s}{2}}} = e\sqrt{\frac{2}{f_s}} \tag{3.2}$$

$E(f)$ has inverse proportional to quantization bits. The more quantization bits, the lower $E(f)$ is and the less noises alias into the signal band. To design high resolution ADC, the number of quantization noise should be increased which lead to the complex circuit design.

We can use this result to analyze oversampling modulators. Consider a signal lying in the frequency band 0≤*f*≤*fB*. The oversampling ratio (OSR), defined as the ratio of the sampling frequency $f_s$ to the nyquist frequency $2f_B$, is given by the integer:

$$OSR = \frac{f_s}{2f_B} \tag{3.3}$$

Hence, the in-band quantization noise will be given by (Jarman 1995; Hsu Kuan 1999):

$$n_\circ^2 = \int_0^{f_B} E^2(f)df = e^2 \cdot \frac{2f_B}{f_s} = \frac{e^2}{OSR} \qquad (3.4)$$

That means higher oversampling ratio, will give higher SNR and also higher resolution in bits.

The MATLAB simulation of the third order sigma delta modulator is shown in Figure 3.6. The multi-stage decimation filter introduced after the modulator, not only removes the noise at higher frequencies but also reduces the frequency of the signal and increases its resolution.

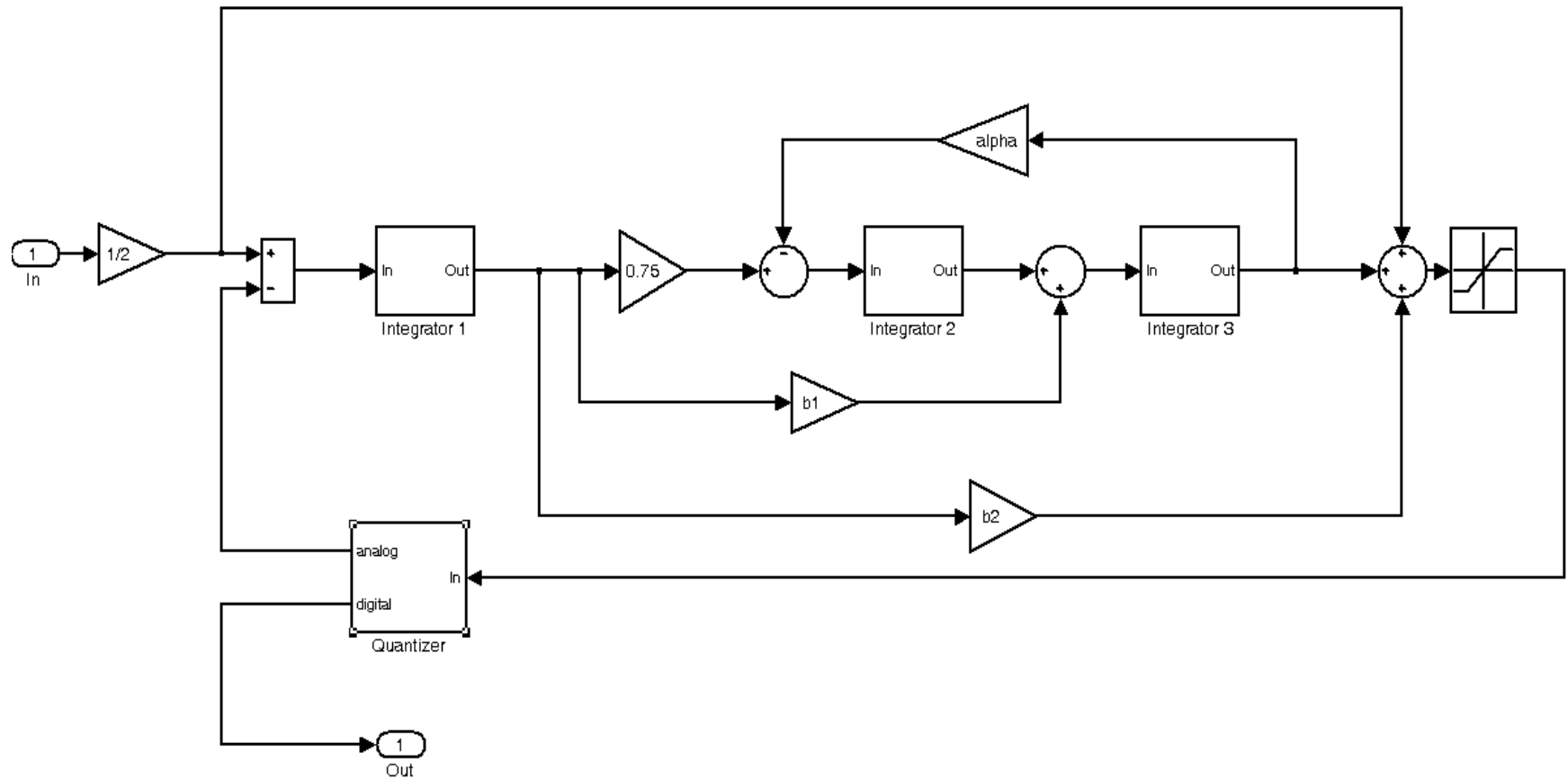

Figure 3.6 MATLAB simulation of the third-order sigma delta modulator

### 3.3.2 CASCADED INTEGRATOR COMB FILTER (CIC)

The fifth-order CIC filter was designed and implemented as a decimator. This filter is supposed to drive down the sampling frequency by the factor of 16 and reject the quantization noise to out of the desired bandwidths. The CIC filter provides 25 bit resolution to increase the SNR more than 140 dB. The architecture of digital CIC filter and its utility will be described in next chapter in detail.

### 3.3.3 FIRST AND SECOND HALF BAND FILTER

The other important components that must be designed are First and second half band filters (Figure 3.4). One issue encountered in designing these filters is the computational complexity. Since the transition band width is inversely proportional to the filter length, the number of multiplier required. To realize a sharp filter is large and thus the corresponding complexity becomes very high. Thus an efficient second stage filter can be selected from a class of filters called the half-band. These are even order linear phase FIR filters having a special property that every odd sample of their impulse response is zero, except for the sample at the centre of the impulse response which has a value of 0.5. Thus, the number of computations required to implement these filters is reduced by approximately half. Due to the known fact that half-band filters reduce the computational complexity of general filters with the same order, half band FIR filters are widely used (Min-Chi Kao & Sau Gee Chen 2000).

The standard architecture of half band filter shown in Figure 3.7. These filters divide the sampling frequency of the discrete time system into two. The FIR filters are often used as half band filters due to their linear phase response. It is used in decimation filtering because half of their time domain coefficients are zero. This means it can achieve the performance of an M-tab FIR filter while only paying the computational price of approximately M/2 multiplications per filter output sample.

The two cascaded half band filters with the filter order eight and eighty respectively are used to reduce the remaining sampling rate reduction to the Nyquist rate. First and second half band filter makes the frequency response more flat and sharper similar to the ideal filter specially second half band filter due to higher order of the filter (N=80), has most effect to make the frequency response sharp. The transition band of second half band filter is less than 5 kHz.

According to Figure 3.7, all the odd coefficients of half band filters are zero with the exception of last one the (0.5). It makes the two filters efficient for 2:1 decimation ratio and reduces the computational complexity by near 50% as compared to direct form filter architecture.

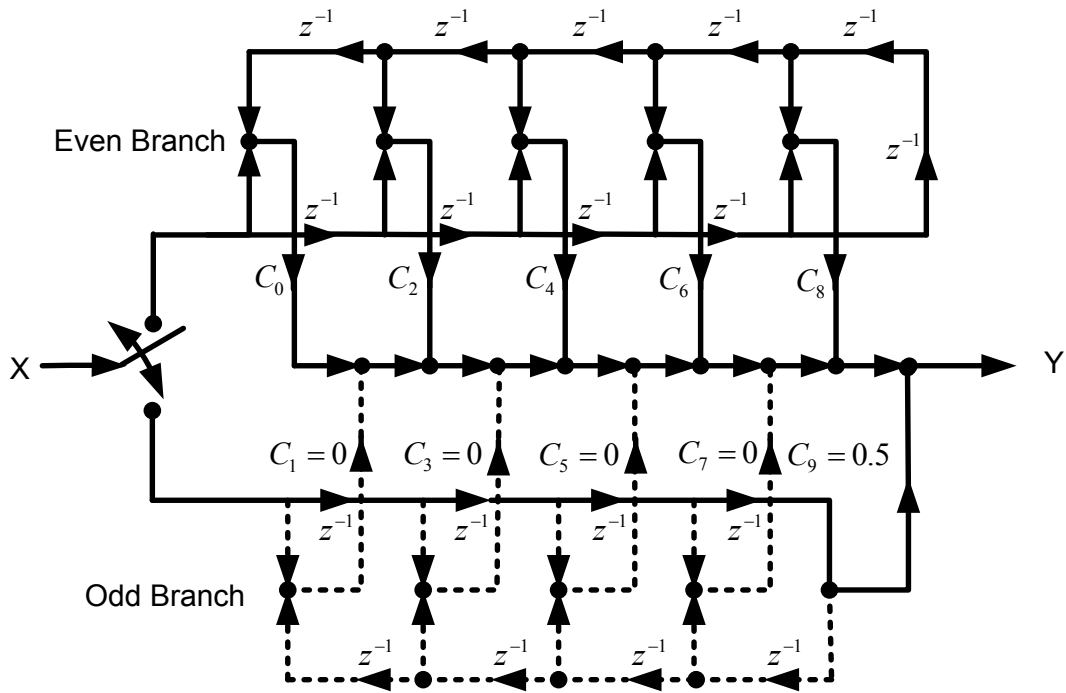

Figure 3.7 The half band filter structure

The broken line in the half band filter structure is removed leaving a very simple odd branch consisting of only delay terms and a single coefficient multiplication.

A half band filter is usually specified by normalized pass band edge frequency, $f_p$, the stop-band edge frequency, $f_s$, the maximum pass-band magnitude ripple $\delta_1$ and the maximum stop-band magnitude ripple $\delta_2$. Figure 3.8 shows the frequency response of the simple half band filter.

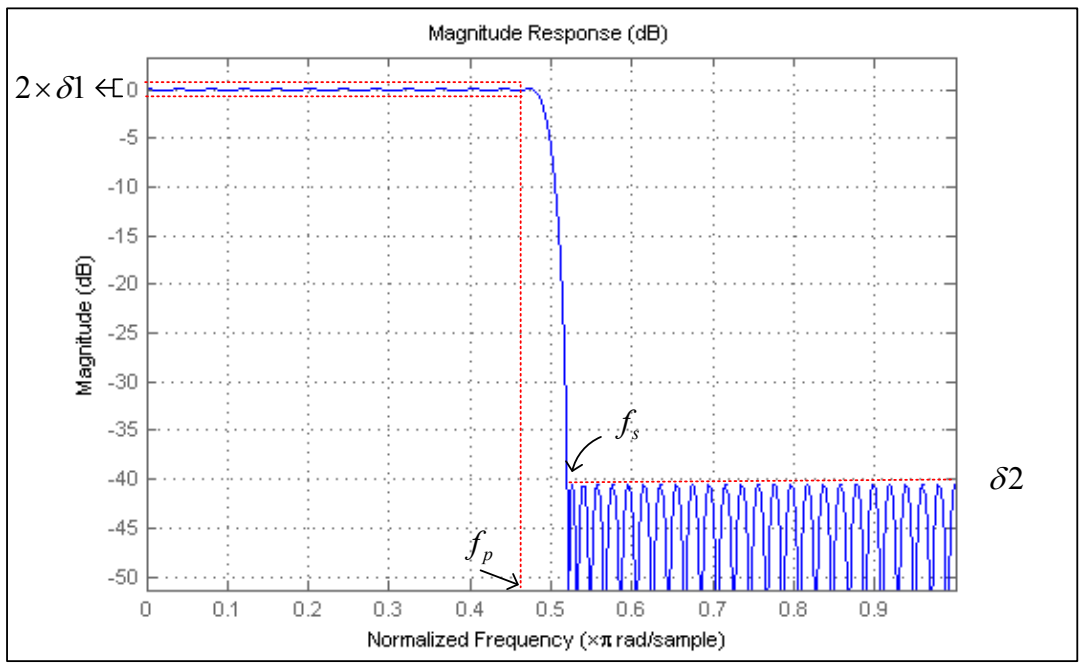

Figure 3.8 Normalized half band filter frequency response

The implementation of the multi-stage decimation filters take the advantage of operating higher order of half band filter in lower sample frequency in the decimation chain. This has resulted in the low power consumption. The transfer function of the first half band filter is given as follow:

$$H(z) = h_0 + h_2 z^{-2} + h_3 z^{-3} + h_4 z^{-4} + h_6 z^{-6} \quad (3.5)$$

The Matlab simulation of the first and second half band filters has been done a as low pass filter decimator to enhance decimation system. Every Y sample is calculated by even and odd coefficients in the half band filters. The tapped-delay lines in the even and odd branches in half band structure doubled back to exploit the symmetric impulse response of the linear phase filter and reduced the number of multiplications by a factor of two (Oppenheim & Schafer 1975). The MATLAB simulation result shows that the pass-band, stop-band frequency and transition band are 32, 170 and 138 kHz for first half band filter and 21.77, 26.53 and 4.76 for second half band filter respectively.

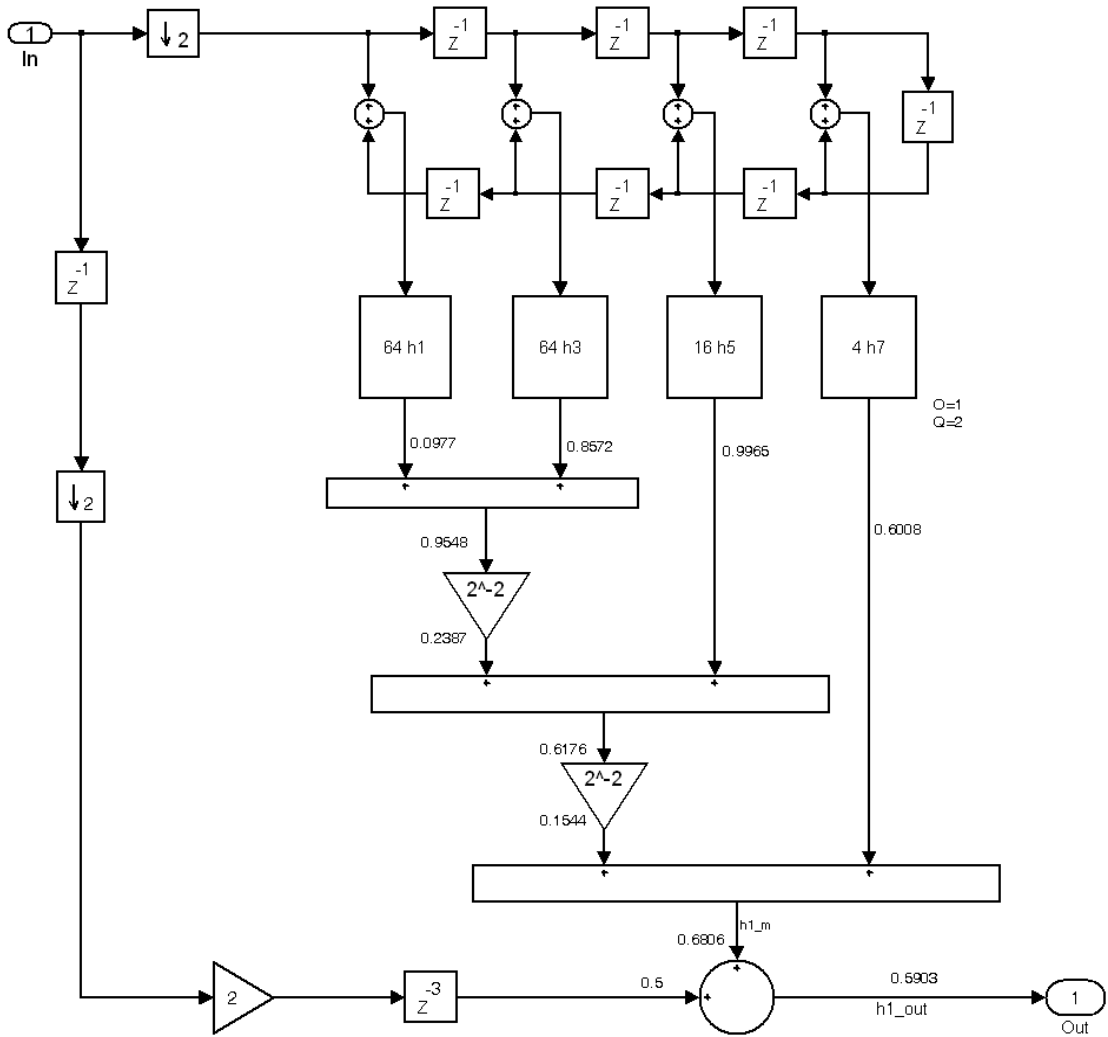

Figure 3.9 MATLAB simulation of the first half band filter

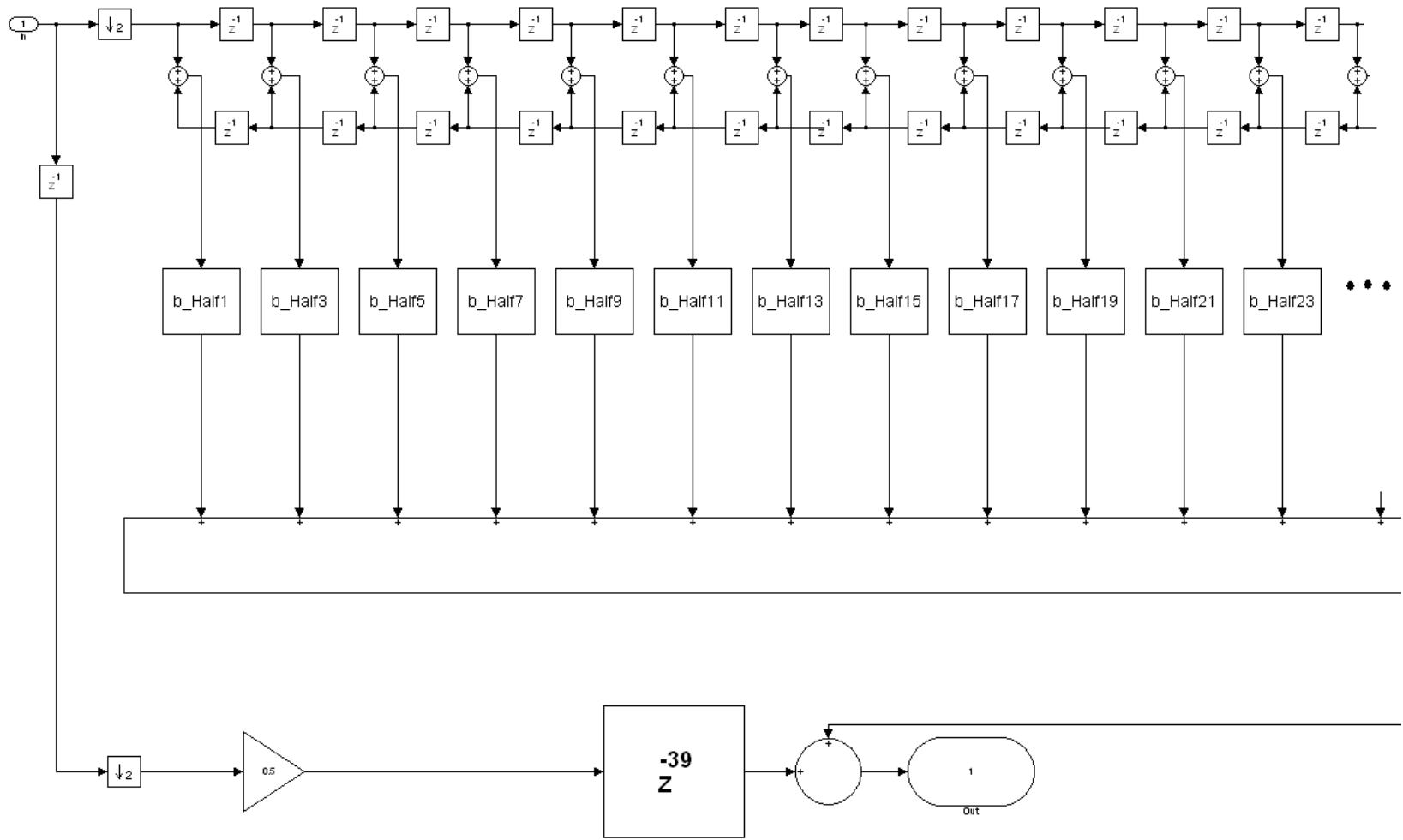

Figure 3.10 MATLAB simulation of the second half band filter

### 3.3.4 DROOP CORRECTION FILTER

The CIC filter in the decimation chain has a desirable attribute on the frequency response which is droop in pass-band of the filter. Typically narrow transition-region and flat passband are not provided by the CIC filter alone, with their drooping pass-band gains and wide transition regions. Figure 3.11 shows the droop which is created by the CIC filter.

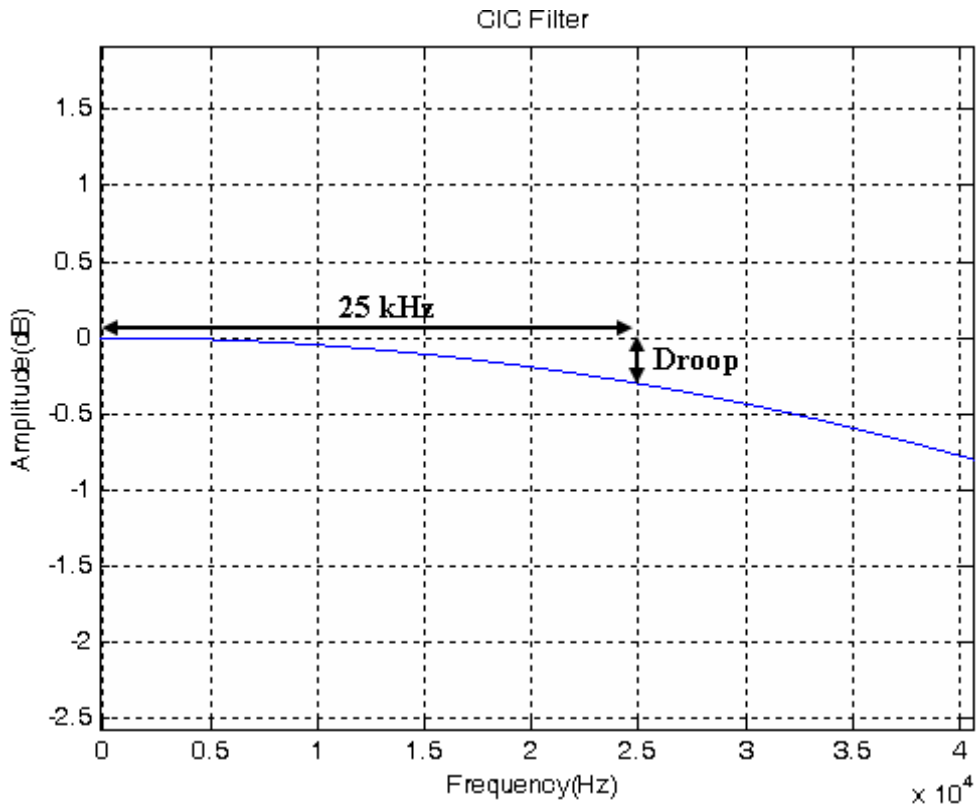

Figure 3.11 Droop in the CIC pass-band frequency

The droop which the CIC filter creates is dependent on decimation factor, R. For compensating this droop a FIR filter placed after the CIC filter. The FIR filter frequency response is ideally an inverted version of the CIC filter pass-band response. So it compensates amplitude droop which is created by the CIC filter and makes overall frequency response flat and desirable. The purpose of using this filter is twofold: 1) to correct the passband droop introduced by three preceding filter and 2) To perform sharp band pass filtering and attenuate remaining out of band quantization noise. The FIR droop correction filter is consisting of down sampler block, double delay tab in each even and odd branches and adder cells. As shown in Figure 3.12, the output of the adder is multiplied by filter coefficient.

In this project, 14-order FIR filter is simulated to decimated and compensate droop in frequency response. The pass-band frequency, stop-band frequency and transition band in droop correction filter are 32 kHz, 70 kHz and 38 kHz respectively. The droop correction filter is simulated by MATLAB software and shown as below.

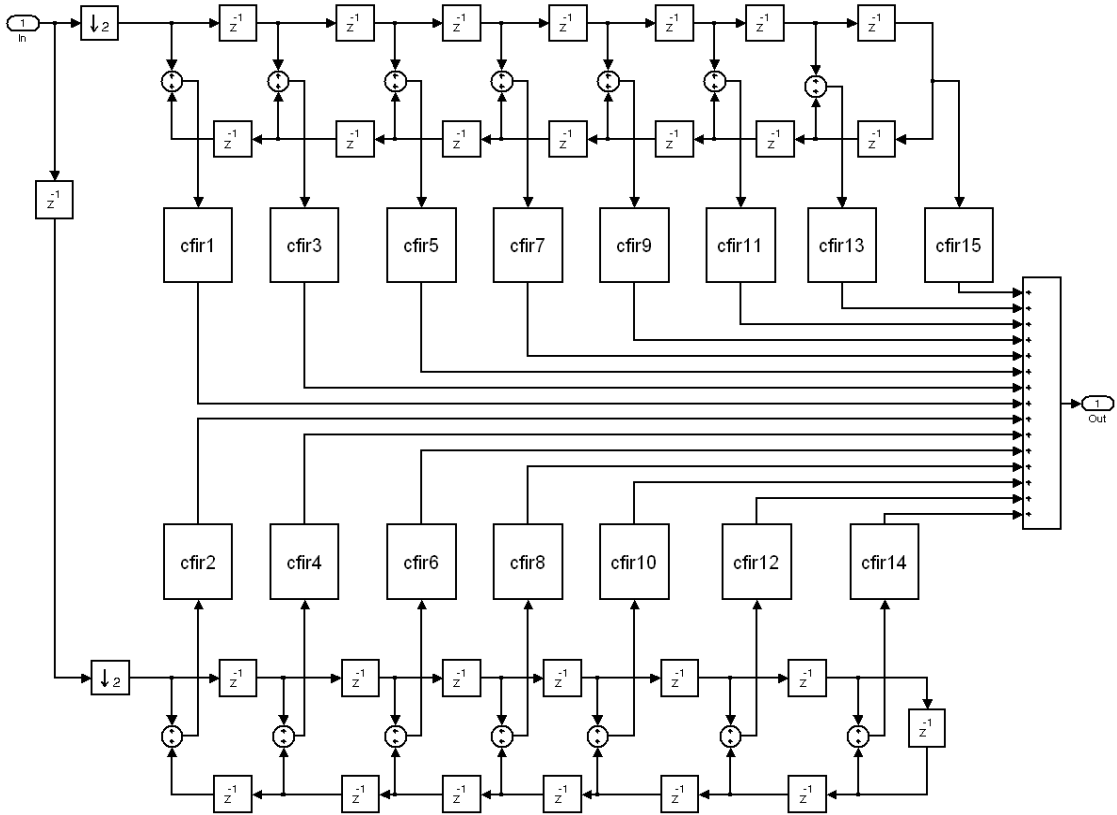

Figure 3.12 MATLAB simulation of the droop correction FIR filter

Several techniques exist for compensating for the droop in the pass-band of the CIC filter frequency response. An FIR filter placed after the CIC filter could be designed to correct the droop for a particular decimation factor. (Kwentus et al. 1997; Kaiser & Hamming 1977; Stephen & Stewart 2004).

The correction droop FIR filter compensates for the -0.25 dB passband droop 22 kHz. This filter operates at 1/32 the input sample rate of 6.144 MHz and decimates by two. Figure 3.13 shows the combined response of the CIC filter coupled with the FIR compensator, indicating the pass-band droop correction of the CIC filter.

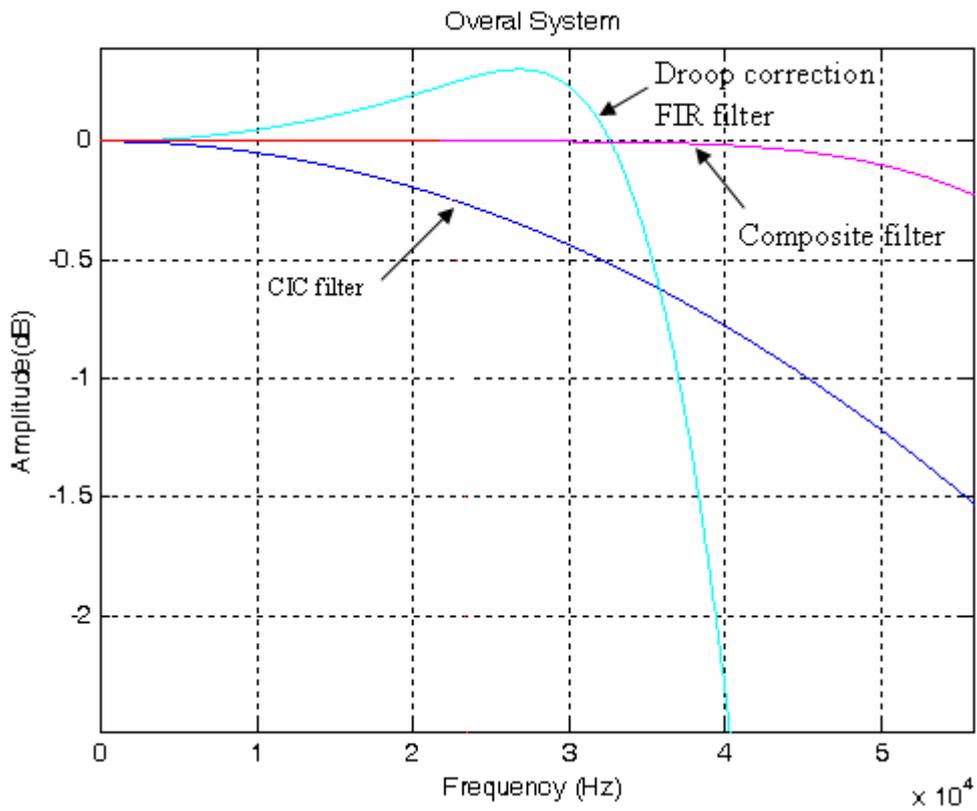

Figure 3.13  Droop compensation by the 14-order FIR filter

**3.4   SUMMARY**

The chapter has presented the research methodologies adhered to in this research including research procedure, system design procedure and tool applied. It also has managed to describe the application of the CIC filter which is the subject of this research. This chapter is continued by the overview of multistage decimation filters compared to single stage decimation. Finally the principle definition of the modulator and the decimation filters was described in brief and simulated by MATLAB software.

# CHAPTER IV

# THE CASCADED INTEGRATOR COMB (CIC) FILTER AS DECIMATOR

After reviewing the overall system in previous chapter, the definition of the digital filter as well as to present the basic principles of the decimation CIC filters with focusing on their properties is described in this chapter. The entire filter components are described separately and analyzed in frequency domain. The comparison is made between a conventional decimator and the CIC filters with the explanation of the CIC filter advantages. Additionally the functionality of a simple CIC filter is evaluated and proposed in chapter IV.

## 4.1 REVIEW OF THE BASIC DIGITAL FILTER

The digital filter operates on digital sampled signal. Implementing using digital filter is preferred as compared to the analogue filter since the digital filter has the advantages of high performance, low implementation cost and long term stability. Additionally, digital filter can be controlled easier than the analog filter.

The digital filter equation is given by:

$$y(n) = \sum_{k=-\infty}^{\infty} h(k) \cdot x(n-k) \qquad (4.1)$$

where $h(k)$ is the filter coefficient. For casual filter, all coefficients all are zero for $k < 0$.

For the FIR filter, the number of the filter coefficients are limited, whilst for the IIR filter, it has unlimited number of the coefficients. The equation (4.1) can be written as:

$$y(n) = \sum_{k=0}^{L-1} h(k) \cdot x(n-k) \qquad (4.2)$$

Figure 4.1 shows the structure of the digital FIR filter which implying equation (4.2).

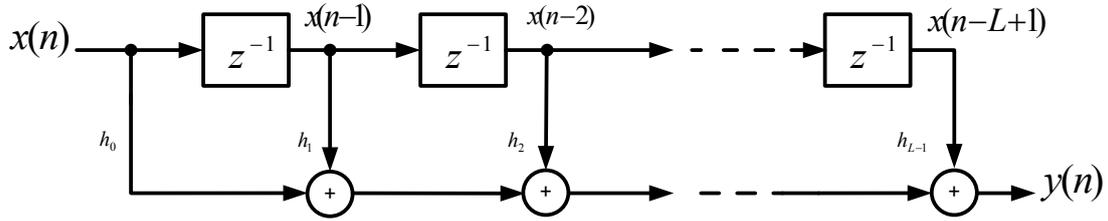

Figure 4.1 The digital FIR filter structure

High value of L will make the filter sharper and close to the ideal filter at the expense of filter cost and a more complex filter structure. The digital filter can be introduced in the z domain as below:

$$Y(z) = H(z) \cdot X(z) \qquad (4.3)$$

where $H(z)$ is the transfer function of the filter and z can be replaced with $z = e^{j\omega\tau}$ in frequency domain given by:

$$H(z) = \sum_{k=0}^{L-1} h(k) \cdot z^{-k} \qquad (4.4)$$

$|H(e^{j\omega\tau})|$ is called the magnitude response of the filter and it is obtained by filter transfer function. The transfer function $H(z)$, will also give the phase information of the filter. The filter is stable when its phase response is linear.

Unlike the FIR filter, the IIR filter has a feedback loop and infinitely long pulse responses. The equation of the IIR filter can be expressed as below:

$$y(n) = \sum_{k=0}^{L-1} h(k) \cdot x(n-k) + \sum_{k=1}^{M} a(k) \cdot y(n-k) \qquad (4.5)$$

Thus the transfer function in the z domain can be written as equation (4.6).

$$H(z) = \frac{\sum_{k=0}^{L-1} h(k) \cdot z^{-k}}{1 - \sum_{k=1}^{M} a(k) \cdot z^{-k}} \qquad (4.6)$$

The block diagram of the IIR filter with feedback loop is shown in Figure 4.2.

Although the length of the IIR filter is infinite, Yiqun Xie (1998) has shown that it can be implemented through an FIR subfilter with appropriate feedback.

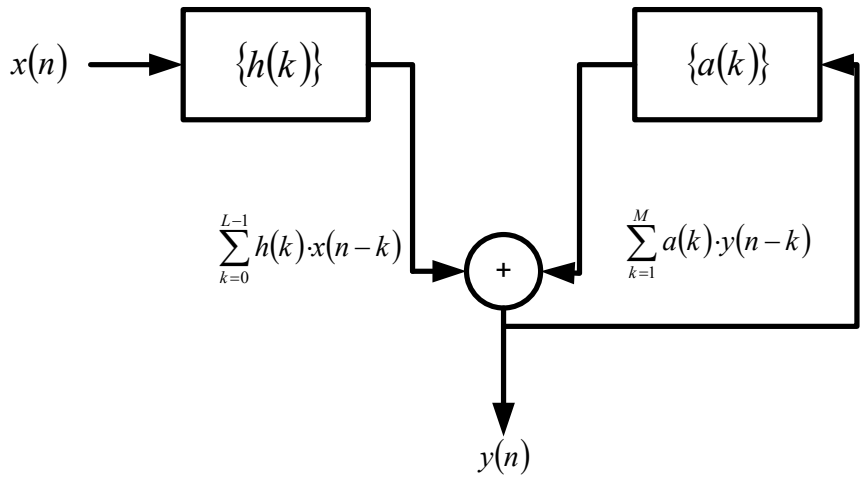

Figure 4.2 The digital IIR filter constructed from the FIR sub-filter

## 4.2 CONVENTIONAL DECIMATION FILTER

As mentioned before, a decimation filter is required after sigma delta modulator to remove out-of-band components of the signal and noise and resample the signal at the nyquist rate. According to Figure 3.4, the output of the modulator provides the input to the decimation filter, with additional quantization noise at higher frequencies.

In the past, a decimation filter was used to reduce the sampling rate. Figure 4.3 shows an example of the traditional decimation filter. These filters were widely used as decimator before the introduction of the decimation filter based on the cascaded integrator comb filter.

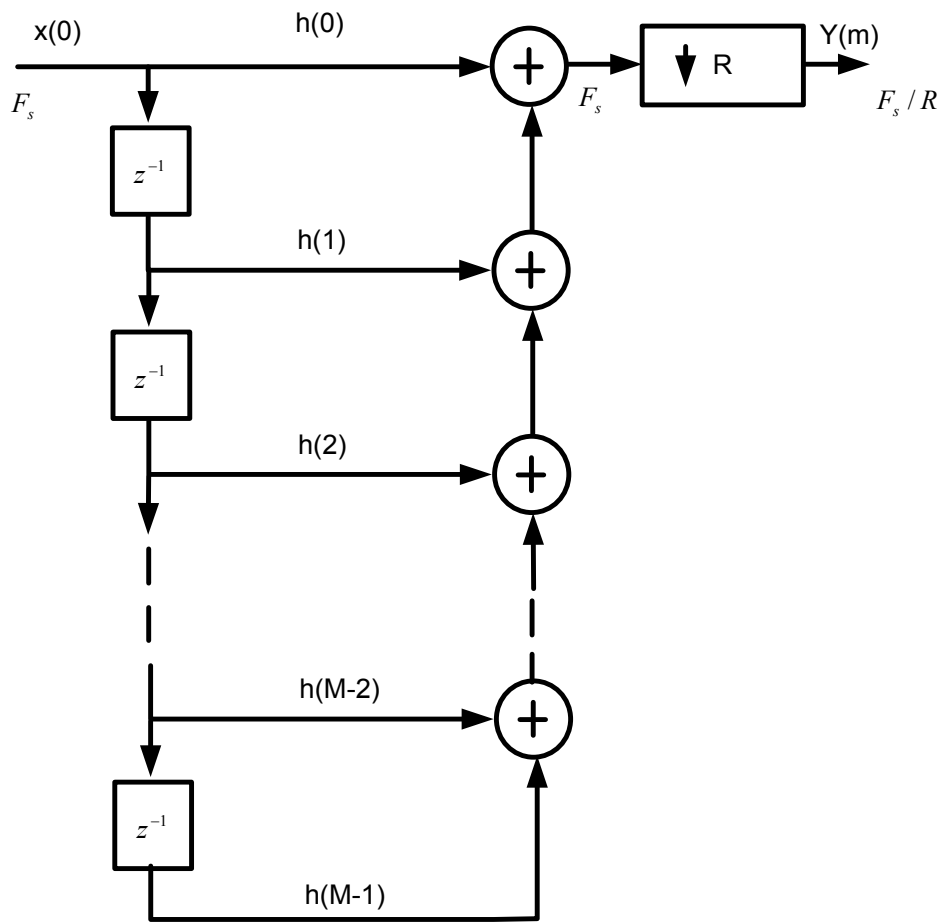

Figure 4.3 The low-pass conventional decimation filter (Yang 2001)

The structure of the decimation filter in Figure 4.3 consists of the low pass filter and a down sampler blocks. In order to produce one sample at the data output when the decimation ratio is R, $R \times M$ multiplications are required. Thus the number of calculation is increased and hence limiting the high speed and significant waste of the power consumption. To improve the decimation system, Proakis & Manolakis (1996) introduced polyphase filter structure for high speed purposes. But the implementation increases the complexity. Then in 1981, Hogenauer has developed an alternative to the polyphase filter that has increased efficiency because it does not required multiplier.

**4.3  STRUCTURE OF CASCADED INTEGRATOR COMB FILTER (CIC)**

The process of down-sampling is called decimation (Dongwon see et al. 2000; Rajic & Babic 2004). On the other hand up-sampling is called interpolation (Schafer & Rabiner 1973; Yaroslavsky & Chernobroadov 2003; Seidner 2005). In audio application, the sampling rate of the sigma delta modulator needs to be decimated and filtered by the filters. The Cascaded Integrator Comb (CIC) filter is used both in the decimation and interpolation. Figure 4.4 shows a complete oversampling system that consists of the sigma delta modulator and the CIC filter. Here the CIC filter is a decimator. In this work the 5-bit output of the modulator produced by the 5 bit quantizer (with the sampling frequency of 6.144 MHz) is inputted to the decimation filter. Figure 4.5 shows the internal block diagram of the CIC filter.

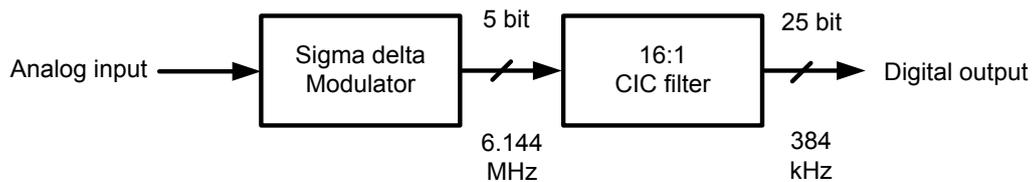

Figure 4.4  Location of the CIC filter with its effect on sampling frequency

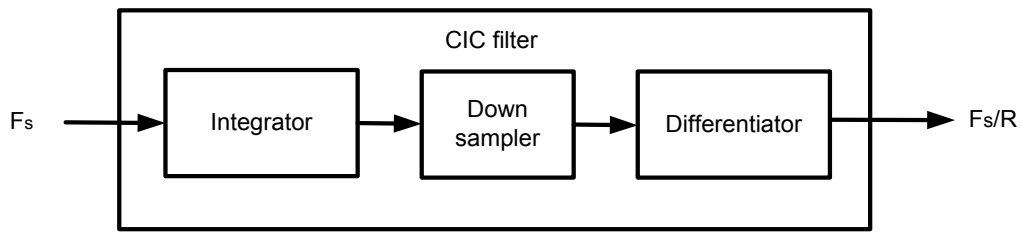

Figure 4.5  Block diagram of the CIC filter

The CIC filter has three basic tasks:

1) Reject the quantization noise to remove out of frequency bandwidth. As described before, sigma delta modulator creates some quantization noise in the baseband. This noise has direct effect on the SNR and can degrade the quality of audio signal.

2) Reducing the sampling frequency or decimation (Babic et al. 2001; Babic & Renfors 2001; Chu & Burrus 1984). Sampling frequency of the sigma delta modulator should decimate down to the Nyquist rate which minimizes the amount of information for digital signal processing. This important task is carried out by the CIC filter.

3) Anti-aliasing (Hogenauer 1981). The last and important task of the CIC filter is to prevent the aliasing that result from the down sampling.

The CIC filter has the advantage that it only has the adder and subtraction and not multiplier. This computational efficiency makes the CIC filter attractive in the DSP applications for low noise. The CIC filters are useful for a narrowband desired signal, where the nulls of the filter are wide enough to protect the desired pass-band from aliasing distortion. Additionally its architecture has a regular structure and little external control or complicated local timing is needed.

Due to all filter coefficients equal to unity, this filter does not require storage for saving filter coefficients. Figure 4.6 shows the block diagram of one stage CIC filter in the z domain.

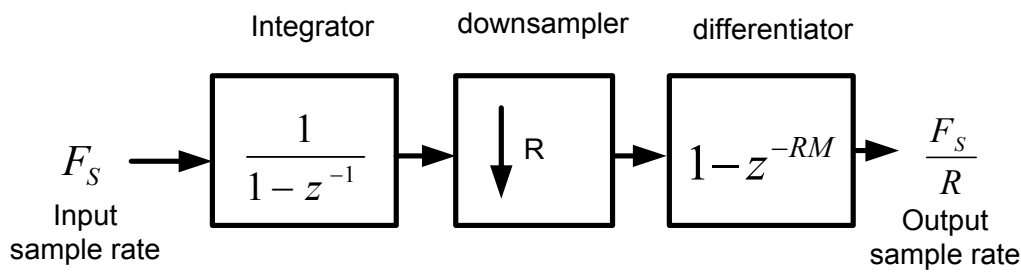

Figure 4.6 CIC filter structure in z domain

The CIC filter consists of three main blocks which are the integrator, the differentiator and the down sampler as shown in the figure. Each block will be described in the next section.

### 4.3.1 INTEGRATOR

The Integrator is the first block of the CIC filter. This block consists of adder and one unit delay register which are connected as one-stage accumulator. The equation of integrator with the assumption of $M$ is one unit delay in discrete time domain is:

$$y[nT_s] = y[(n-1)T_s] + x[nT_s] \qquad (4.7)$$

Figure 4.5 shows the basic structure of integrator in discrete time domain.

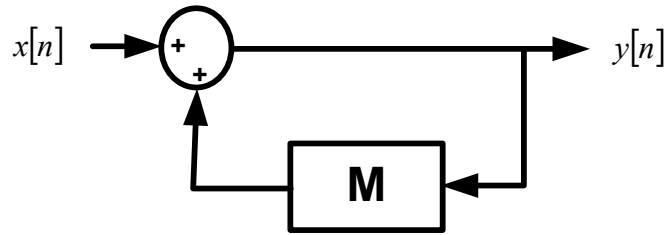

Figure 4.7  Integrator block diagram in discrete time domain

Integrator block can also be defined in the z domain. As for the transfer function 4.8, the integrator has pole in $z=1$ which is the reason for infinite gain at DC (Hogenauer 1981). Due to the single pole at $z=1$, the output can grow without bound that have the tendency to grow without bound that can result in the overflow in the registers. In other words a single integrator by itself is unstable. Figure 4.8 shows the frequency response of the integrator with infinite gain (in DC).

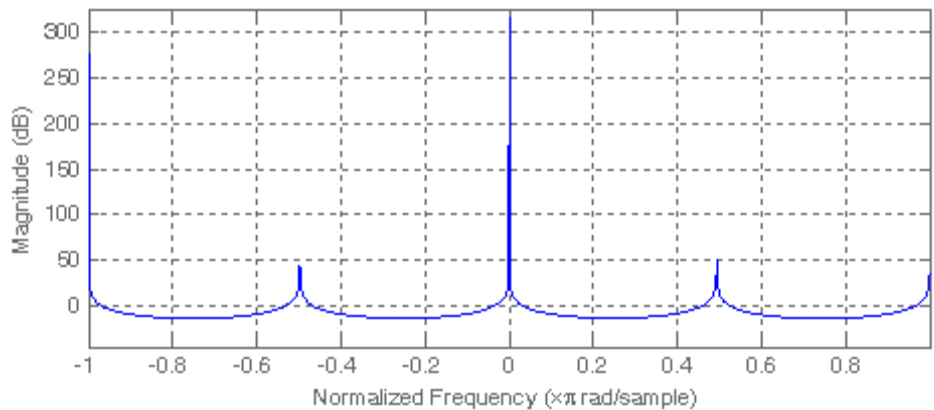

Figure 4.8  Integrator frequency response when in single pole $(z=1)$

Equation 4.8 shows the transfer function of the integrator in the z domain.

$$H_I(z) = \frac{y[z]}{x[z]} = \frac{1}{1-z^{-1}} = \frac{z}{z-1} \tag{4.8}$$

The integrator modelled in the z domain is shown in Figure 4.9. As seen in the Figure, the integrator has low pass structure to push the noise to the high frequency band.

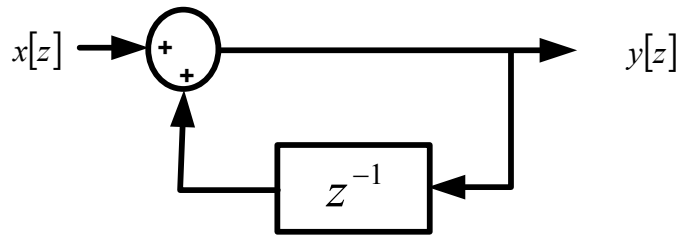

Figure 4.9 Integrator block diagram in z domain

The integrator is an IIR filter with a unity feedback coefficient (Chandel 2005). Using the equations for a single pole integrator system (Oppenheim & Schafer 1989), it can be shown that:

$$\left|H_I(e^{j\omega})\right|^2 = \frac{1}{2(1-\cos\omega)} \tag{4.9}$$

$$Arg\left[H_I(e^{j\omega})\right] = -\tan^{-1}\left[\frac{\sin\omega}{1-\cos\omega}\right] \tag{4.10}$$

$$grd\left[H_I(e^{j\omega})\right] = \begin{cases} \text{Undefined} & \omega = 0 \\ -\frac{1}{2} & \omega \neq 0 \end{cases} \tag{4.11}$$

The pole-zero plots in real and imaginary domain and the magnitude response in frequency domain for the integrator are shown in Figures 4.10 - 4.13.

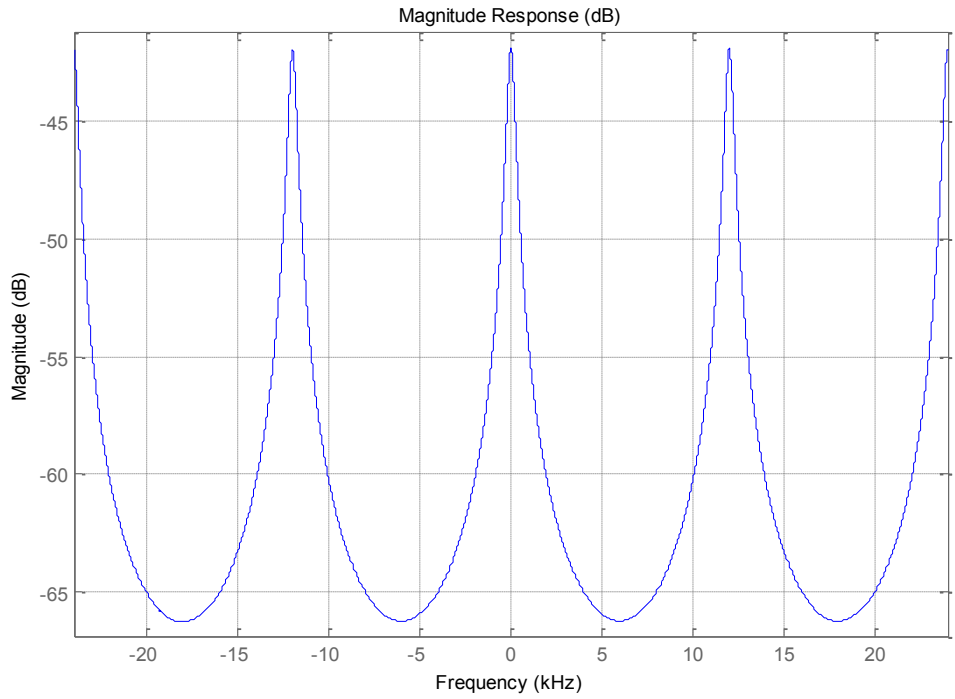

Figure 4.10  Magnitude spectrum of integrator with 4 poles

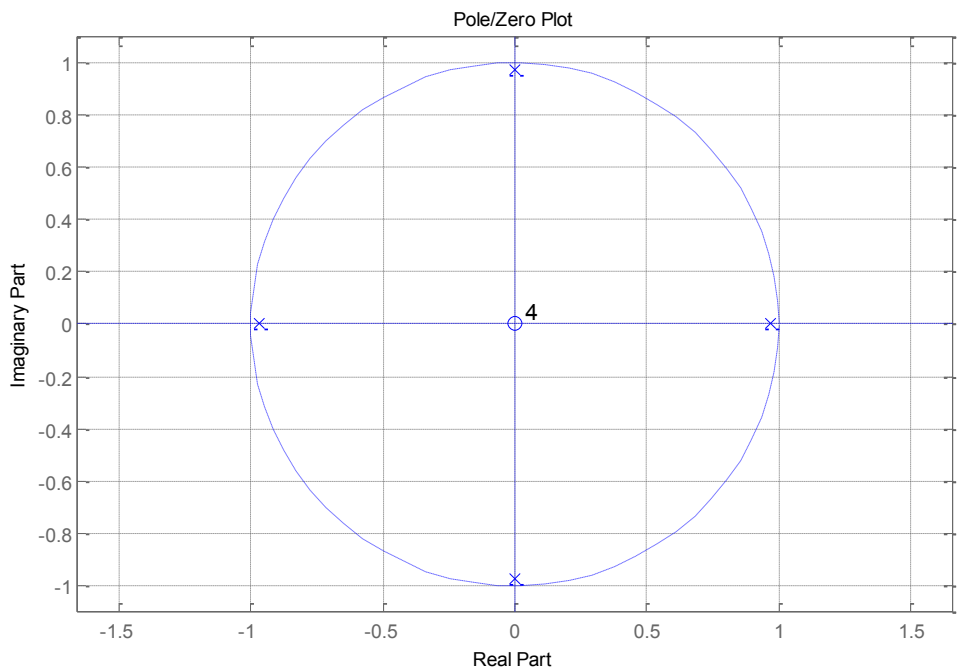

Figure 4.11  Pole-zero plot of integrator with 4 poles

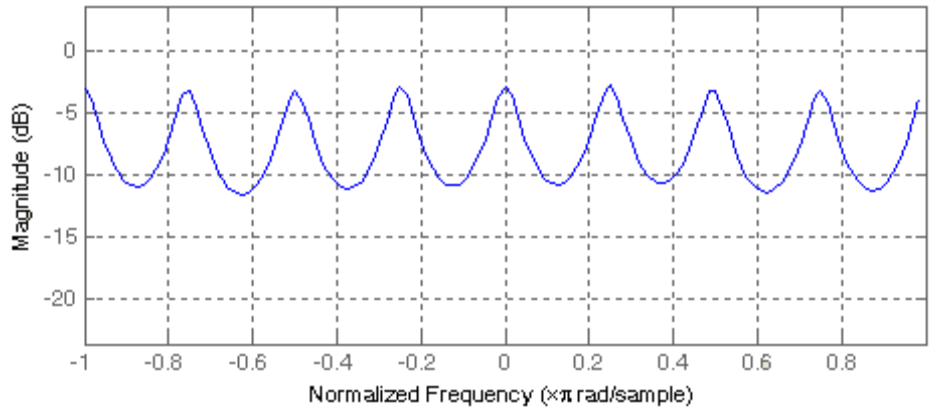

Figure 4.12 Magnitude spectrum of integrator with 8 poles

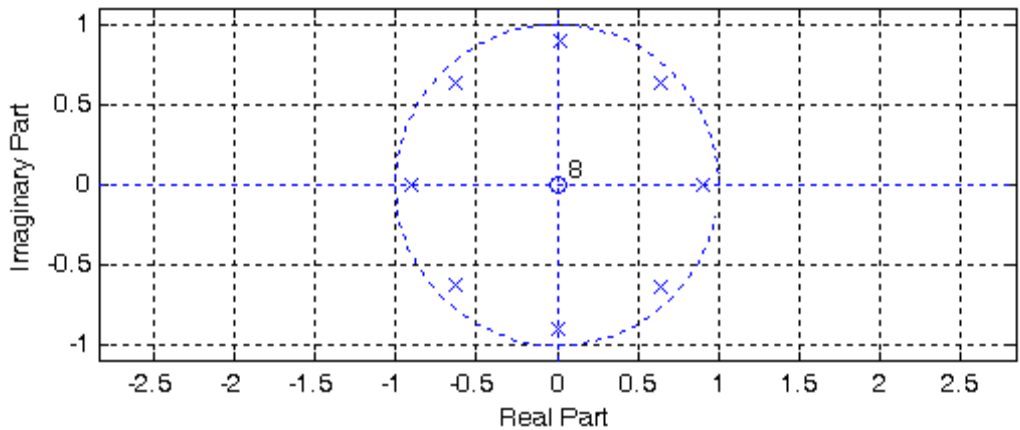

Figure 4.13 Pole-zero plot of integrator with 8 poles

In our research, the CIC filter consists of the N ideal integrator stages operating at the high sampling rate $f_s$. The majority of the power consumed by the CIC is attributed to the integrator as it has to work at the highest sample rate. This stage attenuates high frequency and amplifies the low frequency in the bandwidth.

### 4.3.2 DIFFERENTIATOR

The second main block of the CIC filter is the differentiator. This block is also known as a comb filter. The comb filter running at the lower sampling rate of $\frac{f_s}{R}$, where R is the decimation factor. This stage consists of a subtractor and a time delay unit connected by a forward loop. This filter is Finite Impulse Response (FIR) filter and has no feedback loop. The Comb filter has a high pass structure to attenuate infinite DC component frequency created by the integrator part, making the whole system stable. The equation of the differentiator block in time domain is given as:

$$y[nT_s] = x[nT_s] - x[(n-k)T_s] \qquad (4.12)$$

where $k = M \times R$ ($M$ is the differential delay, $R$ is decimation factor) in the comb structure. $M$ can be any positive integer (normally limited to 1 or 2). The differential delay is a filter design parameter used to control frequency response of the filter.

Figure 4.14 shows the basic structure of differentiator (comb) in discrete time domain.

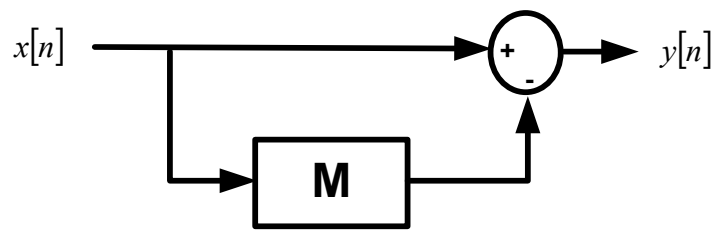

Figure 4.14 Differentiator (Comb) block diagram in discrete time domain

The comb filter can be defined in the z transform domain. Its equation and block diagram are given as:

$$H_C(z) = \frac{y[z]}{x[z]} = 1 - z^{-k} \qquad (4.13)$$

where $y[z]$ is the comb output and $x[z]$ is the comb input in sampling frequency of $\frac{f_s}{R}$.

Again using Oppenheim equations (Oppenheim & Schafer 1989), the phase and complex modulus of $z$ achieve as below:

$$\left|H_c(e^{j\omega})\right|^2 = 2(1-\cos k\omega) \tag{4.14}$$

$$ARG[H_c(e^{j\omega})] = -\frac{k\omega}{2} \tag{4.15}$$

$$grd[H_C(e^{j\omega})] = \frac{k}{2} \tag{4.16}$$

The comb filter has a transfer function that looks much like the rounded teeth of a comb. Comb filter pole-zero pattern and frequency spectrum with different number of poles are shown in Figures 4.15 -4.18.

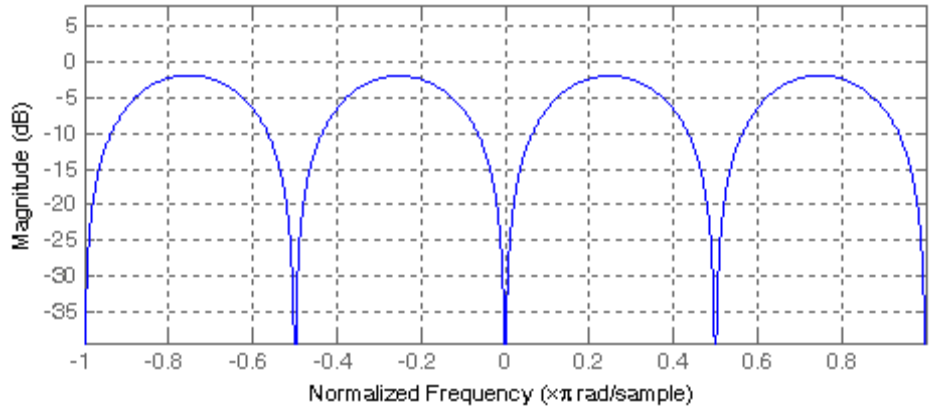

Figure 4.15  Frequency spectrum of the comb with 4 poles

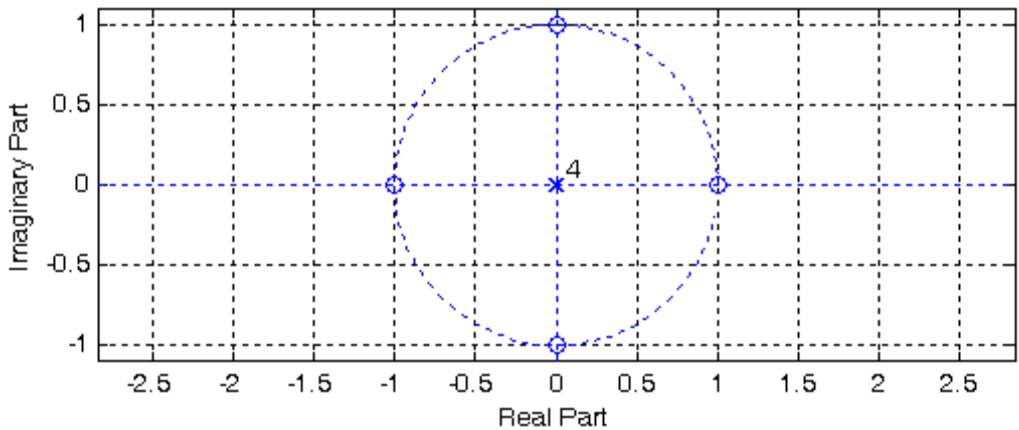

Figure 4.16  Pole-zero plot of the comb with 4 poles

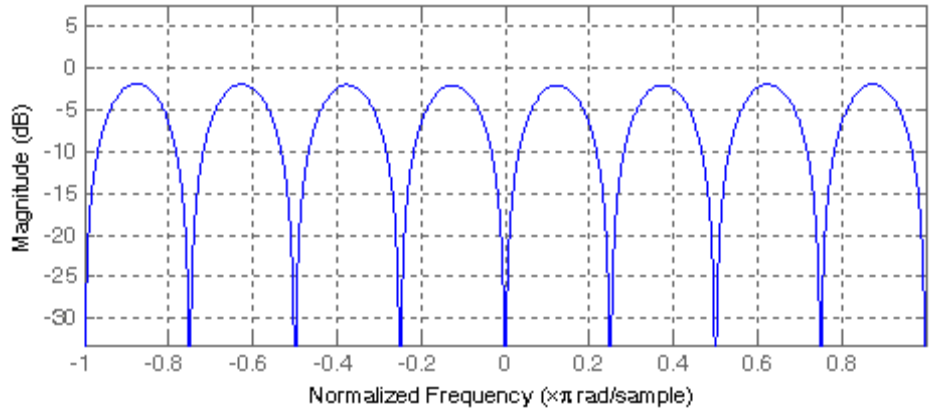

Figure 4.17  Frequency spectrum of the comb with 8 poles

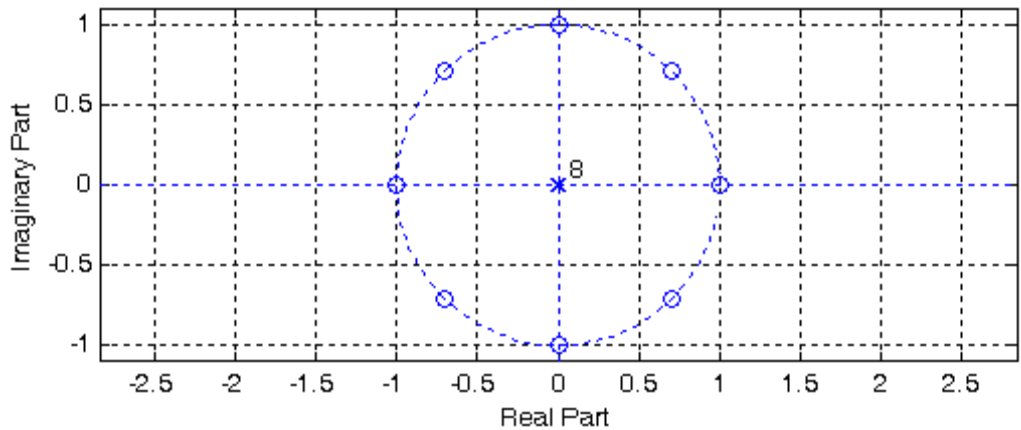

Figure 4.18  Pole-zero plot of the comb with 8 poles

The zeros of $H(z)$ are uniformly spaced $2\pi/RM$ radians apart around the unit circle, starting at $\omega = 0$ and there are $MR$ poles at the origin. For even R, there is also a zero. Being an FIR filter, it is always stable for any R.

The CIC filter consists of the N ideal comb stages operating at the low sampling rate $f_s/R$. While the integrator may produce overflow at the output, a serial connection of comb stages is located after the integrator to avoid overflow in the baseband and guarantee the filter stability.

### 4.3.3 DOWNSAMPLER

Down-sampling is the process of reducing the sampling rate of a signal. This is usually done to reduce the data rate or the size of the data. The operation of down-sampling by factor R, describes the process of keeping every $R^{th}$ sample and discarding the rest. This is denoted by "↓R" in block diagrams, as shown in Figure 4.19.

R which is the decimation factor is an integer greater than unity. It is better to select R as power of two integer digit. This factor multiplies the sampling time or, equivalently, divides the sampling rate.

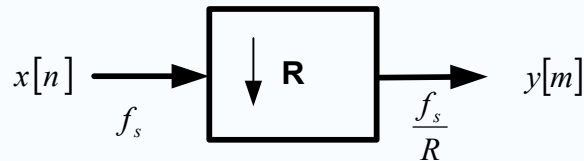

Figure 4.19 Down-sampler block with the decimation factor of R

The relationship between the resulting signals $y[m]$ and the input signals $x[n]$ is as follow:

$$y[m] = x[m \times R] \qquad (4.17)$$

The input signal of the down-sampler is a series. These series can be introduced in z domain as follow:

$$X(z) = \sum_{n=-\infty}^{\infty} x(n) \cdot z^{-n} \qquad (4.18)$$

All the multiple coefficients of R will be kept after passing down-sampler and the rest of the sample will be discarded. To implement the down-sampler, the input must be multiplied with $h(n)$ where $h(n)$ is given by:

$$h(n) = \begin{cases} 1 & n = mR \\ 0 & n \neq mR \end{cases} \qquad (4.19)$$

The output of the down-sampler in z domain described as below:

$$X_\circ(z^R) = \sum_{m=-\infty}^{\infty} x(m.R) \cdot z^{-mR} \qquad (4.20)$$

$$= \sum_{m=-\infty}^{\infty} x(m) \cdot (z^R)^{-m} = Y(z^R)$$

The down-sampling in time domain shown as Figure 4.20 when $R = 4$.

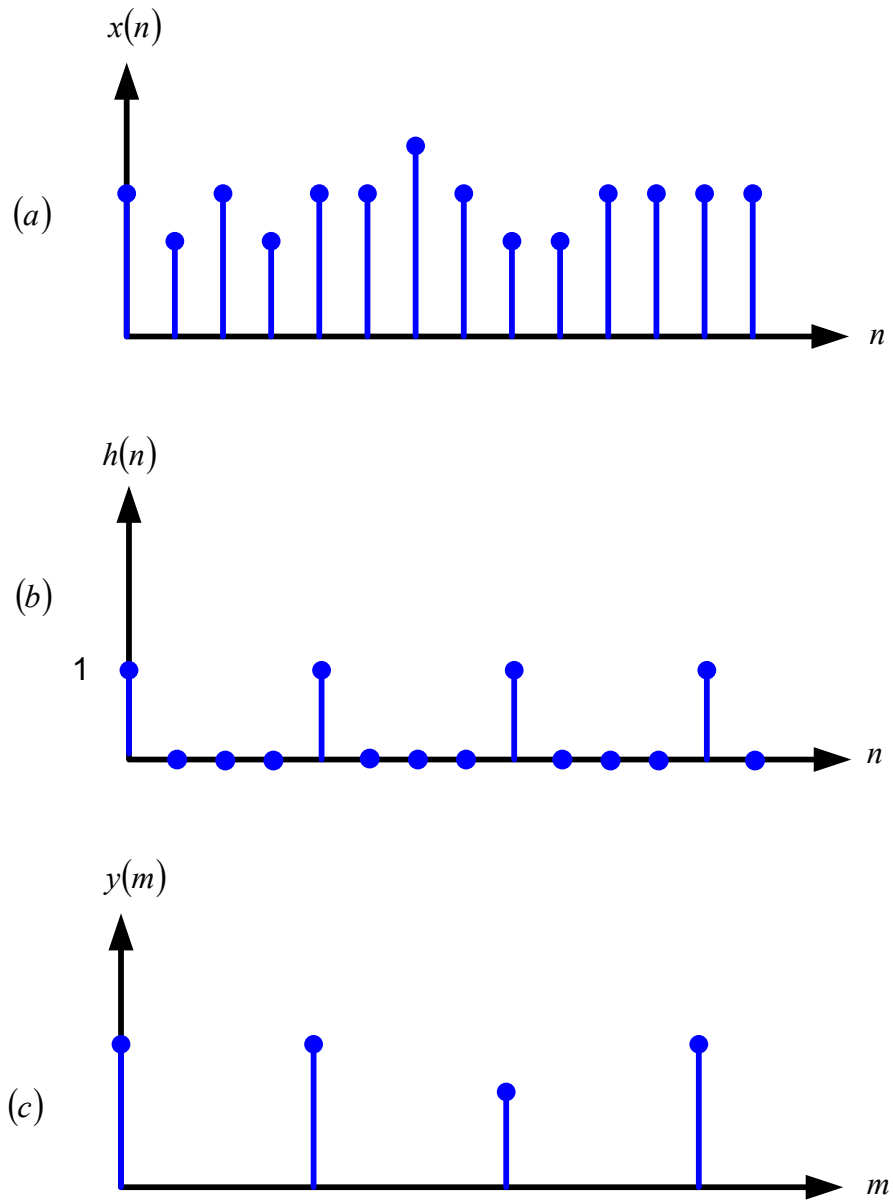

Figure 4.20  Decimation in time domain when $R = 4$. (a) Sampled input signal (b) Decimation filter $h(n)$; (c) the output signal when all the zeros between every R-th sample are discarded

In the above figure, the sampling input signal is passed through the decimation filter which called $h(n)$. The sampling rate of $y(n)$ is R times lower than $x(n)$. The magnitude of the down-sampled signal is shown in Figure 4.21. The down-sampling expends each $2\pi$, periodic repetition of $X(e^{j\omega})$ by a factor of $R$ along the $\omega$ axis.

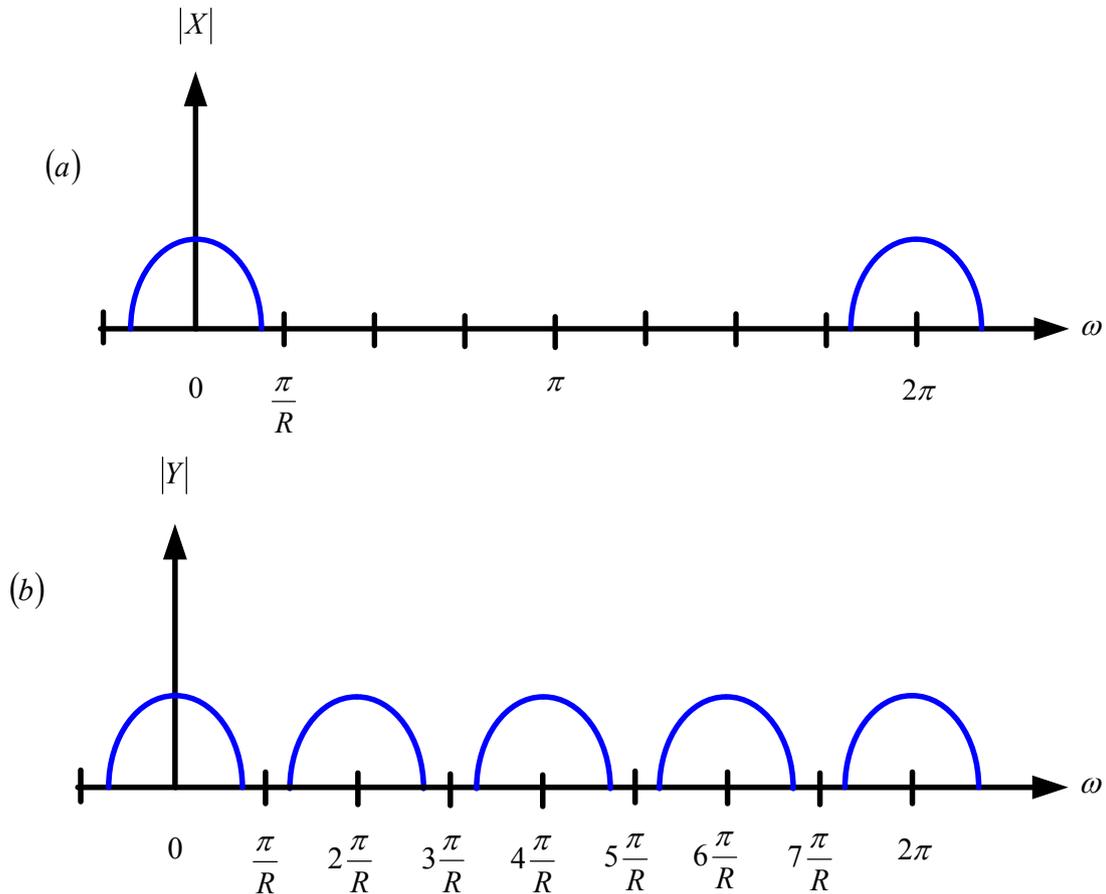

Figure 4.21  Magnitude spectrums of signals (a) before (b) after down-sampling

As shown in Figure 4.22, if $x(n)$ is not band-limited to $\dfrac{\pi}{R}$, aliasing may result from spectral overlapping.

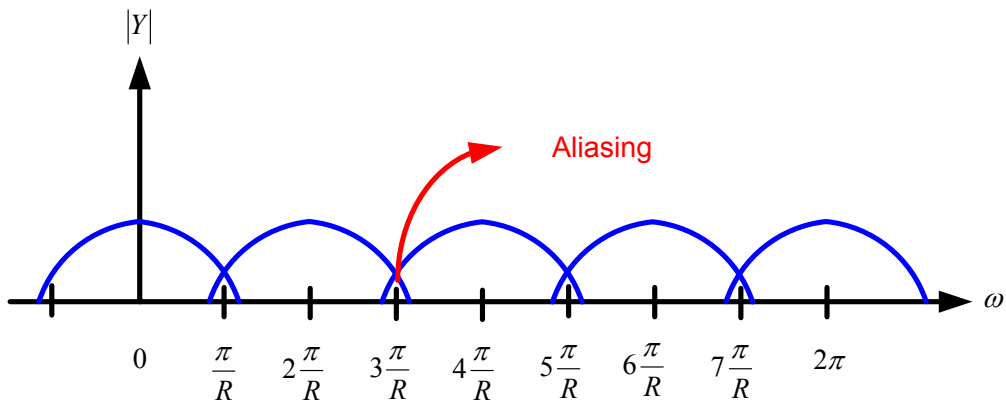

Figure 4.22 Presence of aliasing when the input signal is not limited to $\dfrac{\pi}{R}$

In fact, aliasing occurs when a signal is under-sampled which depends on the sampling frequency. If $f_s$ is too low, aliasing will happened in frequency domain. In general, aliasing by the factor R corresponds to a sampling rate reduction by a factor of R. To prevent aliasing when reducing the sampling rate, an anti-aliasing low-pass filter is generally used. The low-pass filter attenuates all signal components at frequencies outside the interval $\left(-\dfrac{f_s}{2R}, \dfrac{f_s}{2R}\right)$ so that all frequency components which produce alias is first removed.

**4.4　CIC FILTER PROPERTIES**

The CIC filter is a multi-rate filter supporting by the fixed point arithmetic. When a fixed-point filter is considered to design, in fact some data bit is thrown away after the accumulation if the input signal is defined float.

The Hogenauer CIC decimation filter (1981), shown in Fig. 4.23, consists of N cascaded digital integrators (operating at a high input sampling rate, $f_s$) and $N$ cascaded differentiators (operating at at a low rate, $\dfrac{f_s}{R}$).

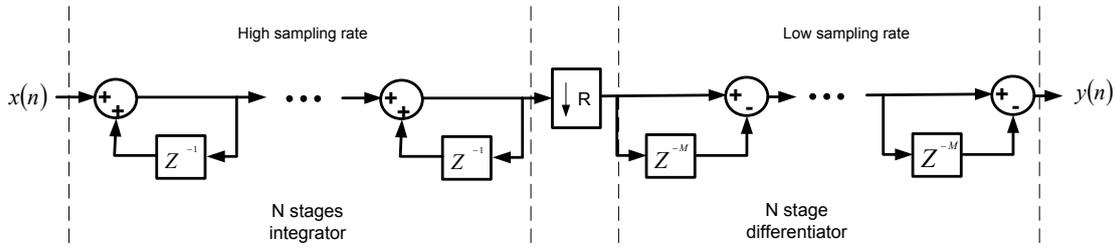

Figure 4.23 N- Order CIC filter structure in z-domain

Since all of the coefficients of this filter are unity, and therefore symmetrical the CIC filter has a linear phase response and constant group delay. The transfer function of the CIC filter is a combination of the integrator and the differentiator as follows:

$$H(z) = H_I^N(z) \cdot H_C^N(z) = \frac{(1-z^{-RM})^N}{(1-z^{-1})^N} = \left[ \frac{(1-z^{-1})(1+z^{-1}+...+z^{-RM+1})}{(1-z^{-1})} \right]^N = \left( \sum_{k=0}^{RM-1} z^{-k} \right)^N \quad (4.21)$$

where $N$ is the filter order, $R$ is decimation factor and M is differential delay.

The differential delay or $M$ is used as a design parameter to control the placement of the nulls in the frequency response. For the CIC decimation filters the region around every $M$-th null is folded in to the pass band causing aliasing errors. This is necessary to prevent signals from aliasing back into the pass-band. The cut-off frequency of the pass-band is adjusted by the differential delay.

After decimation by downs-ampler, the clock frequency of the CIC filter is decreased by the factor of R and the transfer function when $M = 1$ is converted as follow:

$$H(z) = \frac{1}{R} \cdot \left[ \frac{1-z^{-R}}{1-z^{-1}} \right]^N \quad (4.22)$$

As stated in the literature (Hogenauer 1981; Brandth 1991), the frequency response is given by (4.22) evaluated as:

$$z = e^{2\pi f / R} \quad (4.23)$$

$R$, $M$ and $N$ are chosen to provide the acceptable pass-band characteristics over the frequency range from zero to predetermined cut-off frequency. The amplitude or power response of the CIC filter is given by:

$$p(f) = \left[ \frac{\sin \pi M f}{\sin \frac{\pi f}{R}} \right]^{2N} \quad (4.24)$$

It is assumed that the decimation factor has large value and by using the relation, $\sin x = x$ (for small size of $x$), the amplitude response or power response of the CIC filter can be simplified $p_A(f)$ as.

$$p_A(f) \approx \left[ RM \frac{\sin \pi M f}{\pi M f} \right]^{2N} \quad 0 \leq f \leq \frac{1}{M} \quad (4.25)$$

Error between Equations (4.24) and (4.25) is less than 1dB for $R \geq 10, 1 \leq N \leq 7$ and $0 \leq f \leq 255/256$ where $f$ is the frequency relative to the low sampling rate. Since the amplitude response of the CIC filter has a "Sinc" $\left(\sin c(x) = \sin x / x\right)$ function, the CIC filter is also called a Sinc filter. The magnitude response is shown in Figure 4.24.

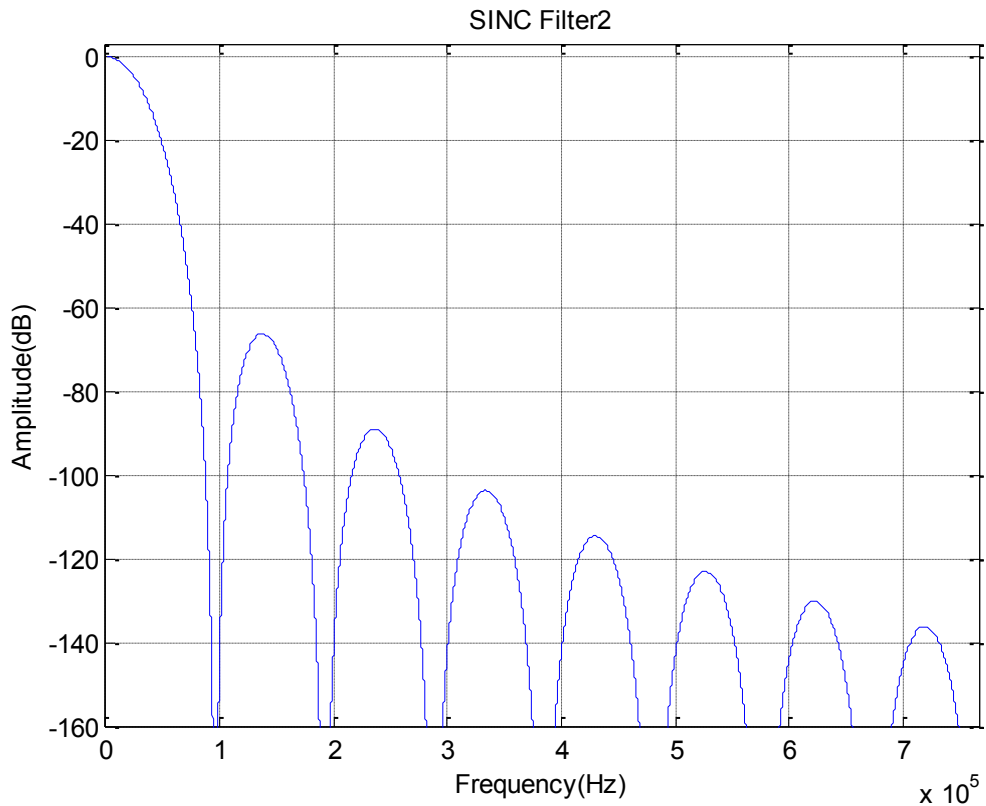

Figure 4.24 Magnitude response of the CIC filter

The CIC filters are efficient since they provide attenuation only in frequencies required by using cascaded integrator and comb filters to form higher order CIC. This will be ensured a better attenuation over wider frequency range around zero. From Figure 4.26, the integrators allowing the low frequency content to pass then attenuate the high frequency content.

$N$ which is the number of stage determines the pass band attenuation. Increasing the $N$ will improve the filter ability to reject aliasing and imaging but it also increases the droop (or roll off) in the filter pass band. For a sigma-delta modulator of order L, a cascade of $N$ order CIC filter is needed to adequately attenuate the quantization noise that would alias into the desired band (Candy 1986).

$$N = L + 1 \qquad (4.26)$$

Hence, for a third order modulator, a forth order CIC filter should be used. To ensure filter capability of reject aliasing in the baseband, the fifth order of CIC filter is used in this project. Figure 4.25 shows the effect of increasing filter order on the frequency response.

Figure 4.25 shows the CIC filter frequency response when filter order is changed whereas the decimation factor $(R=16)$ and differential delay $(M=1)$ are fixed. The sampling frequency is $f_s = 6.144 MHz$. It is clear that at least 4 stages of the CIC filter is necessary to provide sufficient attenuation. In this research, 5 stage of the CIC filter has been considered to design and implement.

Figure 4.26 shows the CIC filter frequency response when decimation factor is variable, whereas the filter order and differential delay is fixed, $N=5$ and $M=1$.

The phase response of the CIC filter is shown in Figure 4.27. As is evident phase is linear in the baseband.

Figure 4.28 presents the impulse response of the five stage CIC filter. The zero-pole pattern of the CIC filter is also shown. As clear in the figures, The CIC filter has nulls at each multiple of the output sampling frequency $f = f_s/R$.

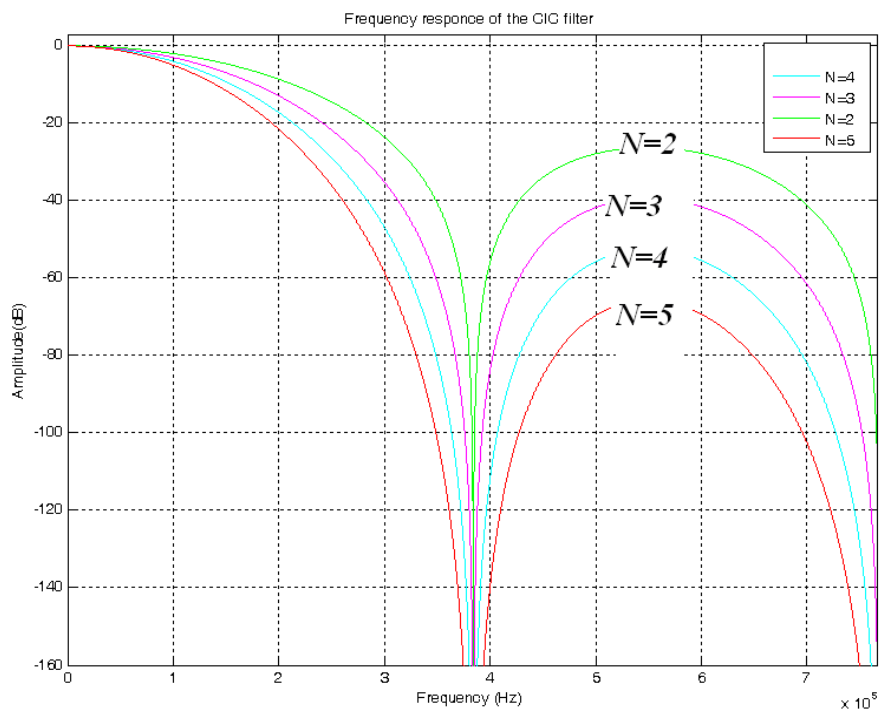

Figure 4.25  Amplitude response of the CIC filter with different filter order

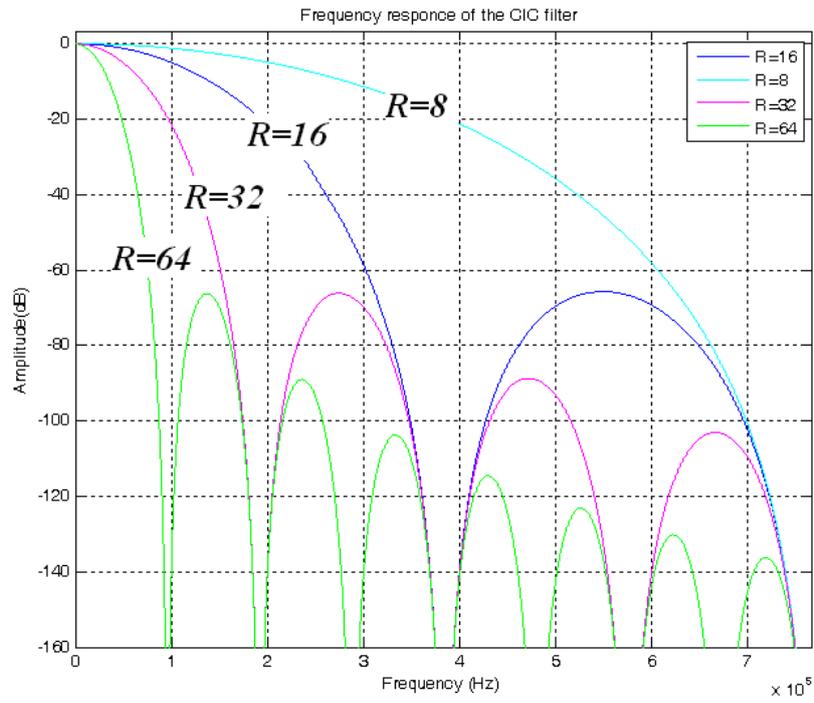

Figure 4.26  Amplitude response of the CIC filter with different decimation factor

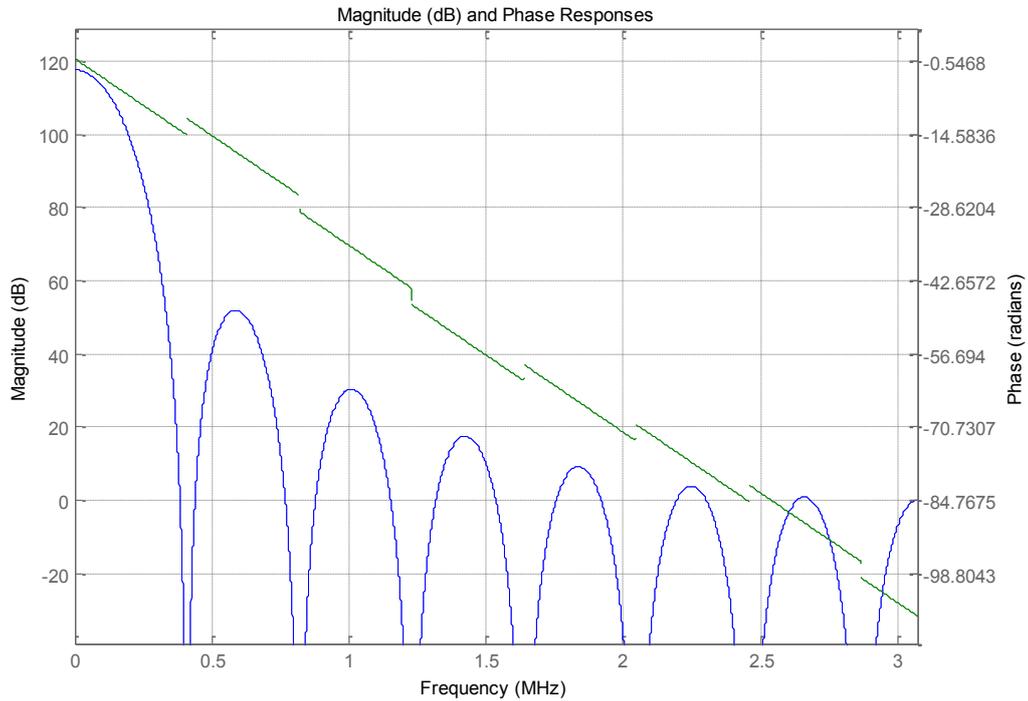

Figure 4.27  Frequency and phase response of the CIC filter

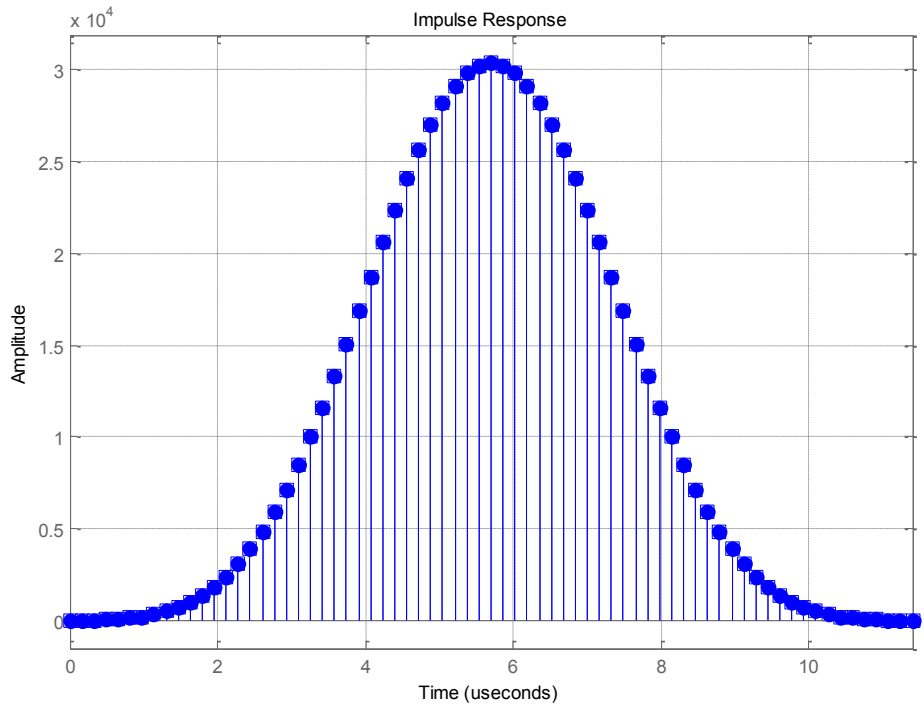

Figure 4.28 Impulse response of the CIC filter

As shown in figure above, the CIC filter is stable and its impulse response decays to 0 as n goes to infinity and its energy is limited.

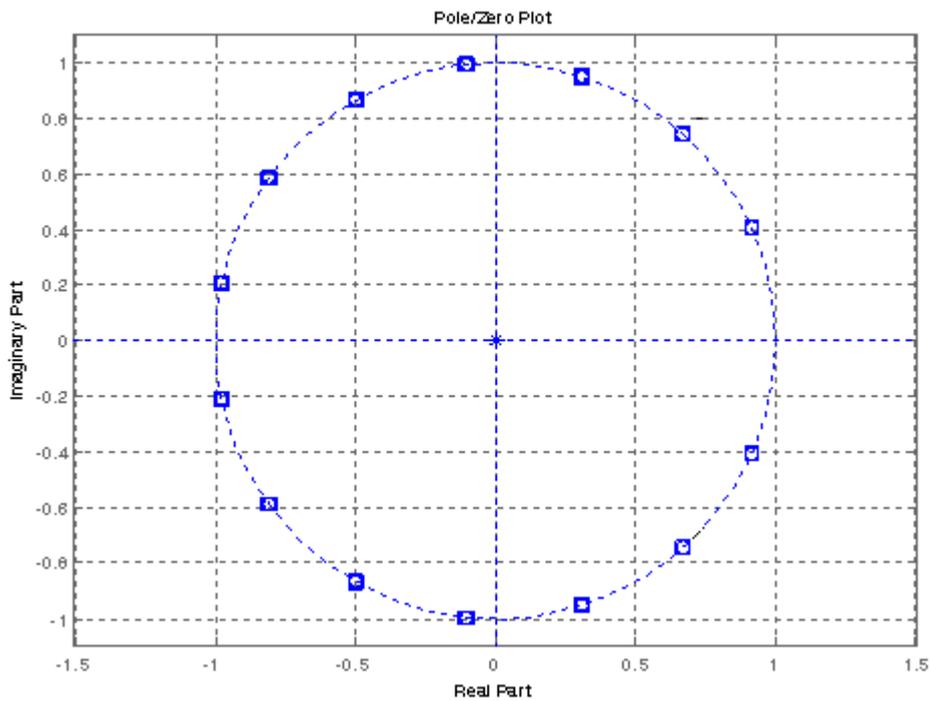

Figure 4.29 Pole-zero plot of the CIC filter

## 4.5 REGISTER GROWTH

The number of bit must be determined in the CIC filter structure in order to avoid overflow in the integrator and the comb stage. This necessitates determining the maximum magnitude. In other word, $G_{max}$ is the maximum register growth and a function of the maximum output magnitude due to the worst possible input conditions (Hogenauer 1981). N, M and R are parameters to determine the maximum register growth requirements necessary to assure no data loss and can be expressed as equations (4.27) and (4.28).

$$H(z) = \sum_{k=0}^{(RM-1)N} h(k)z^{-k} = \left[\sum_{k=0}^{RM-1} z^{-k}\right]^N \leq \left|\sum_{k=0}^{RM-1} z^{-k}\right|^N \leq \left(\sum_{k=0}^{RM-1} |z|^{-k}\right)^N = \left(\sum_{k=0}^{RM-1} 1\right)^N = (RM)^N \quad (4.27)$$

$$G_{max} = (RM)^N \quad (4.28)$$

If $B_{in}$ is the input data word length the maximum number of bit in the CIC filter obtained by:

$$B_{max} = [N\log_2 R + B_{in} - 1] \quad (4.29)$$

$B_{max}$ is maximum word length of the filter output that is a function of the parameters, $R$, $N$ and the Input bit width. Since the output of the integrators can grow without bound, $B_{max}$ represents the maximum number of bits which can be propagated through the filter without loss of data. $B_{max}$ is not only the maximum word length at the filter output; it is also the maximum word length for all sections in the CIC filter. $B_{max}$ is determined to avoid overflow in the CIC filter structure.

The CIC decimation filter puts a constraint of the bit-size or the register width for the integrator and comb filter. Firstly the word lengths of the filter stages should not be increased. It means the each stage of the CIC filter must has greater size of the word length compared to next stage. Secondly the number of bit in the first filter stage which is first integrator must be greater than or equal to the quantity of B$_{max}$.

## 4.6 INTERPOLATION CIC FILTER

Unlike the decimation, interpolation (Yong Ching Lim & Rui Yang 2005; Goodman & Carey 1977; Brandt & Wooley 1993) means increasing the sample rate by the interpolation factor of $R$. The CIC filters as interpolator are a class of linear phase FIR filters comprise of a comb and an integrator and an up-sampler. Exchanging the integrator cascade with the differentiator cascade in the CIC decimation as shown in Figure 4.30, produces a CIC interpolator. The cascade differentiator operates at the low sampling frequency of $f_s/R$. The up-sampler in the filter structure increases a rate of sampling by factor of $R$. It is done by inserting $R-1$ zero valued samples between consecutive samples of the comb section output. Then the interpolated samples lead to the cascade integrator which is operated at high sampling rate of $f_s$.

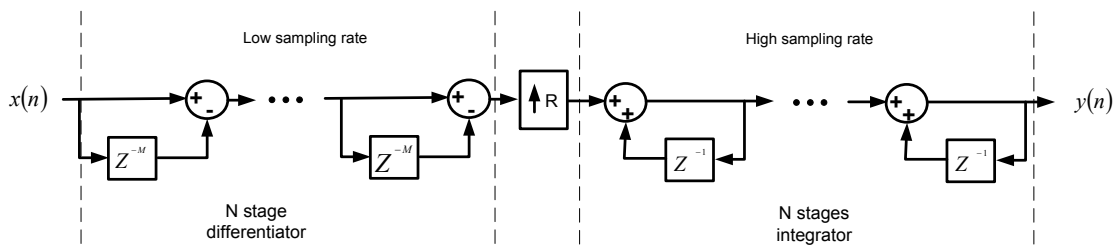

Figure 4.30 The interpolation CIC filter structure

## 4.7 NON-RECURSIVE COMB FILTER

When the decimation ratio and filter order are high the CIC filter use high power consumption since the integrator stage works at the highest oversampling rate with a large internal word length. The circuit speed will also be limited by the recursive loop of the integrator stage. Thus the Non-recursive comb filter is introduced to replace when the decimation ratio and filter order are high. This filter is a decimation filter with regular structure. This property makes it suitable for VLSI implementation. It is investigated and developed (Farag et al. 1997; Candy & Temes 1992; Yonghong Gao et al. 1999) because of its advantages such as no recursive loop and using low power consumption due to computations are performed at lower sampling rate. The non-recursive comb filter has ability of wide range of rate change. Its transfer function is shown as follow:

$$H(z) = \left(\sum_{i=0}^{R-1} z^{-i}\right)^N = \left(\sum_{i=0}^{2^M-1} z^{-i}\right)^N = \prod_{i=0}^{M-1}\left(1+z^{-2^i}\right)^N \qquad (4.30)$$

where R is the decimation factor, $N$ is the filter order and $M$ is the number of stage. Note that $R$ should be a power of 2.

The non-recursive comb filter structure is shown in Figure 4.31. Compared to the CIC filter structure the Non-recursive comb filter has the specification as:

1. During decimation process and decreasing sampling frequency, the number of bit increases and it is the reason of the saving in power.

2. The comb decimator using the Non-recursive algorithm can achieve higher speed since the first stage always has small word length.

3. When decimation ratio increases, the silicon size of the Non-recursive comb filter design algorithm increases faster compared to the CIC filter design algorithm.

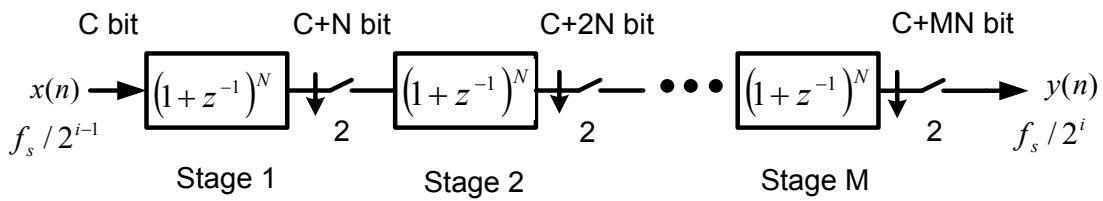

Figure 4.31 Block diagram of the $M$ stage non-recursive Comb filter

Figure 4.32 shows that each stage of $M$ is included of $N$ blocks non-recursive comb filter.

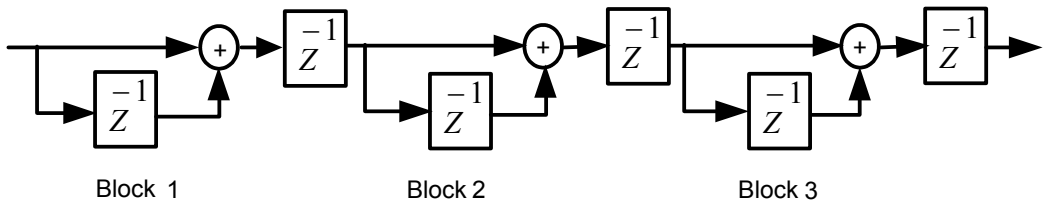

Figure 4.32 One stage of the comb filter when $N = 3$

## 4.8 SUMMARY

In this chapter principle of the decimation and interpolation was introduced. The conventional decimation filter was described and the comparison has been done with the CIC filter structure. The CIC structure and its advantages were explained in detail. The frequency analysis of the CIC decimation filter was also presented in this chapter. The CIC filter has three important components that is the integrator, the differentiator and the down sampler. The integrator has low-pass structure to attenuate high frequency component and the differentiator has high-pass structure to attenuate DC component and makes whole system stable. Finally the structure of the non-recursive comb filter and its specification was described. The Non-recursive comb filter is replaced as decimator when the decimation ratio and filter order in the CIC filter are high.

# CHAPTER V

# CIC FILTER IMPLEMENTATION

This chapter discusses on the implementation details of the CIC filters. After the functional testing of the CIC filter and its system simulation were performed, the filter implementation is embarked up on. The design and implementation of the CIC filters verified by Verilog HDL, is presented before the Verilog codes were loaded into the field programmable gate array (FPGA) chip. The FPGA chip is the integrated circuit that is field configurable by the end user for a logic design implementation (Ashok 1998).

Besides, the CIC filter system is synthesized using SilTerra $0.18\,\mu m$ and MIMOS $0.35\,\mu m$ technologies, separately. The comparison results between two technologies are given that shows the design enhancement by different technologies.

In order to reach high throughput and low power consumption, some design innovations were considered. Thus, these techniques were introduced to overcome the problems. A new architecture is utilized to achieve the design objectives. The high speed filter is achieved by three methods i.e. using MCLA in adder, truncation and applying pipeline architecture. Furthermore, the power consumption is reduced by minimizing the number of calculations and optimizing the electronic elements.

## 5.1 STAGE REALIZATION

The overall stage realization of the CIC filter is redrawn in Figure 5.1. This section describes the logic blocks that function in architecture of the CIC filter such as cascaded integrator, downsampler and cascaded comb filter. The complete process flow for the IC design using Verilog HDL (i.e. top-down approach) will be described in future (chapter V & VI).

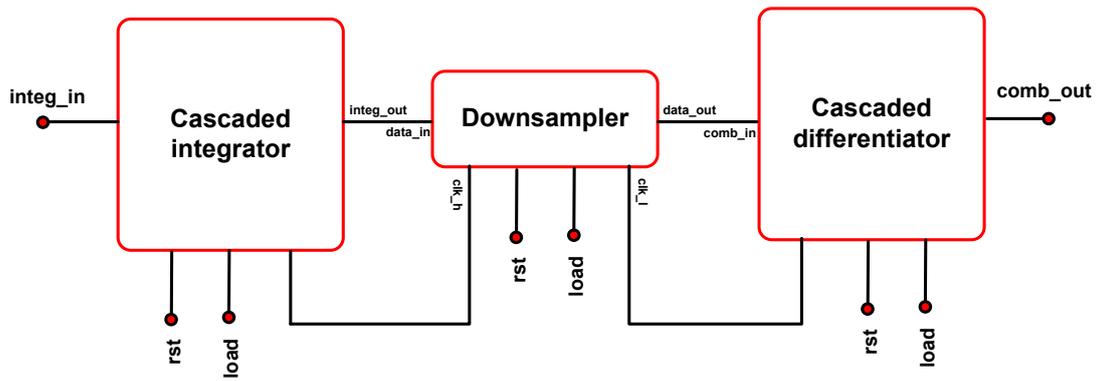

Figure 5.1  Block diagram of the CIC filter

In order to implement the CIC filter above, the following functional logic blocks are required.

- Adder for the cascaded integrator.
- Subtractor for the cascaded differentiator
- Register delay
- Adjuster
- Counter

From the equations (4.28) and (4.29), when $N=5$, $R=16$, $M=1$ and $B_{in}=5$; the maximum number of bit selected is 25-bit for each stage of the integrator and the comb filter. The bit size will avoid the overflow as mentioned in chapter IV. Using the above chosen parameter, results the CIC architecture is as shown in Figure 5.2.

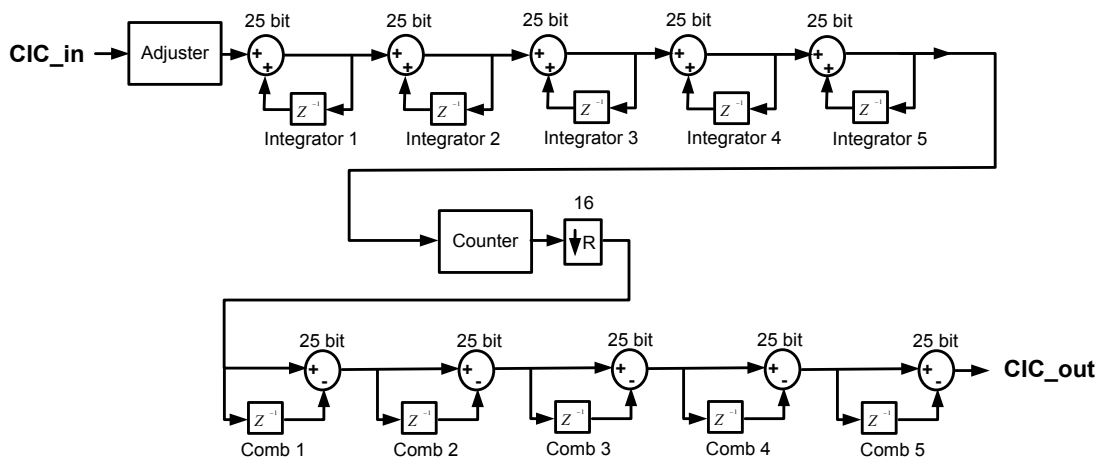

Figure 5.2  Implementation of the CIC filter

The input signal which is fed to the filter is in the two's complement format to avoid the overflow effect on the data. Thus an adjuster is required to convert the 5-bit input data to the 25-bit and to make the input data as two's complement. Then input sample is sent to the first stage integrator filter for further computation. The integrator output is accumulated with the first input using the feedback loop and an adder, i.e. the result is passed from 1-bit register (delay) before adding with the input samples in the loop. At each clock cycle, the content of the shift register is shifted to complete the integration process. The same procedure is repeated in the five-stage integrator. The output of the $5^{th}$ integration is transferred to the down-sampler (Counter and $R$). Based on the decimation ratio (by using an up-counter), $R-1$ samples are discarded and multiple coefficients of $R$ is kept before sent to the comb filter.

The subtractor in the comb structure subtracts the down-sampler output from the delayed input. Basically the subtractor is an adder with one input inverted. The comb output signal is passed through the five order comb structure. At the end of the process we obtain the decimation input signal.

In order to verify the functionality of the CIC filters, we have downloaded the Verilog codes into the FPGA so that we can test the CIC filter. Xilinx and Modelsim EDA tools were used to synthesize and to simulate the design. Table 5.1 shows the summary of Xilinx synthesis for the CIC filter.

Table 5.1 CIC filter summary of Xilinx synthesis

| HDL Synthesis Report | QTY (Unit) | Timing Summary | |
|---|---|---|---|
| 25-bit adder | 5 | Minimum period (ns) | 9.280 |
| 25-bit subtractor | 5 | Maximum Frequency (MHz) | 107.759 |
| 4-bit down counter | 1 | Minimum input arrival time before clock (ns) | 3.924 |
| 1-bit register | 1 | Maximum output required time after clock(ns) | 14.995 |
| 25-bit register | 11 | Total equivalent gate count for design | 5216 |
| 4-bit register | 1 | | |
| 25-bit 2-to-1 multiplexer | 1 | | |
| 25-bit 2-to-1 multiplexer | 1 | | |
| Number of 4-input LUT cells | 253 | | |
| Number of external IOB | 32 | | |
| Flip Flop Latch | 280 | | |

## 5.2 SPEED AND POWER IMPROVEMENT

The result obtained as shown Table 5.1. The implementation of the CIC filter made use of the standard components in the library. In order to improve the CIC filter functionality, further speed and power optimization was implemented by using proposed new architecture. The new architecture proposes the following methods:

1. Faster adder algorithms
2. Truncation process
3. Pipeline structure

## 5.2.1 IMPLEMENTATION OF THE FAST ADDER

Fast adder is one of the approaches to increase the throughput of the CIC filter. There are different methods to do addition. The straight forward method of adding *m* numbers with *n* bits is to add the first two, then add that sum to the next and so on. This technique requires a total of *m-1* addition and high number of gate delay and is normally known as ripple carry adder (RCA). In RCA (Ciletti 2003; Doran 1988), the most significant bit addition has to wait for the carry to ripple through from the least significant bit addition. This approach of carry calculation makes the adder very slow. To reduce the delay other techniques chosen as follow.

### 5.2.1.1 CARRY SAVE ADDER (CSA)

The carry save adder (CSA) has proved to a powerful technique to improve the adder timing. This technique adds 3 numbers (*x, y* and *z*) and then converts them into 2 numbers (*s, c*) in one unit delay. The following example shows the procedure of the technique.

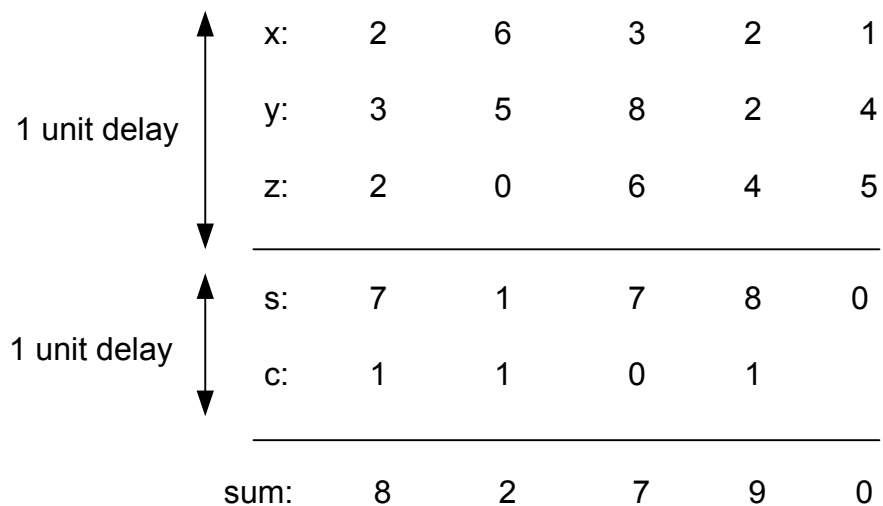

The carry, *c* is added to *s* with consideration of carry grade. The advantage of CSA is that *c* and *s* can be calculated in dependently.

The CSA structure is shown in Figure 5.3. The *n* bit CSA consists of *n* disjoints of full adders (FAs). It uses three *n*-bit input vectors and produces two outputs, *n*-bit sum, *S,* and *n*-bit carry, *C*. Unlike other type of adders such as ripple carry adder (RCA) and carry look-ahead adder (CLA), the CSA contains no carry propagation. Consequently the CSA has the same propagation delay as only one FA delay and delay is constant for any value of *n*, where n is number of bit. For sufficiently large number of *n*, the CSA implementation becomes much faster and relatively smaller in size than implementation of normal adders (Waste & Eshraghian 1985; Kim et al. 1998).

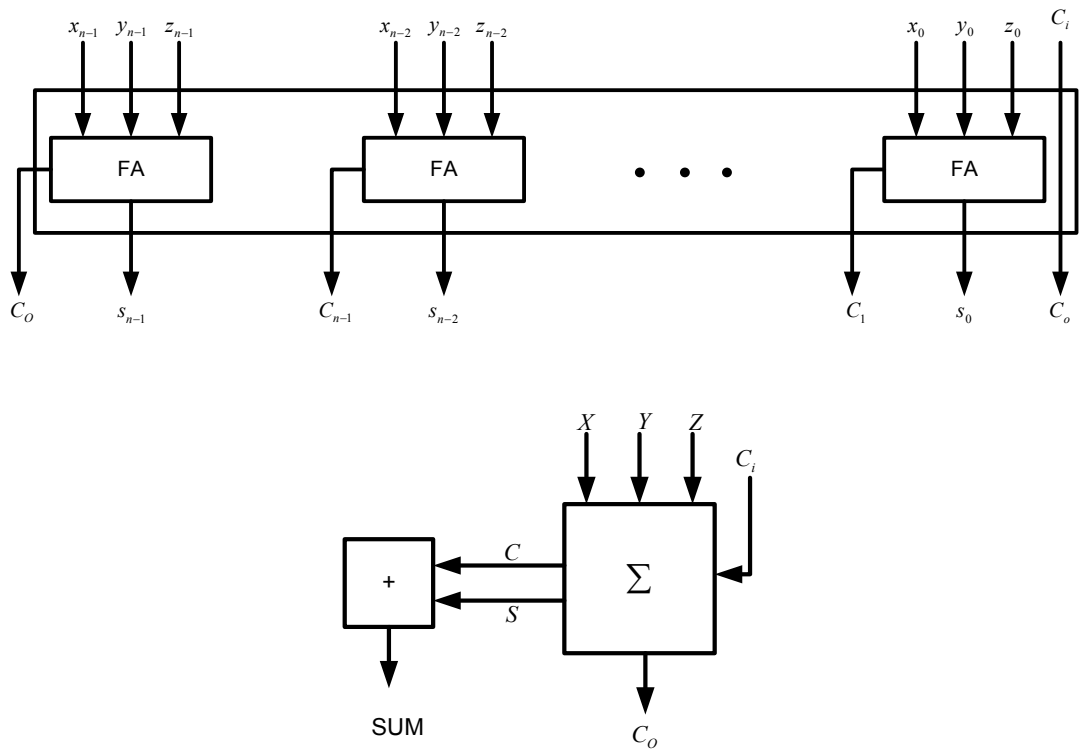

Figure 5.3  CSA structure

The integration of CSAs is used in the integrator sections of the CIC filter. The synthesized result shows the maximum clock frequency of 97 MHz in the CIC filter. Although the above clock frequency is sufficient for normal audio application, we carried further research in order to increase the clock frequency due to using this filter in receiver applications. In digital receivers digitizing IF signals is at the high frequencies. In wide band communications signal, the bandwidth has increased to tens of MHz so oversampling frequency of A/D converters will be hundreds of MHz (Brandt & Wooley 1994; Jensen et al. 1995). In order to isolate the signal of interest from the digitized signals and decimation it to lower sampling rate, high speed decimation filters are required. Not only does this decimator operate at high speed but it also requires high resolution as well. Thus this research work design and implement high speed and high resolution CIC filter use for audio application and suitable for IF digital receiver.

### 5.2.1.2 MODIFIED CARRY LOOK-AHEAD ADDER (MCLA)

Modified carry look-ahead adder (MCLA) is known as a solution to improve the CIC filter speed. The carry look-ahead adder (CLA) is the fast adder (Naini et al. 1992; Tsujihashi et al. 1994) which can be used for speeding up purposes. However the disadvantage of the CLA adder is that the carry logic is getting complicated for more than 8 bits. Consequently, we developed the enhanced version of the CLA introduced to replace the CLA adder. Improvement in speed in MCLA based on CLA is due to the carry calculation. Therefore, the carry of the MCLA adder has become a focus of the study in speeding up the adder circuits. The 8-bit MCLA architecture is shown in Figure 5.4.

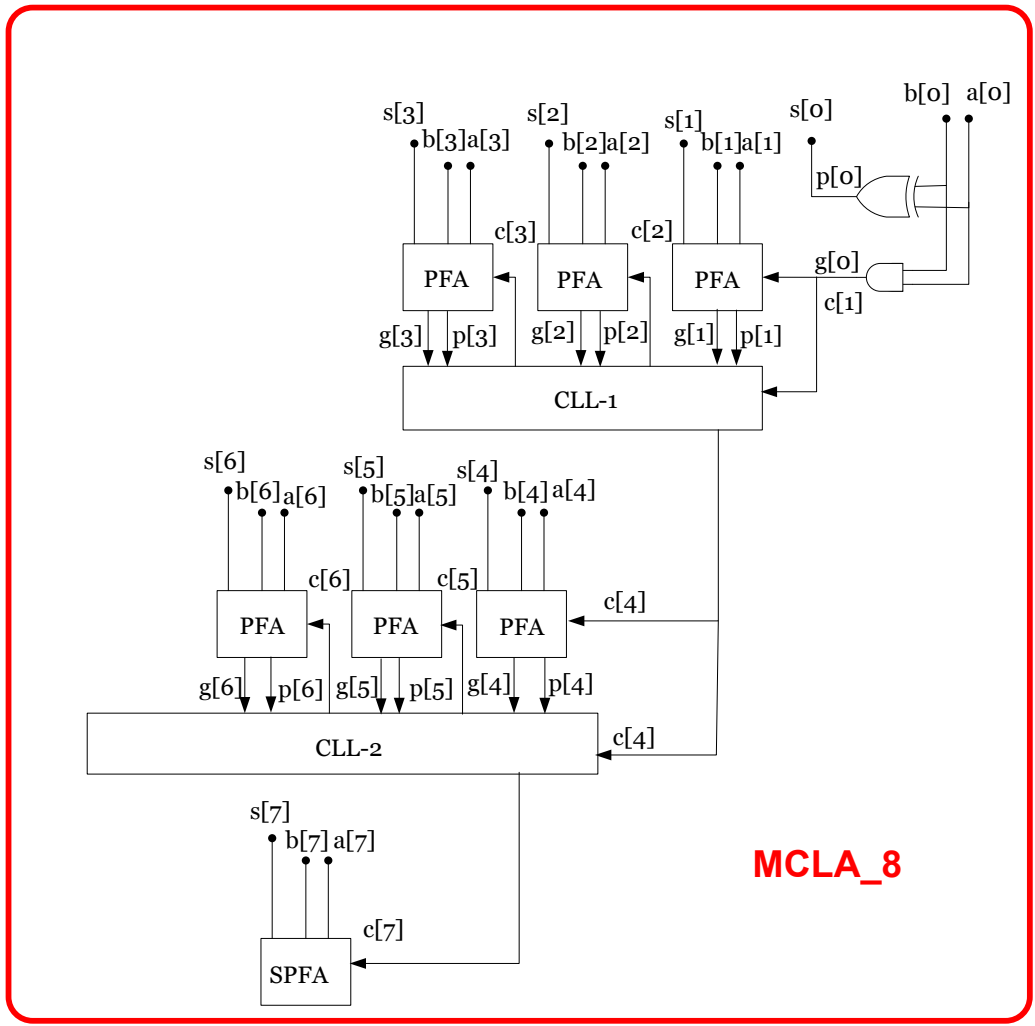

Figure 5.4 8-bit MCLA architecture

This block diagram consists of 2, 4-bit modules which are connected and each previous 4-bit calculates carry out for the next carry. The CIC filter of this work involves with the five MCLA in integrator parts. The maximum number of bit is 25 thus the adder should be designed accordingly. The operation of the MCLA is based on CLA equations to calculate the carry out.

For MCLA adder, if there are three input data $a_0$, $b_0$ and $carryIn_0$, the $carryIn_1$ (Ciletti 2003) is defined as:

$$carryIn_1 = (a_0 \cdot carryIn_0) + (b_0 \cdot carryIn_0) + (a_0 \cdot b_0) \qquad (5.1)$$

The $carryIn_2$ for next two bit, $a_1$ and $b_1$ when these two bits are added with $carryIn_1$ can be expressed as:

$$carryIn_2 = (a_1 \cdot carryIn_1) + (b_1 \cdot carryIn_1) + (a_1 \cdot b_1) \tag{5.2}$$

These equations can be minimized as:

$$c_1 = (a_0 \cdot c_0) + (b_0 \cdot c_0) + (a_0 \cdot b_0) \tag{5.3}$$

$$c_2 = (a_1 \cdot c_1) + (b_1 \cdot c_1) + (a_1 \cdot b_1) \tag{5.4}$$

According to the above equations, $c_2$ can be written as:

$$\begin{aligned}c_2 &= (a_1 \cdot a_0 \cdot b_0) + (a_1 \cdot a_0 \cdot c_0) + (a_1 \cdot b_0 \cdot c_0) + (b_1 \cdot a_0 \cdot b_0) \\ &+ (b_1 \cdot a_0 \cdot c_0) + (b_1 \cdot b_0 \cdot c_0) + (a_1 \cdot b_1)\end{aligned} \tag{5.5}$$

It is clear that the equation is too complicated and for higher order of bits, the equation grows exponentially. This complexity is reflected in the cost of the hardware for fast carry. The carry equation can be defined in general form as:

$$c_{i+1} = (a_i \cdot c_i) + (b_i \cdot c_i) + (a_i \cdot b_i) = (a_i \cdot b_i) + (a_i + b_i) \cdot c_i \tag{5.6}$$

The Generate $(g_i)$ and Propagate $(p_i)$ are defined:

$$g_i = a_i \cdot b_i \tag{5.7}$$

$$p_i = a_i + b_i \tag{5.8}$$

$$c_{i+1} = g_i + p_i \cdot c_i \tag{5.9}$$

In terms of generate and propagate the carry out is given as:

$$c_1 = g_0 + (p_0.c_0) \tag{5.10}$$

$$c_2 = g_1 + (p_1.g_0) + (p_1.p_0.c_0) \tag{5.11}$$

$$c_3 = g_2 + (p_2.g_1) + (p_2.p_1.g_0) + (p_2.p_1.p_0.c_0) \tag{5.12}$$

$$c_4 = g_3 + (p_3 \cdot g_2) + (p_3 \cdot p_2 \cdot g_1) + (p_3 \cdot p_2 \cdot p_1 \cdot g_0) + (p_3 \cdot p_2 \cdot p_1 \cdot p_0 \cdot c_0) \quad (5.13)$$

Notice that each 4-bit adder provides a group propagate and generate signal, which is used by the MCLA Logic block. The group Propagate $P_{out}$ and Generate $G_{out}$ of a 4-bit adder will have the following expressions:

$$P_G = p_3 \cdot p_2 \cdot p_1 \cdot p_0 \quad (5.14)$$

$$G_G = g_3 + p_3 \cdot g_2 + p_3 \cdot p_2 \cdot g_1 + p_3 \cdot p_2 \cdot p_1 \cdot g_0 \quad (5.15)$$

The most important equation to obtain carry for next stage is defined as below:

$$c_4 = G_{out} + (P_{out} \cdot c_0) \quad (5.16)$$

There are half adders (HA) and partial full adders (PFA) in MCLA structure to add each 2 bit data. Table 5.2 and Figure 5.5 show the truth table and the logic circuit for the HA respectively while table 5.3 and Figure 5.6 show truth table and the logic circuit for the PFA respectively.

Table 5.2  Truth table of the HA

| A | B | S | C |
|---|---|---|---|
| 0 | 0 | 0 | 0 |
| 0 | 1 | 1 | 0 |
| 1 | 0 | 1 | 0 |
| 1 | 1 | 0 | 1 |

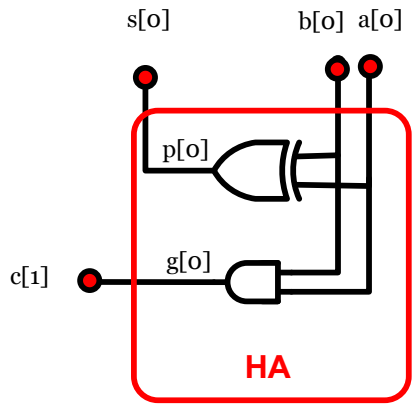

Figure 5.5 Structure of the HA

The HA normally is used to add the LSB of the 2 n-bit data when carry is equal to zero or $c_0 = 0$.

Table 5.3 Truth Table of the PFA

| A | B | Ci | S | Ci+1 |
|---|---|----|---|------|
| 0 | 0 | 0  | 0 | 0    |
| 0 | 0 | 1  | 1 | 0    |
| 0 | 1 | 0  | 1 | 0    |
| 0 | 1 | 1  | 0 | 1    |
| 1 | 0 | 0  | 1 | 0    |
| 1 | 0 | 1  | 0 | 1    |
| 1 | 1 | 0  | 0 | 1    |
| 1 | 1 | 1  | 1 | 1    |

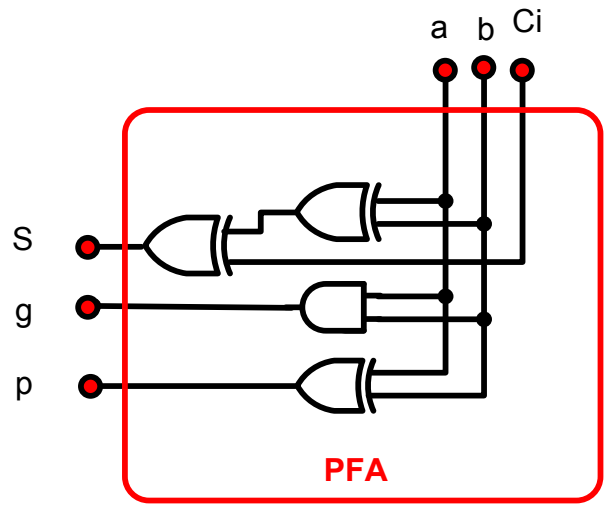

Figure 5.6  PFA structure

The sum of partial full adder (SPFA) can be defined in Figure 5.7 when it is used to calculate the most significant bit (MSB).

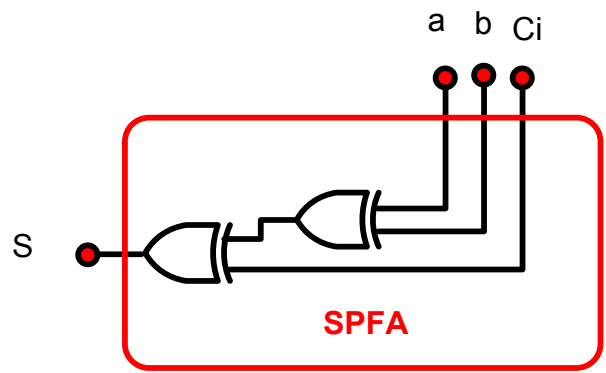

Figure 5.7  SPFA structure

Verilog HDL codes for the MCLA are written to implement the 25-bit adder for the CIC filter using the same design procedure as discussed before. Comparison of the results obtained is compared with the CLA adder and is presented in Table 5.3. Generally the MCLA adder perform better than the CLA adder, however when the bit size is increased, the throughput of the MCLA shows a slight decreased in performance. We have chosen the MCLA because of the easier configuration and fast calculations compared to others fast adders.

Table 5.4  Xilinx ISE synthesis summary for MCLA and CLA

|  | MCLA | CLA |
|---|---:|---:|
| Minimum period (ns) | 3.701 | 4.583 |
| Maximum Frequency (MHz) | 270 | 220 |
| Minimum input arrival time before clock (ns) | 1.634 | 1.678 |
| Maximum output required time after clock(ns) | 4.570 | 4.575 |
| Total equivalent gate count for design | 90 | 128 |

Table 5.4 shows better performance in the MCLA compared to the CLA.

### 5.2.2  TRUNCATION

Truncation improves the speed of the system as well as decrease the power consumption. Truncation means estimating and removing least significant bit (LSB) that helps to reduce the area requirements on the chip. Although this estimation and removing introduces additional error, this error can be made small enough to be acceptable for DSP and audio applications.

The error *(E),* created by the truncation process can be described in equation (5.17).

$$E_j = 2^{-b} \qquad (5.17)$$

where *b* is the number of bit in LSB that can be discarded (Rabiner & Gold 1975) and *j* is the number of stage in the CIC structure.

As stated in the literature (Hogenauer 1981), when the truncation is applied for a system with *M* stages, *M* noise sources are created. For the *N*-order CIC filter, *2N+1* truncation noise sources are contributed to the output data if all stages need to be truncated. Thus there are a total of *2N+1* error sources. The first *2N* sources are caused by the truncation at the inputs to the *2N* filter stages. The "1" error source is truncation into output register. Since the error has uniform distribution then the mean of the error is:

$$\mu = \frac{1}{2}E = 2^{-b-1} \tag{5.18}$$

Thus the variance of the error is:

$$\partial_j^2 = \frac{1}{12}E_j^2 \tag{5.19}$$

By multiplying the error mean and variance in *k*-th filter coefficient, the total mean and variance in the output will be:

$$\mu_{Tj} = \mu_j D_j \tag{5.20}$$

where

$$D_j = \begin{cases} \sum_k h_j(k) & j = 1,2,....,2N \\ 1 & j = 2N+1 \end{cases} \tag{5.21}$$

The total variance can be described as (5.22).

$$\partial_{Tj}^2 = \partial_j^2 F_j^2 \tag{5.22}$$

where

$$F_j^2 = \begin{cases} \sum_k h_j^2(k) & j = 1,2,....,2N \\ 1 & j = 2N+1 \end{cases} \tag{5.23}$$

$D_j$ is simplified from (4.27) as following:

$$D_j = \begin{cases} (RM)^N & j = 1 \\ 0 & j = 2,3,....,2N \\ 1 & j = 2N+1 \end{cases} \tag{5.24}$$

From (5.24), the first and the last stages of the CIC filter create the error mean for the whole system. However, the truncation error is too small and negligible.

$$\mu_T = \sum_{j=1}^{2N+1} \mu_{Tj} = \mu_{T1} + \mu_{T2N+1} \tag{5.25}$$

$$\partial_T^2 = \sum_{j=1}^{2N+1} \partial_{Tj}^2 \tag{5.26}$$

The number of bit which is neglected in the integrator stages is determined by Matlab software. Figure 5.8 shows the SNR reduction when CIC filter is truncated to lower number of bit. Note that audio application required the SNR of at least 80 dB.

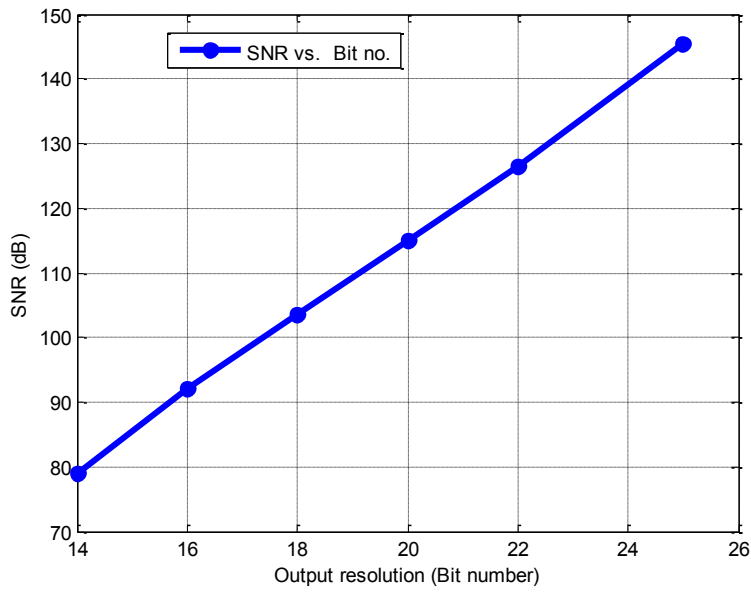

Figure 5.8 Estimated SNR vs. output resolution of the CIC filter

Figure 5.9 illustrates five stages of the CIC filter when $B_{max}$ is 25 bit. Thus the truncation is applied to reduce the width of register and to 16 bit output. Whilst the truncation process can be applied at the integrator stage, the truncation process can not be applied to the comb stage due to SNR reduction in the filter output.

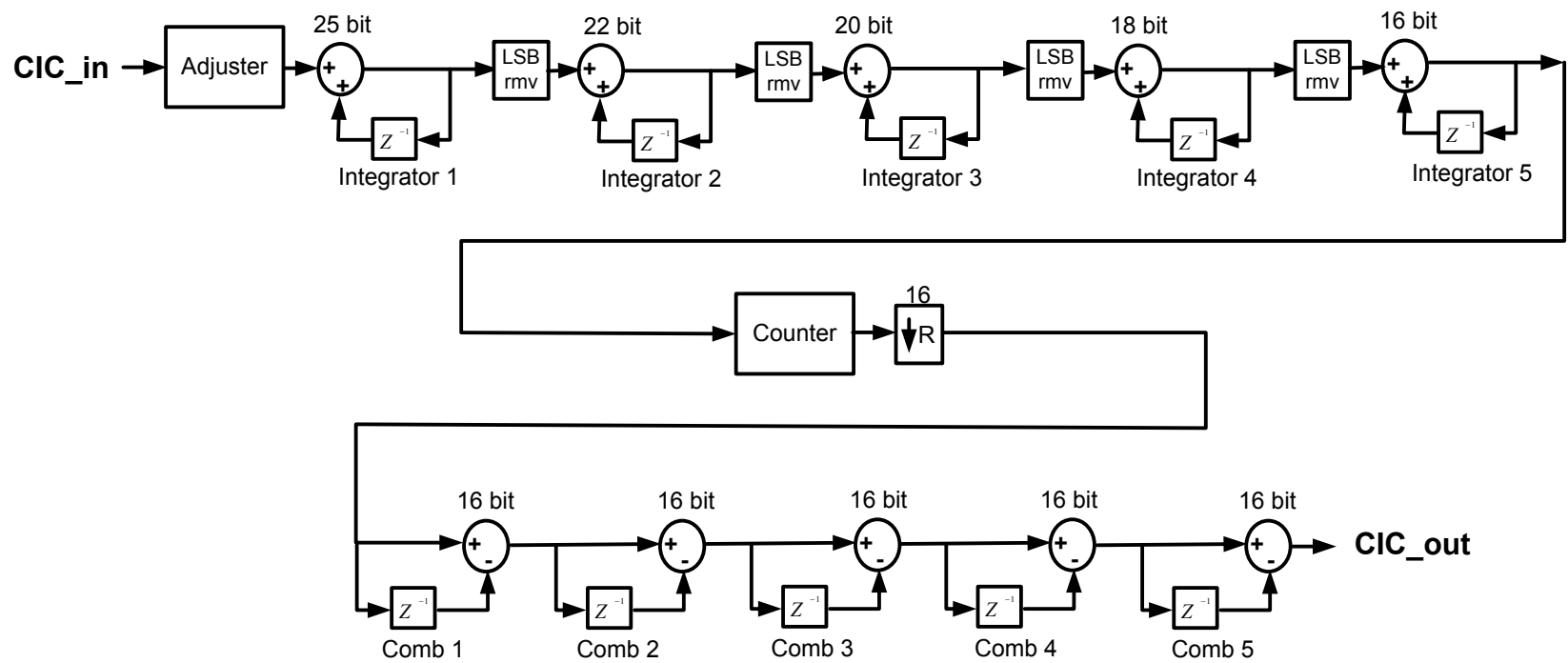

Figure 5.9  5-order CIC filter with truncation

## 5.2.3 PIPELINE CIC FILTER STRUCTURES

Pipeline structure is a technique to increase the throughput of the CIC filter. This structure improves the digital filter efficiency. The outputs of the adder and subtractor are stored into the pipeline register located between the sub-components in the filter unit. By incorporation these internal registers, the pipelined unit operates faster. This reason stands for the concurrent operation of the all stages. However there is also a latency introduced by these pipelining structures. Fortunately, this is not an issue in most applications (Srivastava 1995). This means that the number of clock cycles between the first input data and the appearance of valid output data is increased in pipeline architectures. It is due to the location of the register delay after adder in the CIC structure.

The latency depends on $N$, where $N$ is the number of stages in either the comb or the integrator section of the filter. Thus, higher number of $N$ increases the system latency. The pipeline adder is implemented for the MCLA as shown in Figure 5.10. The load and reset pins are the external pins to provide a better control for the CIC filters.

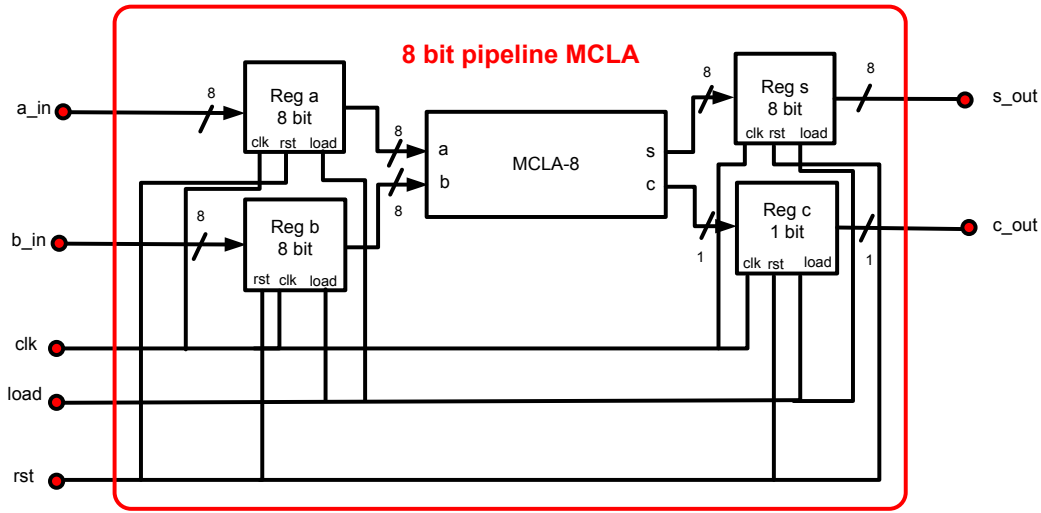

Figure 5.10  8-bit pipeline adder in CIC structure

The pipeline architecture can be expanded for the CIC filter as shown in Figure 5.11. The pipeline registers were inserted in between every comb structure.

Putting additional pipelined registers between integrators is not a recommended solution for low latency and low power consumption. Instead, the pipeline integrator is implemented by moving the feedback register to the location of the pipeline register (Djadi et al. 1994). This results the pipeline implementation of the integrator but with less one register for every integrator stage which is similar to the structure of the non-pipeline integrator. The CIC decimation filter clock rate is determined by the first integrator stage that causes more propagation delay than any other stage due to the maximum number of bit. Thus the high clock rate is used at the integrator of the CIC filter, to obtain high throughput. The clock rate in integrator section is R times higher than in the comb section. As shown in Figure 5.11, pipeline architecture is used for the integrator, comb and down sampler that results in a further increase of the CIC filter throughput by 20 MHz.

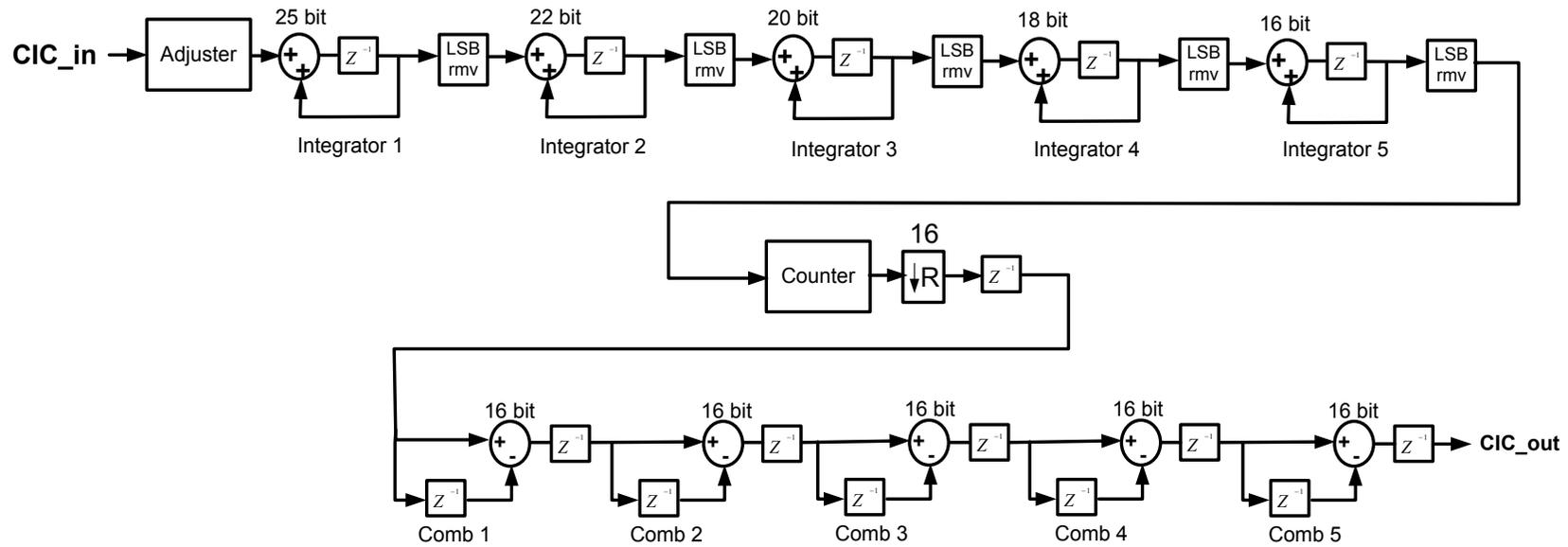

Figure 5.11 Pipeline CIC filter architecture with truncation

## 5.3  PIPELINE RIPPLE CARRY ADDER/SUBTRACTOR (RCAS)

The CIC filter structure has five subtractors in the 5 cascaded differentiators. To perform subtraction, it is necessary to take the one's complement of the bits of *B* and add 1 to the LSB (Stine 2004). In this research work, ripple carry adder subtractor (RCAS) is designed to carry out the subtraction. This structure is shown in Figure 5.12. Note that the "Sub" pin should be 1; otherwise the RCSA is converted to the adder.

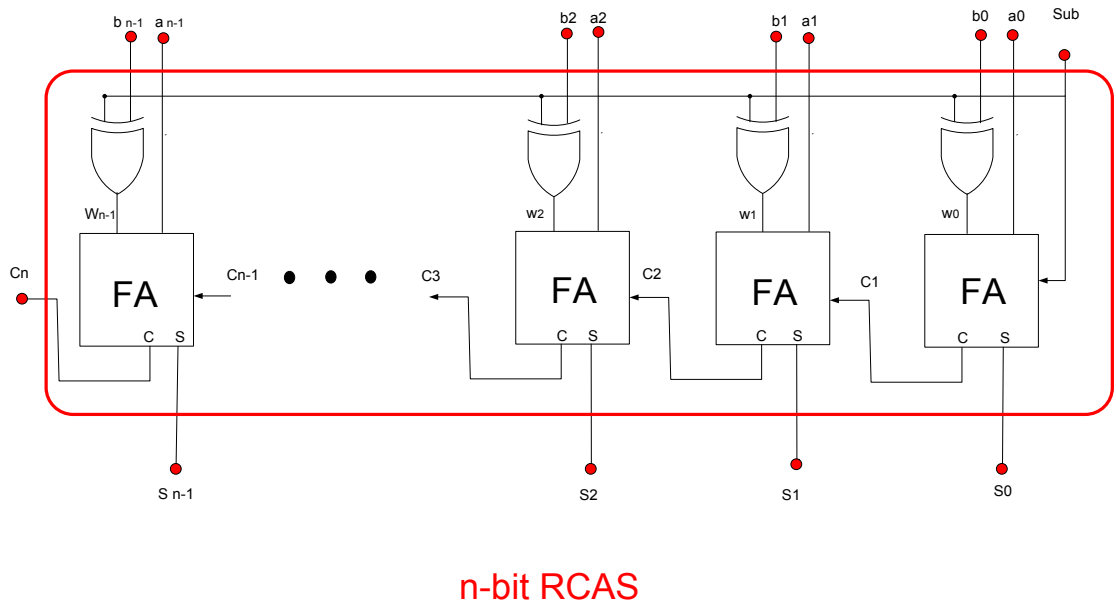

n-bit RCAS

Figure 5.12  N-bit  RCAS subtractor architecture

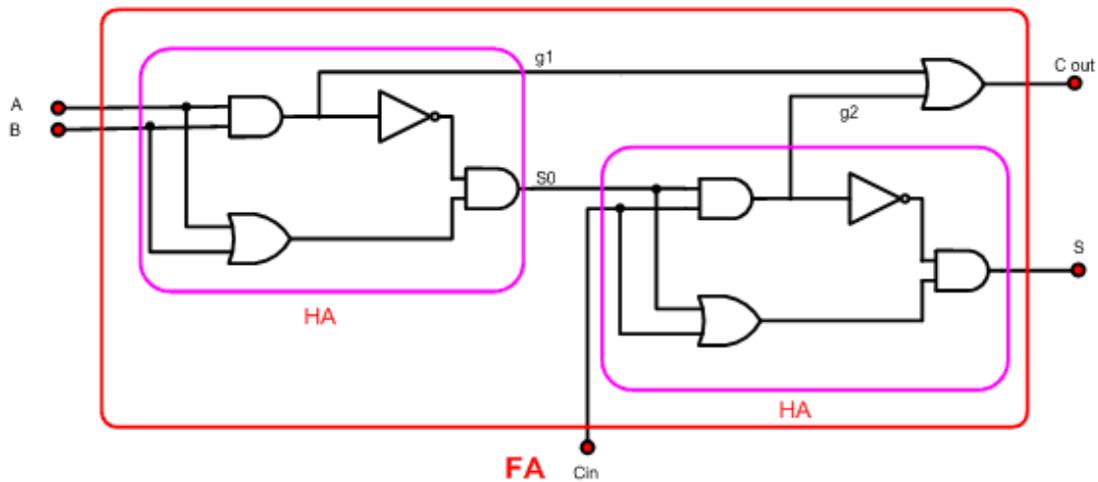

Figure 5.13  Full adder architecture of the RCAS

To achieve pipeline subtractor pipeline registers are added in RCAS structure as shown in Figure 5.14.

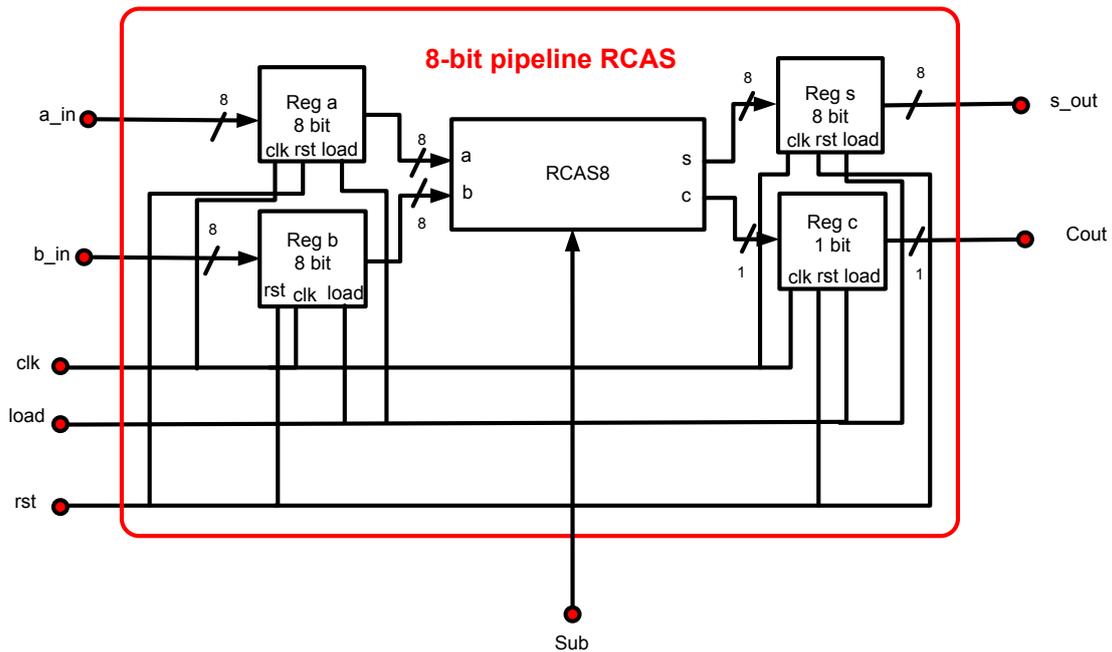

Figure 5.14 Pipeline RCAS structure

Xilinx ISE software synthesized the RCAS design for FPGA implementation. Table 5.5 shows Xilinx ISE synthesis result after RCAS implementation. The Table shows that the maximum throughput of the 8-bit subtractor is 219 MHz.

Table 5.5  Xilinx ISE synthesis summary for RCAS

|  | RCAS |
|---|---|
| Minimum period (ns) | 219.010 |
| Maximum Frequency (MHz) | 4.566 |
| Minimum input arrival time before clock (ns) | 5.021 |
| Maximum output required time after clock(ns) | 4.575 |
| Total equivalent gate count for design | 96 |

## 5.4  DOWNSAMPLER

As mentioned before, the down sampler block re-samples input data R times lower than the input rate. Offset sample must be less than *R-1*. To avoid any data lost at the beginning of the sampling, it is better to initiate the counter of the down-sampler by zero value.

There are two down sampler types which are multirate down sampler and single rate down sampler. Down sampler is multirate when the output sample period is $R$ times longer than the input sample period $(T_{SO} = R \cdot T_{SI})$. Down sampler is single rate when the output sample rate is the same period with input sample rate $(T_{SO} = T_{SI})$ by repeating every $K^{th}$ input sample at the output.

In this project, multirate down sampler is selected to generate the output at a slower rate than the input. The schematic implementation of down-sampler is shown in Figure 5.15. The down-sampler is designed and implemented using Verilog HDL. The decimation factor is selected to be 16. The synthesis report of the Xilinx ISE software shows the maximum throughput of the down-sampler is 352.72 MHz. Decimation factor is defined by down sampler block.

From the MATLAB simulation shown in Figure 5.16, decimation factor has reverse relation with the CIC filter throughput. Thus decimation process is designed in multistage decimation filter and CIC filter is implemented with the decimation factor of 16.

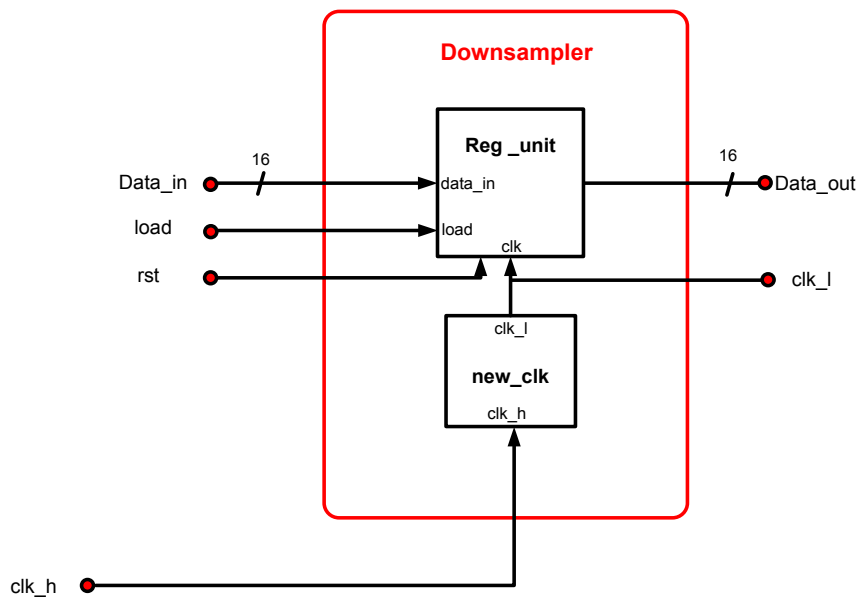

Figure 5.15 Multirate down-sampler structure

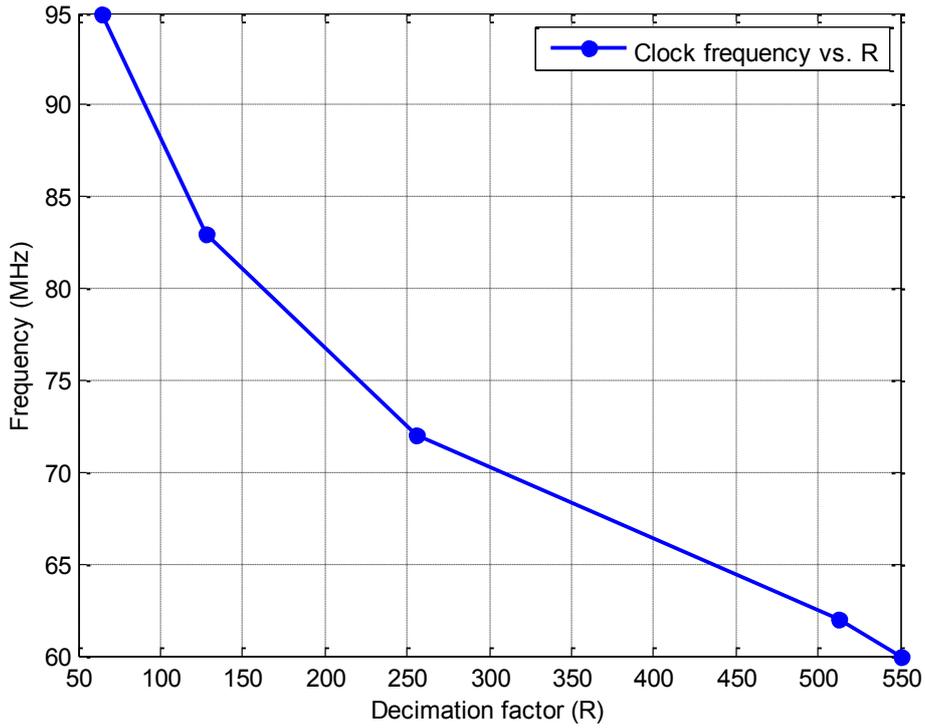

Figure 5.16  Decimation factor versus clock frequency in CIC filter

## 5.5  HIGH SPEED CIC IMPLEMENTATION

Based on the previous discussion, the pipeline (truncated) CIC filter is implemented using MCLA adder and RCAS subtractor. Then follow the standard procedure for FPGA, the CIC filter was downloaded into the FPGA chip.

Top level of the pipeline CIC filter is shown in figure 5.17.

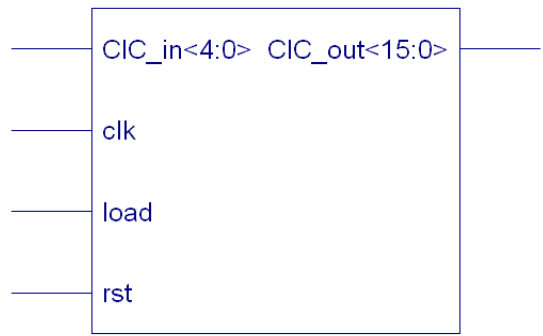

Figure 5.17  CIC filter top level

Figure 5.18 shows overall implementation of the CIC filter which includes the cascaded integrator, the cascaded comb, the down sampler and the adjuster. The output signal is truncated to 16 bit resolution.

Figure 5.19 illustrates the implementation of the $5^{th}$-order cascaded integrator. This stage which operates in high sampling rate is truncated to 25, 22, 20, 18 and 16 bit. The Integrator limits the CIC filter throughput due to the maximum number of bit in the first stage.

The $5^{th}$ – order cascaded differentiator / comb is shown in Figure 5.20. This stage operates in low sampling rate of 16 bit.

Figure 5.21 shows the down-sampler with the decimation ratio of 16. Down-sampler creates low clock frequency for cascaded differentiator.

Figure 5.22 and 5.23 show implementation of the one stage integrator and comb filter respectively in the CIC structure.

Figure 5.24 and 5.25 show implementation of the RCSA and MCLA respectively. The RCSA is the subtractor when the "Sub" pin is connected to the voltage supply.

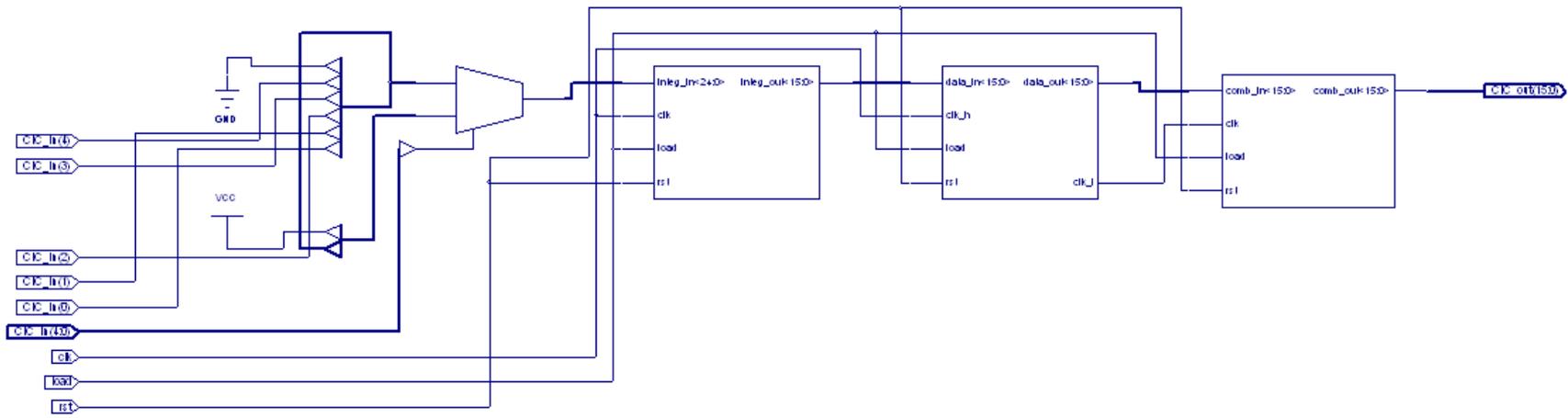

Figure 5.18  5-order CIC filter block

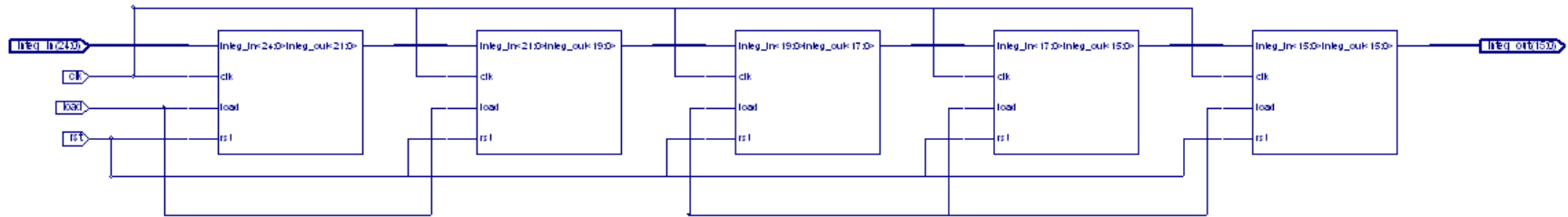

Figure 5.19  5-order cascaded integrator block

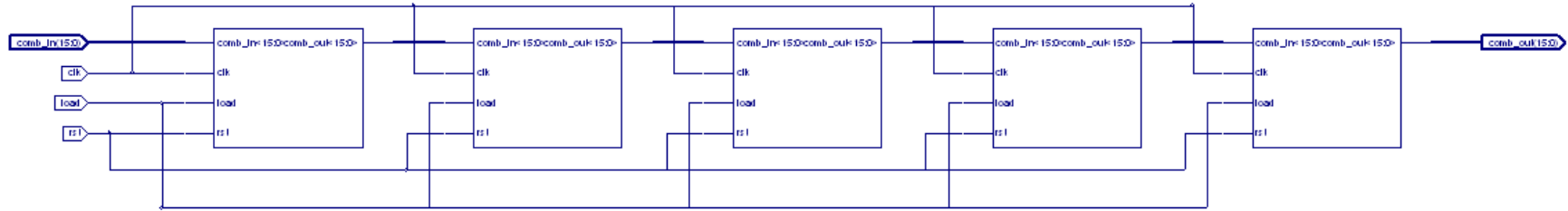

Figure 5.20  5-order cascaded differentiator block

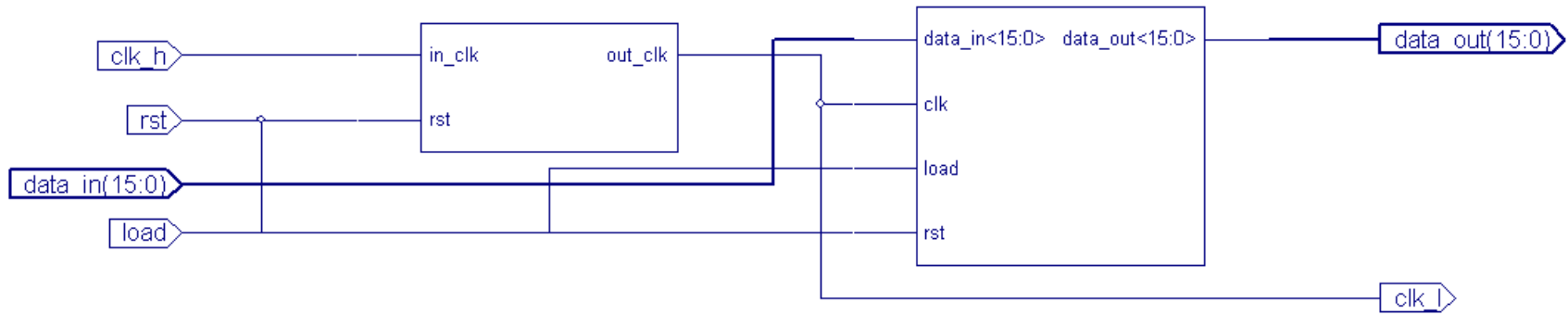

Figure 5.21  Down-sampler block

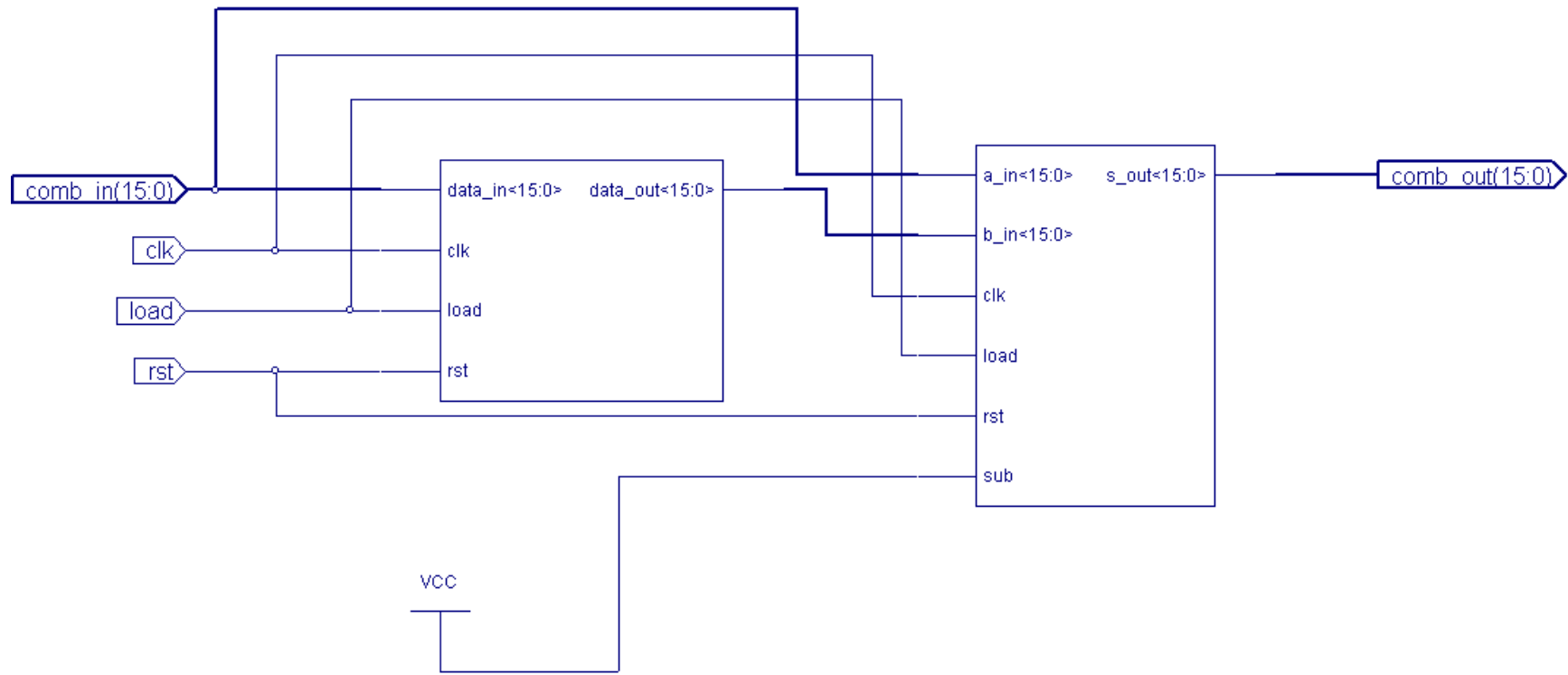

Figure 5.22 One stage differentiator

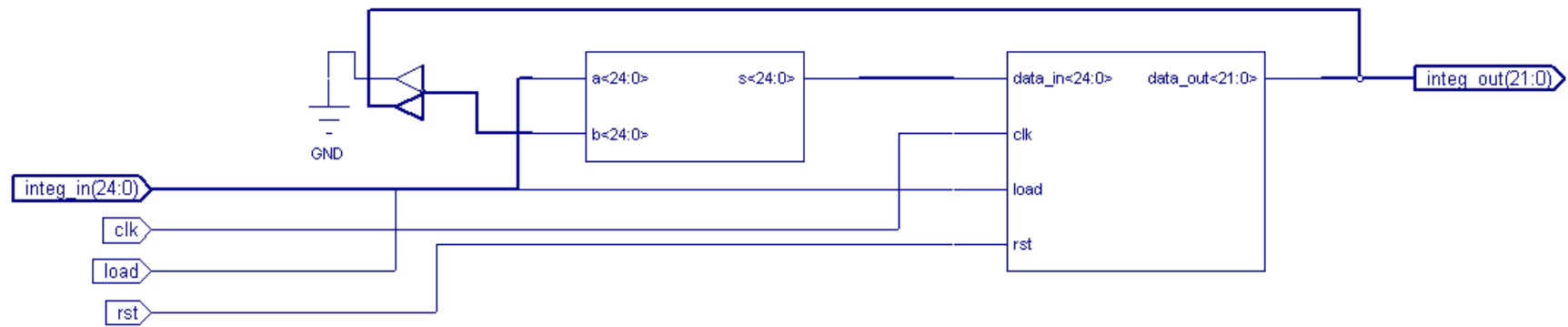

Figure 5.23 One stage integrator implementation

Figure 5.24  16-bit RCAS

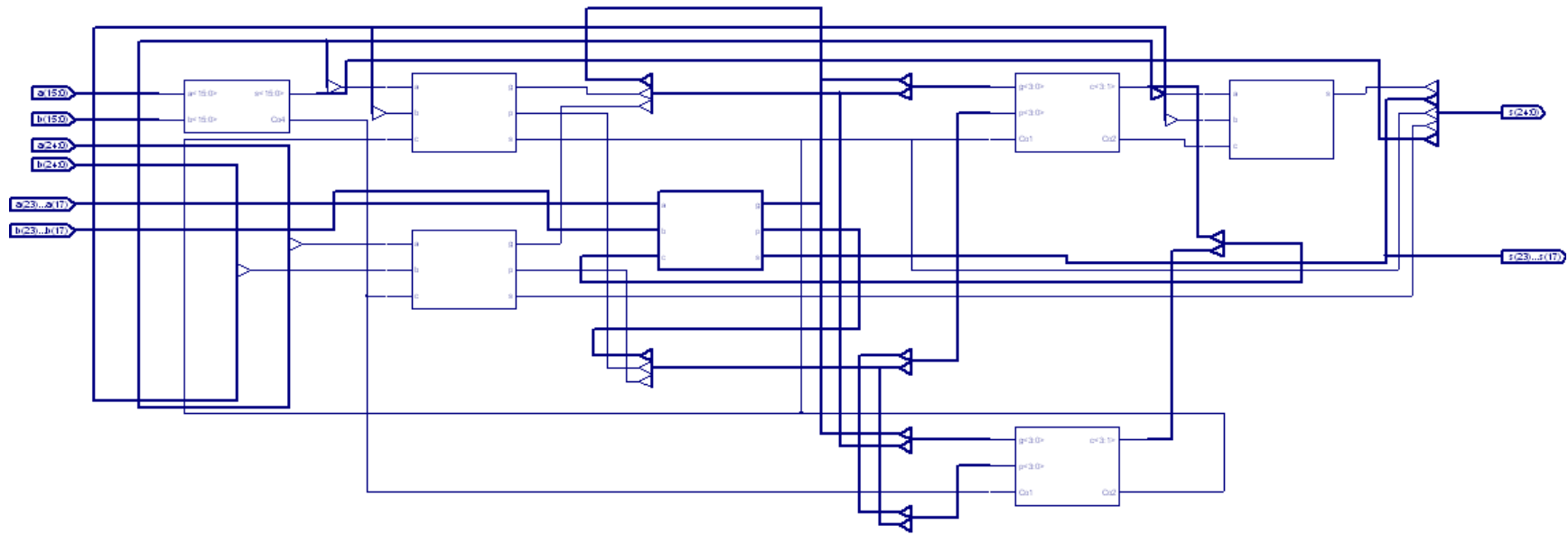

Figure 5.25   25-bit MCLA

The synthesis report of the Xilinx ISE is shown in Table 5.6.

Table 5.6 Xilinx synthesis summary of the high speed CIC filter

| HDL Synthesis Report | | Timing Summary | |
|---|---|---|---|
| | No. | | |
| MCLA (adder) | 5 | Minimum period (ns) | 5.293 |
| RCAS (subtractor) | 5 | Maximum Frequency (MHz) | 188.947 |
| 4-bit down counter | 1 | Minimum input arrival time before clock (ns) | 5.127 |
| Register | 18 | Maximum output required time after clock(ns) | 4.575 |
| Multiplexer | 2 | Total equivalent gate count for design | 6519 |
| Number of 4-input LUT cells | 1344 | | |
| Number of external IOB | 33 | | |
| Flip Flop Latch | 273 | | |

The table shows the CIC filter can operate up to a maximum frequency of 189 MHz. The FPGA power consumption depends on the amount of configurable blocks which are used for the filter implementation. This filter was implemented using $0.18\,\mu m$ and $0.35\,\mu m$ technologies which are expected to have better performance the FPGA chip. Results are explained in the later sections.

## 5.6  FPGA DOWNLOADING

Additional blocks such as clock divider, memory and up-counter are required for downloading the high speed CIC filter design on the FPGA board. As shown in Figure 5.26, memory is used to load the input data. Then, data is transferred to the CIC filter by memory-up-counter connection. The counter is trigged by every positive edge of the clock cycle. The 13-bit counter is connected to the output of the clock divider. The clock divider provides lower rate of clock pulse for the counter. Reducing the clock frequency, will make visible on the LED of the FPGA board. Since the resolution is more important than the speed in audio application, this research work does not consider truncation to reduce the output number of bit. Thus the output of the CIC filter was chosen to be 25-bit.

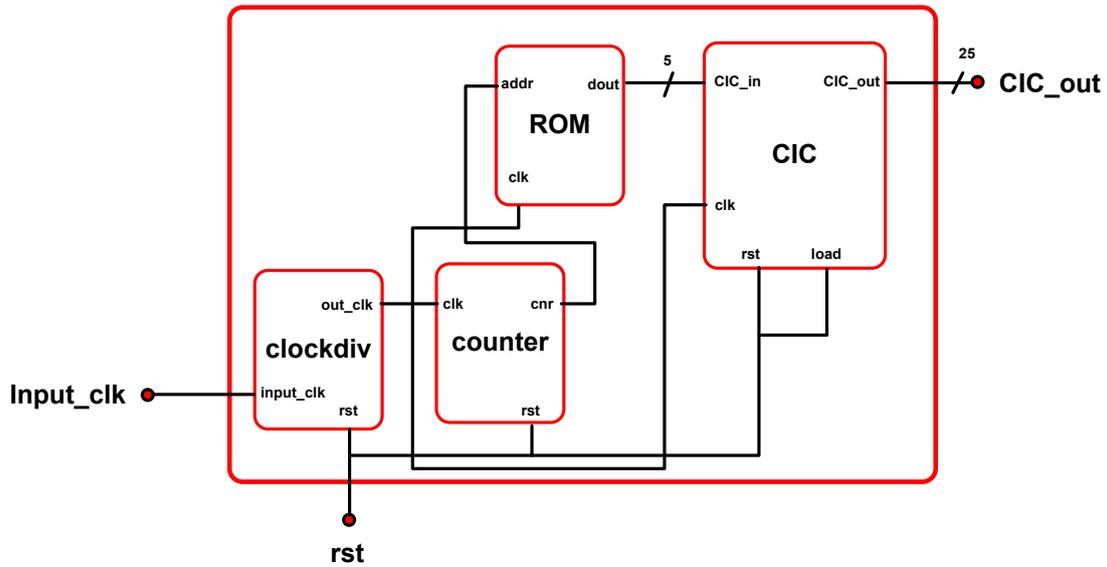

Figure 5.26  Schematic design of the CIC filter using ROM on FPGA board

The total block was synthesized by Xilinx software as shown in Figure 5.28. The measured performance of the CIC filter after place and route (PAR) process is shown in Figure 5.27. These measured values were shown in Table 5.5.

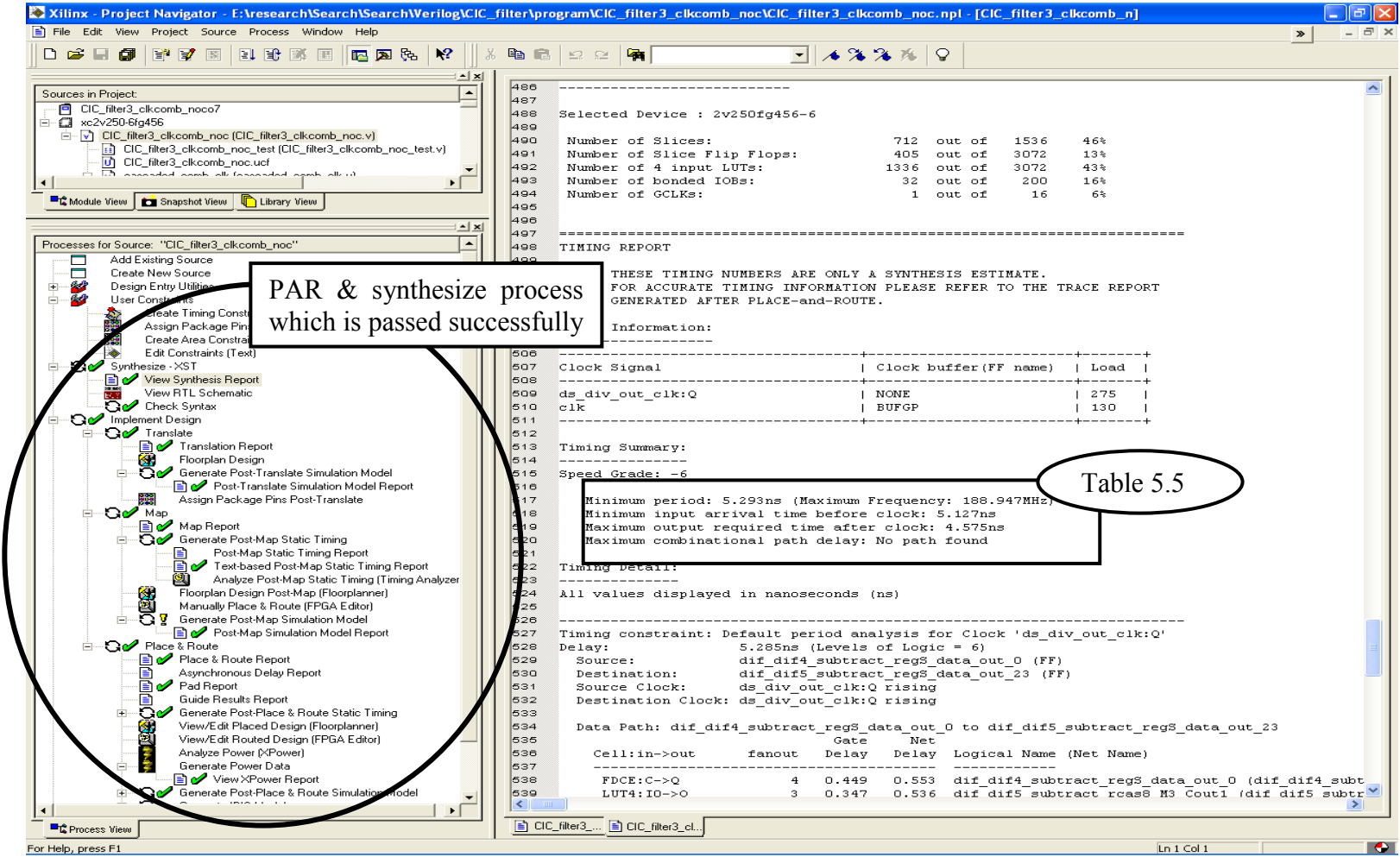

Figure 5.27  Successful implementation of place and route (PAR) the CIC filter in Xilinx synopsis ISE

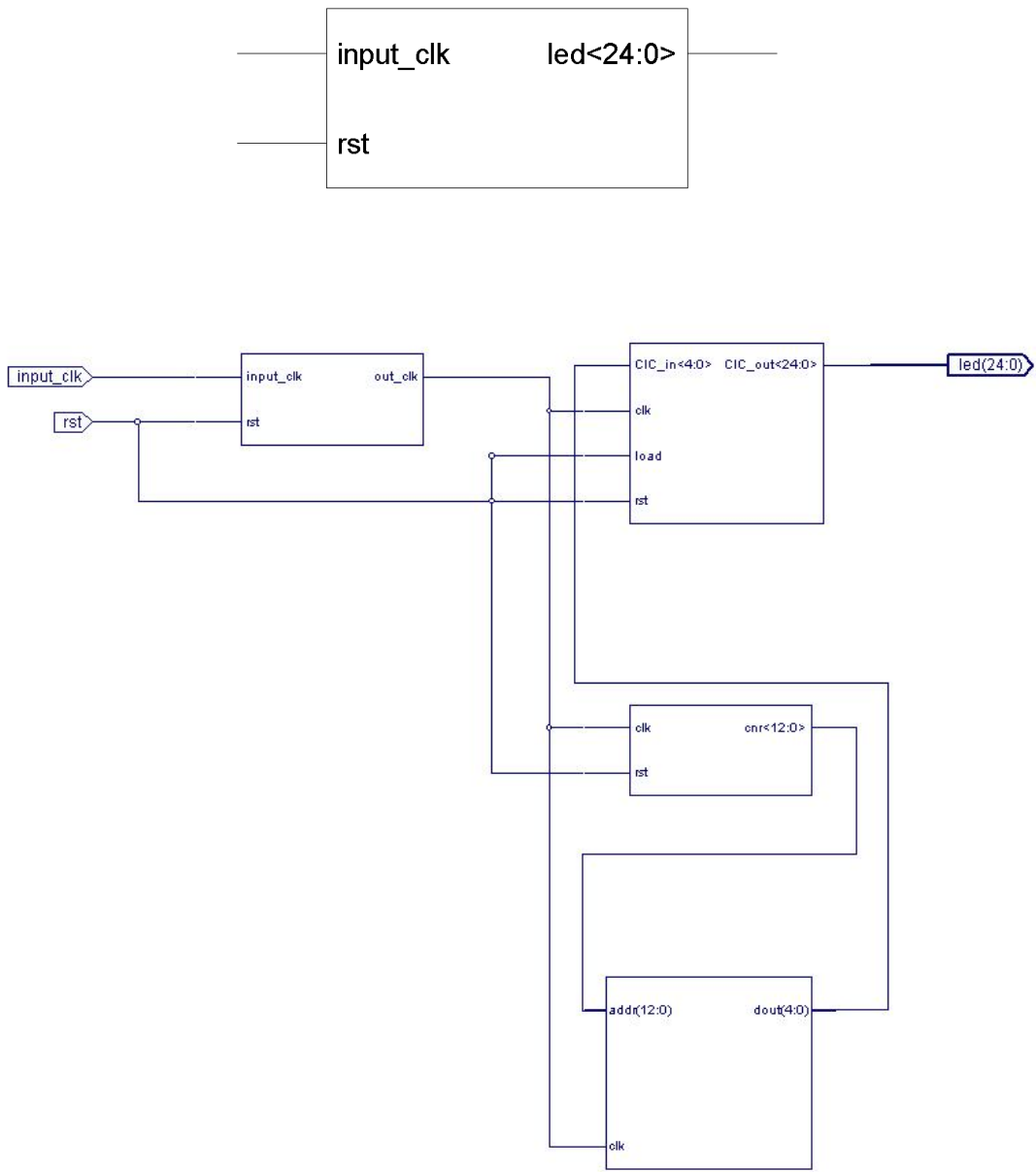

Figure 5.28    Implementation of the CIC filters with ROM and clock divider and counter on FPGA

## 5.7 ASIC OPTIMIZATION (GATE LEVEL SYNTHESIS )

An application specific integrated circuit (ASIC) is an integrated circuit (IC) modified for particular use, rather than intended for general purpose use. ASIC is often termed a system on a chip (SOC) and bring about great saving of cost and lower power consumption. Minimal propagation delays can be achieved in ASICs versus the FPGA. To describe the system functionality on ASIC, hardware language such as Verilog and VHDL are used.

From above explanation, the CIC filter is synthesized in ASIC to optimize the area and power consumption by using Synopsys and Cadence tools. The technology libraries to do optimization are SilTerra 0.18 $\mu m$ and MIMOS 0.35 $\mu m$. In this section the design is synthesized to convert from RTL code into the gate level. Timing constraints is defined to meet the final chip result. The RTL design code is divided into two parts, core and chip. The chip contains the top wrapper module that instantiates this core and puts I/O buffer pads along ports. In fact, analysis and simulation are required to ensure that the power supply is adequate. In this section the functionality of the Verilog behavioral CIC design is verified by Cadence Verilog simulation. In addition, the RTL model of the CIC design is converted into the Netlist (schematic).

Using design compiler (Synopsys tool), the CIC filter is compiled and executed. Firstly, all the module and submodule in the CIC design are analyzed to check the Verilog syntax of the design. It is converted into intermediate format and stored in the specified library. Secondly, the top design is elaborated to convert the CIC into generic gates and logic blocks. The elaboration reports the memory elements in the CIC architecture. After elaboration process the overall system is checked for debugging the error. All errors are corrected to precede design to the next stage. Figure 5.29 and 5.30 show the symbol and schematic view of the CIC filter, respectively.

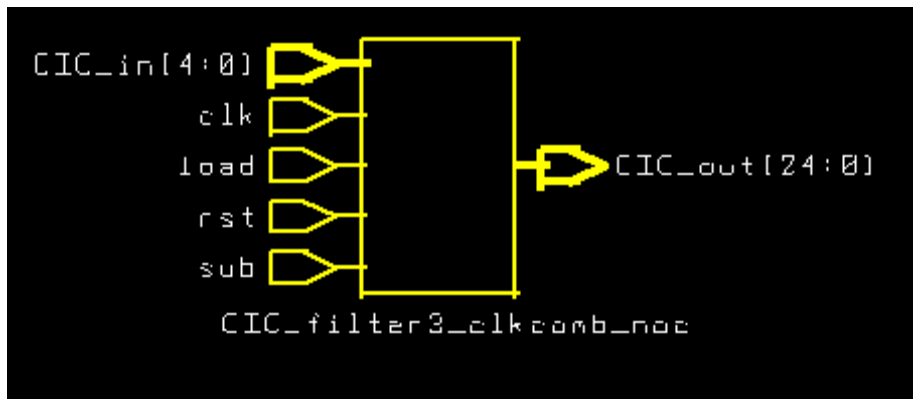

Figure 5.29  Symbol view of the CIC filter

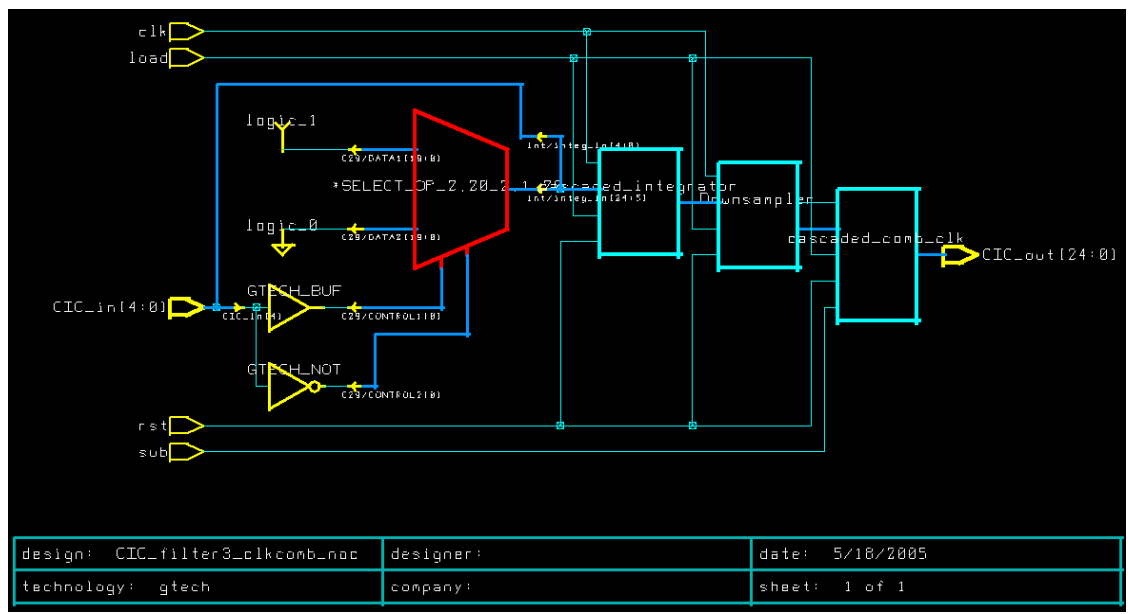

Figure 5.30  Schematic view of the CIC filter

Now the top module with suffix .db is created which is the I/O pad wrapper. Some constraints must be defined for the CIC filter design. These constraints are necessary conditions that have to be met for a proper functionality of the circuit. The most important constraint is related to the signal timing of the system. Constraint for clock includes the period and skew. It is necessary to specify the clock period, to show the maximum speed of the CIC design. All the constraints are fixed to the design prior compiling. These are typically applied to the top module and propagated down to the lower level module as separated step. For defining constraints, the input and output pins had to be set before compiling the CIC design. The maximum fan-out and timing are determined for driving pins to change logic values. Within Synopsys there is a power analysis tool which estimate power consumption under various conditions and configurations. After defining constraint, the design is checked to ensure that there is no any driving of multiple ports by a single net and also all constrained is defined on the top module.

Compiling the CIC core performs gate-level synthesis and optimization. After compiling the CIC filter, the netlist is created. The estimated area and power consumption result of the high speed CIC filter in SilTerra 0.18 $\mu m$ and MIMOS 0.35 $\mu m$ technology in maximum clock frequency is shown in Table 5.7.

Table 5.7  The estimated power and area result in different technology libraries

| The CIC specification | Silterra 0.18 $\mu m$ technology | Mimos 0.35 $\mu m$ technology |
| --- | --- | --- |
| Active core area $(mm^2)$ | $0.308 \times 0.308$ | $1.148 \times 1.148$ |
| Power consumption $(mW)$ | 3.14 | 6.03 |

The schematic gate level of different components in the CIC filter is shown below. The filter in gate level is simulated by NCLaunch under Cadence software.

Figure 5.31 shows the top level implementation of the CIC filter. All the filter components are located in the top level block (Figure 5.32). Figure 5.33-5.35 show the filter sub-blocks consist of differentiator, down-sampler and integrator. The CIC Filter in gate level after creating the netlist is shown in Figure 5.36 and 5.37.

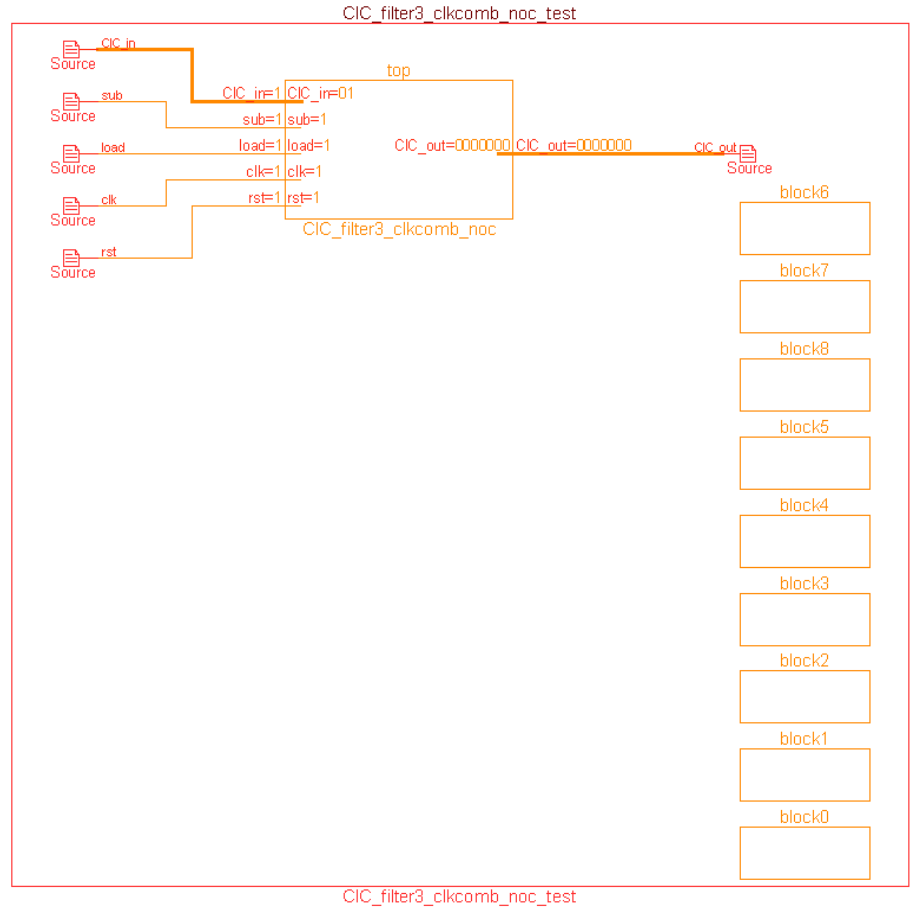

Figure 5.31  Top level implementation of the CIC filter

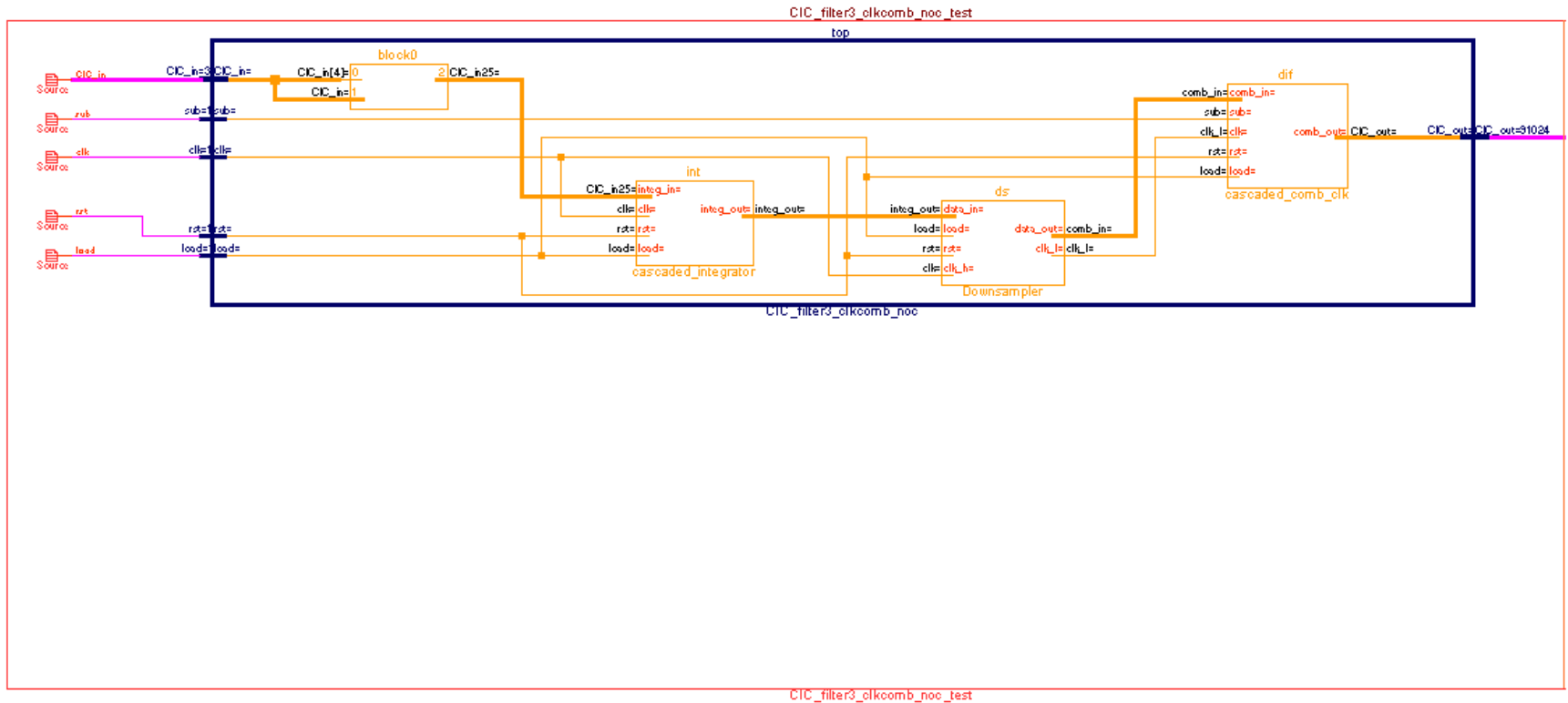

Figure 5.32 CIC filter Sub-blocks

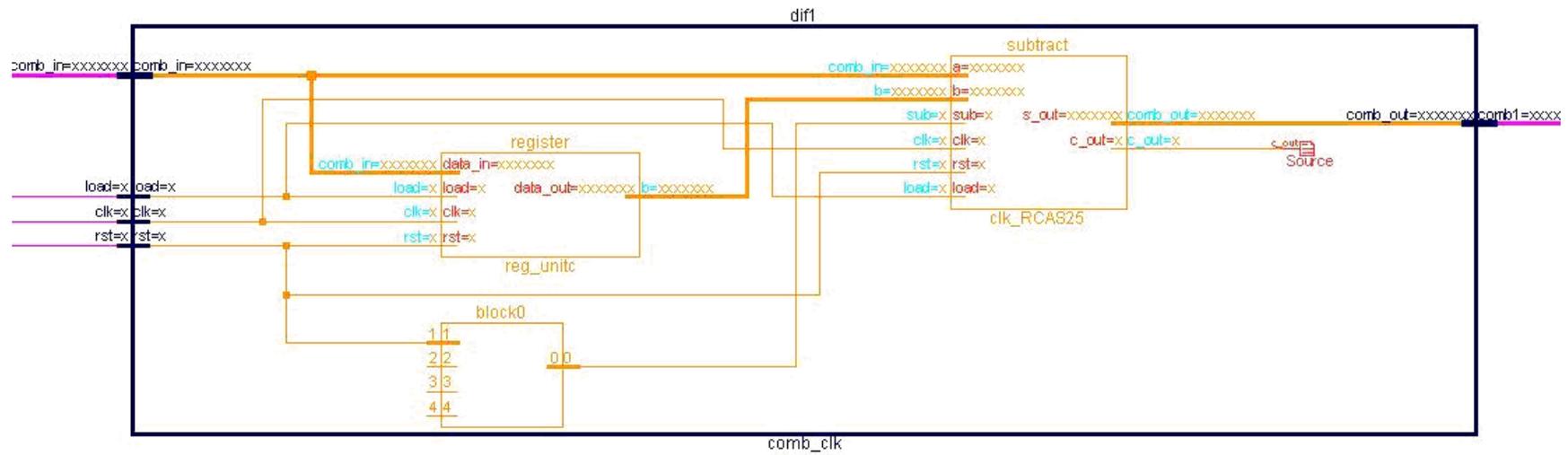

Figure 5.33 Differentiator stage

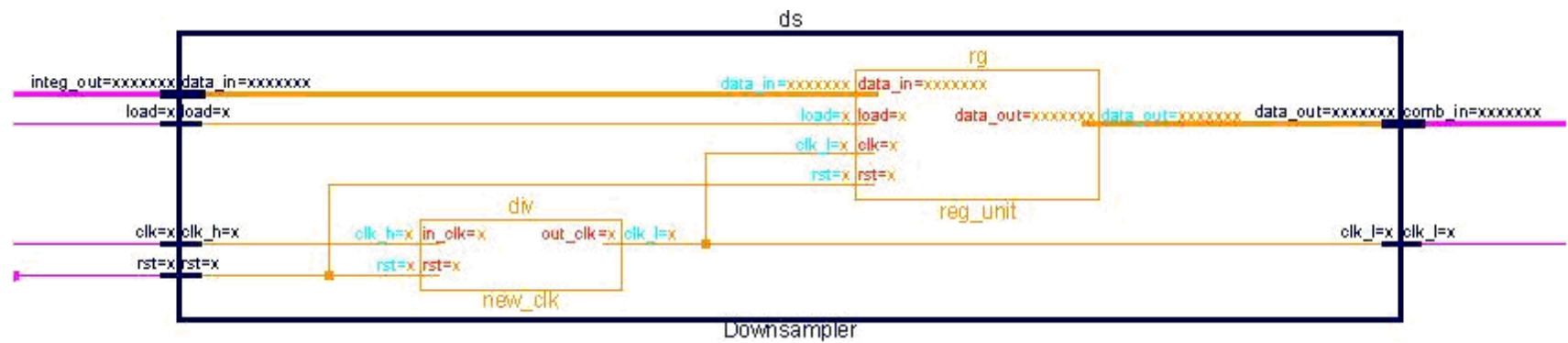

Figure 5.34 Down sampler stage

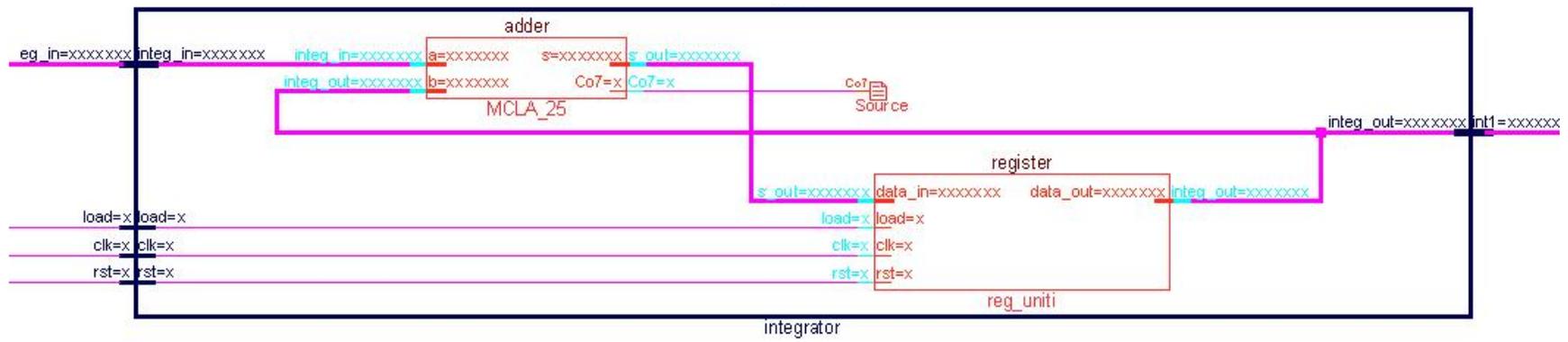

Figure 5.35 Integrator stages

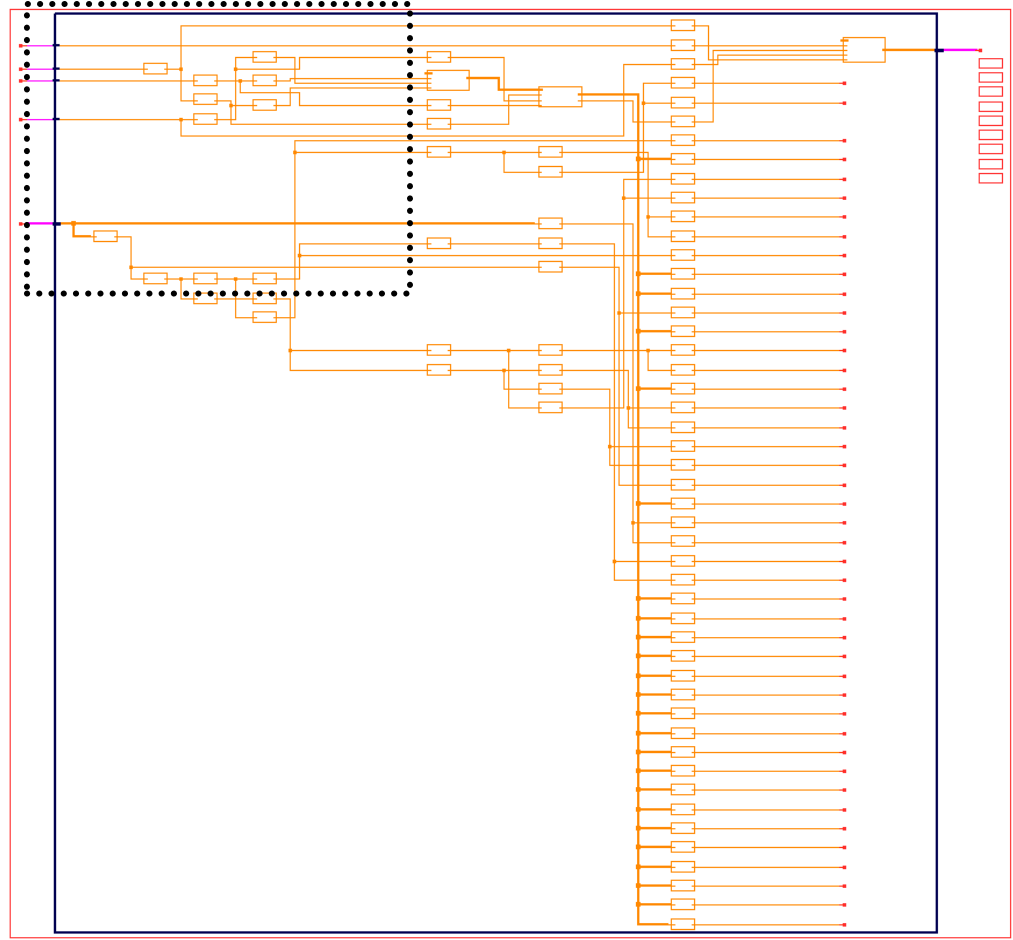

Figure 5.36  Gate level implementation of the overall CIC filter

Figure 5.37  Design extension of dotted line of figure 5.36

## 5.8 SUMMARY

In this chapter, the digital decimation CIC filter architectures were discussed and implemented. Several methods were explained and performed to increase the throughput of the filter. Firstly the standard CIC filter is implemented and then the adder was first replaced with the CSA. The system was synthesized and shown that the CIC filter could operate with the maximum frequency of 97 MHz. Then the MCLA architecture is introduced to perform addition. Implementation result shows that maximum throughput of the 8-bit MCLA is 270 MHz. The new architecture of pipeline truncated high speed CIC filter was designed and programmed to the FPGA board. The Xilinx synthesis result shows that the filter can work with the maximum clock frequency of 189 MHz.

Additionally, the CIC decimation filter was also synthesized in SilTerra $0.18\,\mu m$ technology and the MIMOS $0.35\,\mu m$ technology. After synthesizing and constrained definition, the CIC filter netlist (gate level) was created to produce ASIC layout. The design compiler result shows the filter power consumption and die area of 3.14 mW and $0.308 \times 0.308\ mm^2$ respectively in Silterra $0.18\,\mu m$ technology and 6.03 mW and $1.148 \times 1.148\ mm^2$ respectively in Mimos $0.35\,\mu m$ technology.

# CHAPTER VI

# RESULTS AND DISCUSSIONS

The chapter presents the simulation results of the designed CIC filter. The simulation process is first conducted for entire system that form the oversampling and decimation system. Once it is verified, then the implementation results of the CIC filter will be demonstrated. The FPGA and ASIC implementation of the CIC filter will be presented. A comparison is made between different technology libraries (Mimos and Silterra technology) in terms of power consumption and area die size. Finally the layout of the CIC filter in 0.18 technology and FPGA will be presented.

## 6.1　FUNCTIONAL VERIFICATION OF THE OVERSAMPLING AND DECIMATION PROCESS USING MATLAB

As an initial step to verify the functionality of the designed CIC filters, Matlab software was used. The oversampling and decimation process is shown in Figure 6.1. Thus the selection of inputs to the block diagram must be properly chosen so that the input to the CIC filter is corrected.

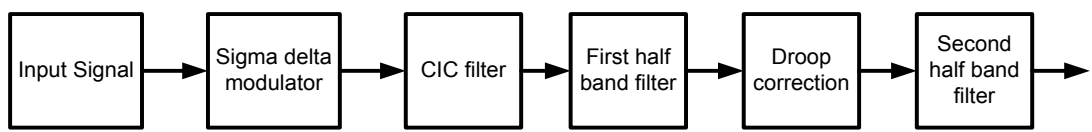

Figure 6.1　Oversampling and decimation process

### 6.1.1 AUDIO INPUT SIGNAL

As shown in Figure 6.1 the input is passed to the sigma delta modulator before it enters into the CIC filters and its decimation process. Thus the value of input must be properly chosen so that the signal arrives at the CIC filters must have certain characteristic that enable it to be used to test the CIC filter. As stated in the reference (Caban & Allen 1996), the SNR of the sigma delta modulator has the properties as shown in the Figure below:

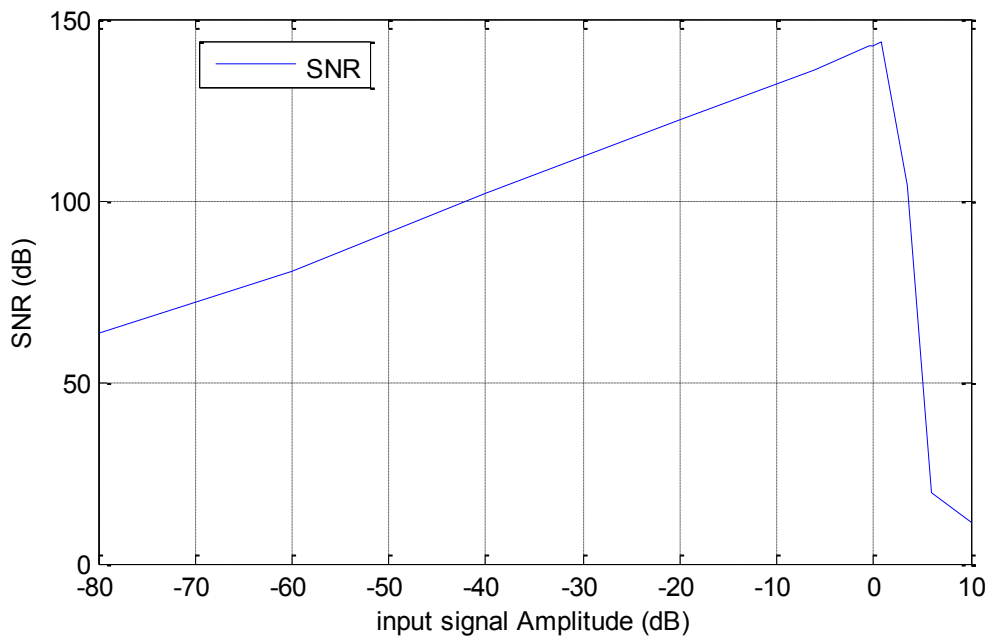

Figure 6.2 SNR as a function of input signal amplitude

Thus SNR of the modulator will drop to zero if the input is greater than 1 V. This necessity is for the input signal to be about 1 V when the SNR is at its map value. The input bandwidth is also chosen to be less than 24 kHz, which is the required bandwidth for the audio application. Figure 6.3 shows the input audio signal used for the testing of our CIC filter.

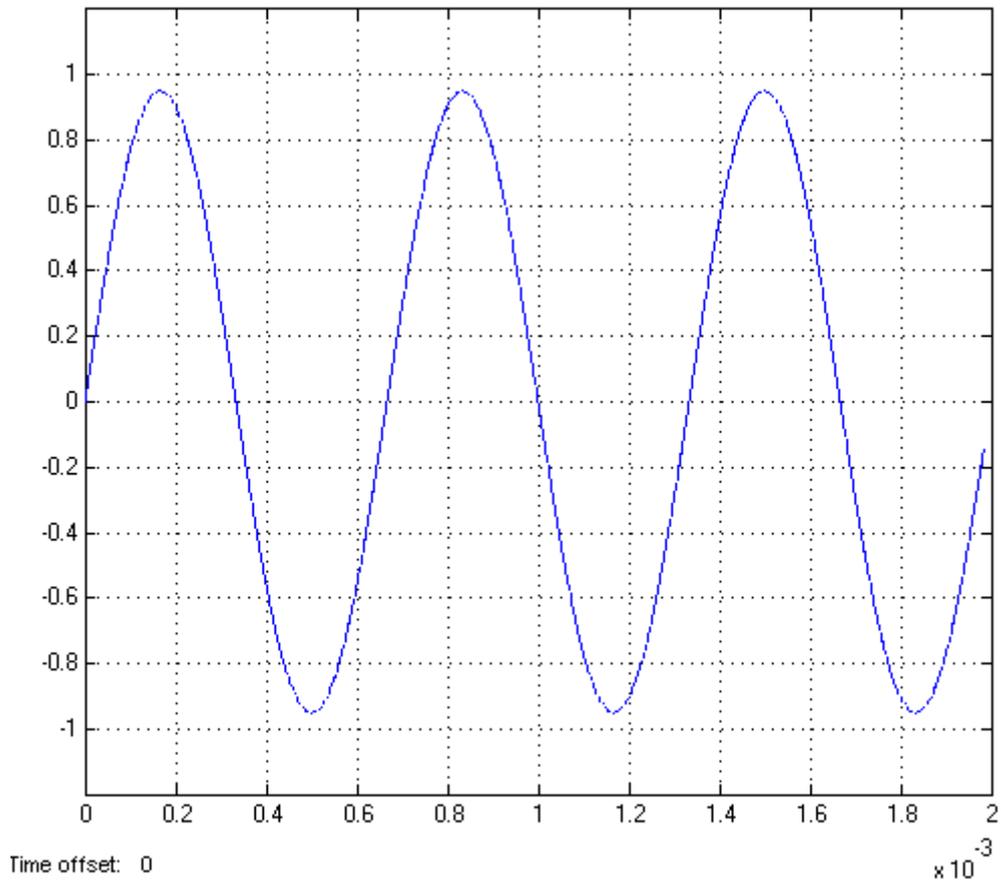

Figure 6.3  Audio input signal

### 6.1.2  SIGMA DELTA MODULATOR SIMULATION RESULTS

Third-order sigma delta modulator converts the analog audio signal to the digital data with the sampling frequency of 6.144 MHz. This signal is quantized by 5-bit quantizer that produces modulator with 5-bit resolution. The modulator output is shown as Figure 6.4.

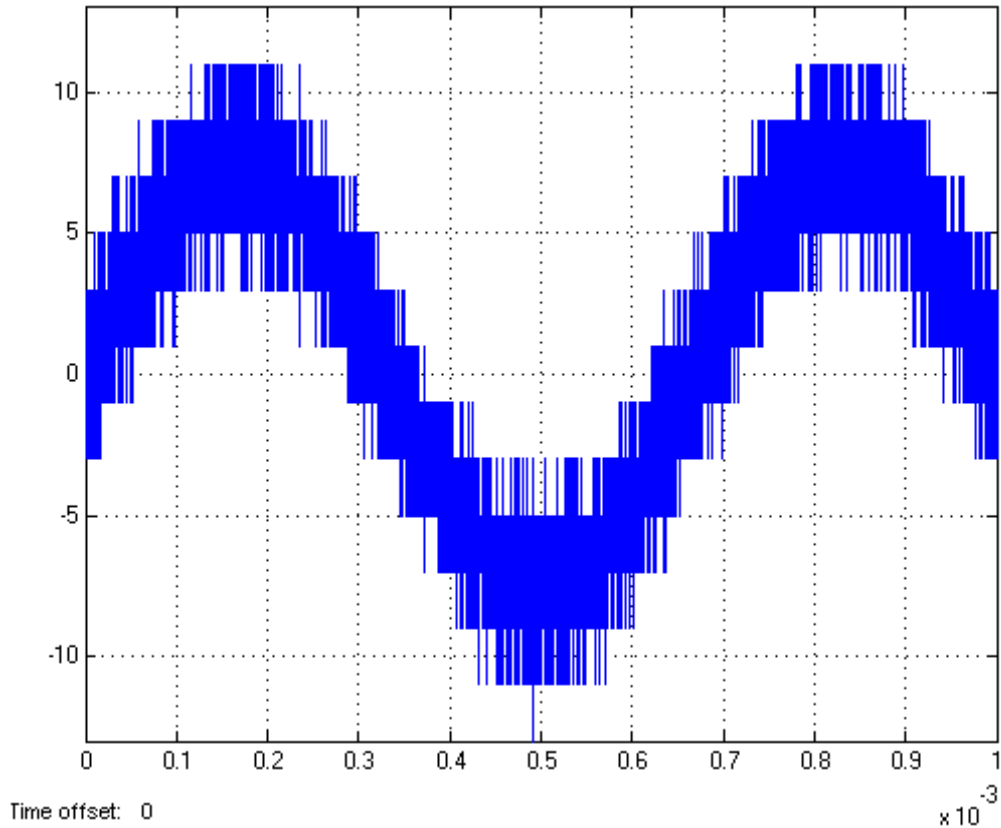

Figure 6.4  Third-order sigma delta modulator output

Figure 6.4 shows the output of the 5-bit sigma delta modulator. As the value of the analog input increases, the corresponding digital value increases and at the peak of the analog sine wave, the value of the digital data is the maximum value. When the input decreases, the digital data value decreases and when the input reaches zero the digital data equal to zero too. As the analog sine wave decreases below zero the data of the output decreases to reach a maximum negative value of digital data. It can be shown that the average value of the digital output equals to the value of analog input.

As mentioned in chapter II, because of the use of the integrator and the oversampling technique, the noise is pushed outside of the audio bandwidth resulting in the increased SNR. From the Matlab simulation, SNR is found to be 141.56 dB in the sigma delta modulator output which is acceptable for audio applications.

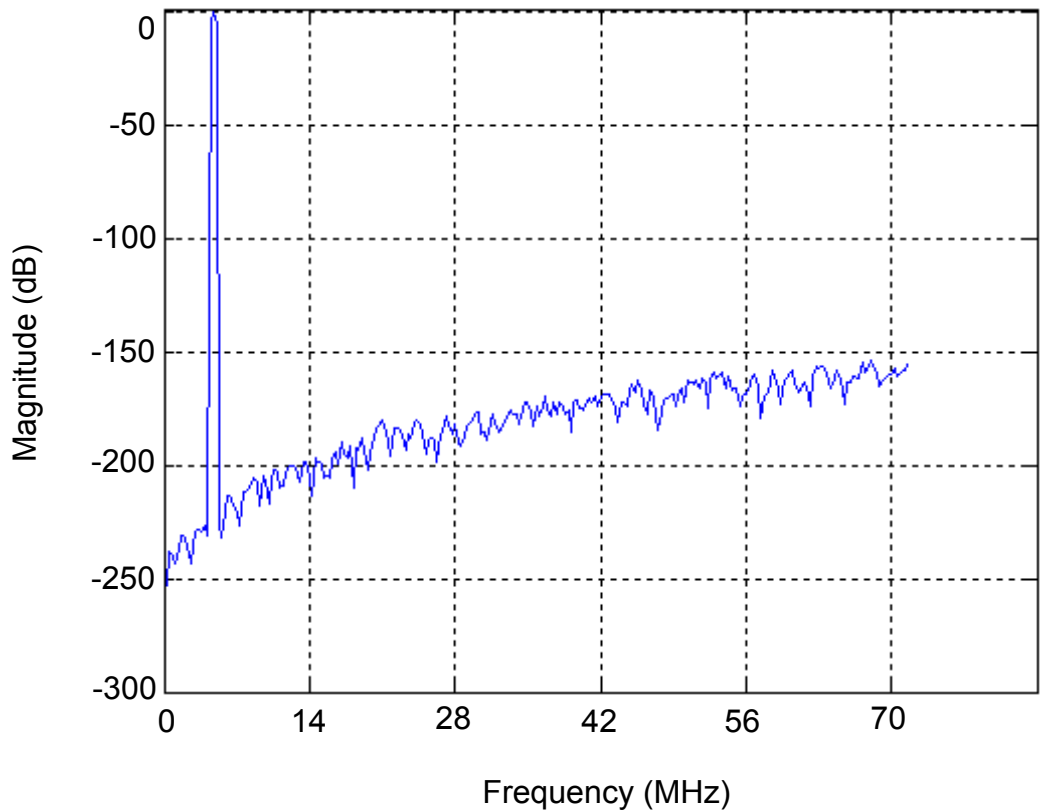

Figure 6.5  Signal to noise ratio of the sigma delta modulator

### 6.1.3  CIC FILTER SIMULATION RESULTS

The 5-bit output for the sigma delta is fed to the fifth order CIC decimation filter. That results in the reduction of the sampling frequency from 6.144 MHz to 384 kHz and as mentioned by the CIC filter, output has 25 bit width resolution. The frequency response of the 5 order CIC filter for *M=1* is shown in Figure 6.6.

As illustrated in the Figure 6.6, the CIC filter has an abrupt droop in the passband (0-384 kHz) that is considered a wide bandwidth. Figure 6.6 shows that the pass band, stopband and transition band of the CIC filter are 7 kHz, 384 kHz and 377 kHz respectively.

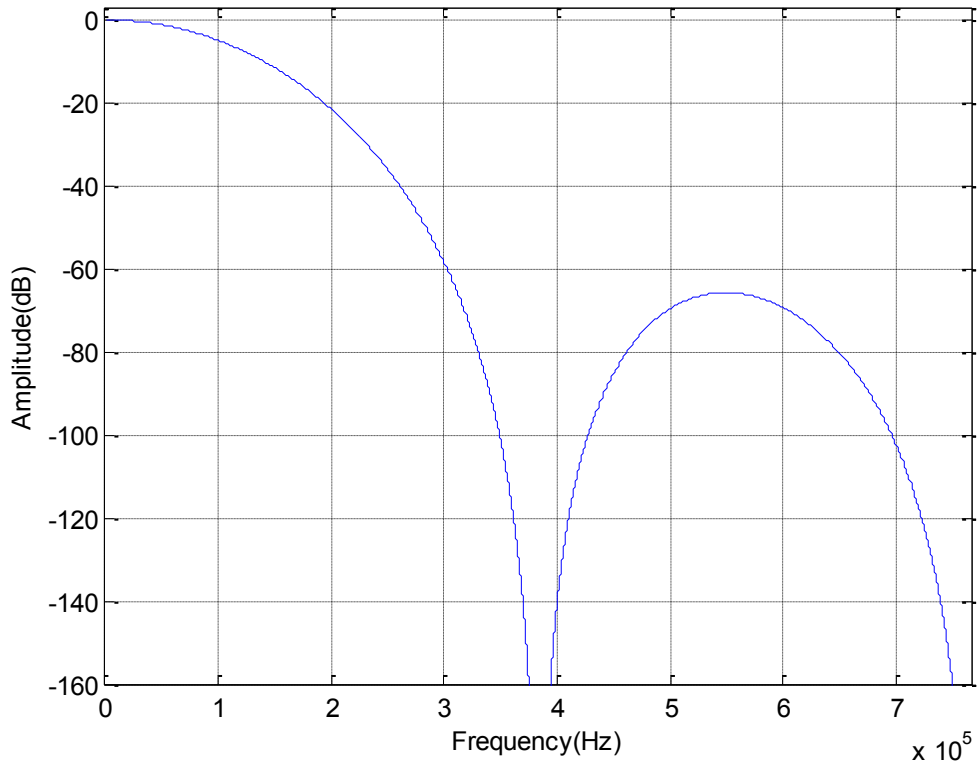

Figure 6.6  Frequency response of the fifth order CIC filter

As a result of the digitized signal passing through the CIC filter, the output of the CIC filter now has the SNR value of 145.35 dB which is a slight improvement over the SNR for the sigma delta modulator.

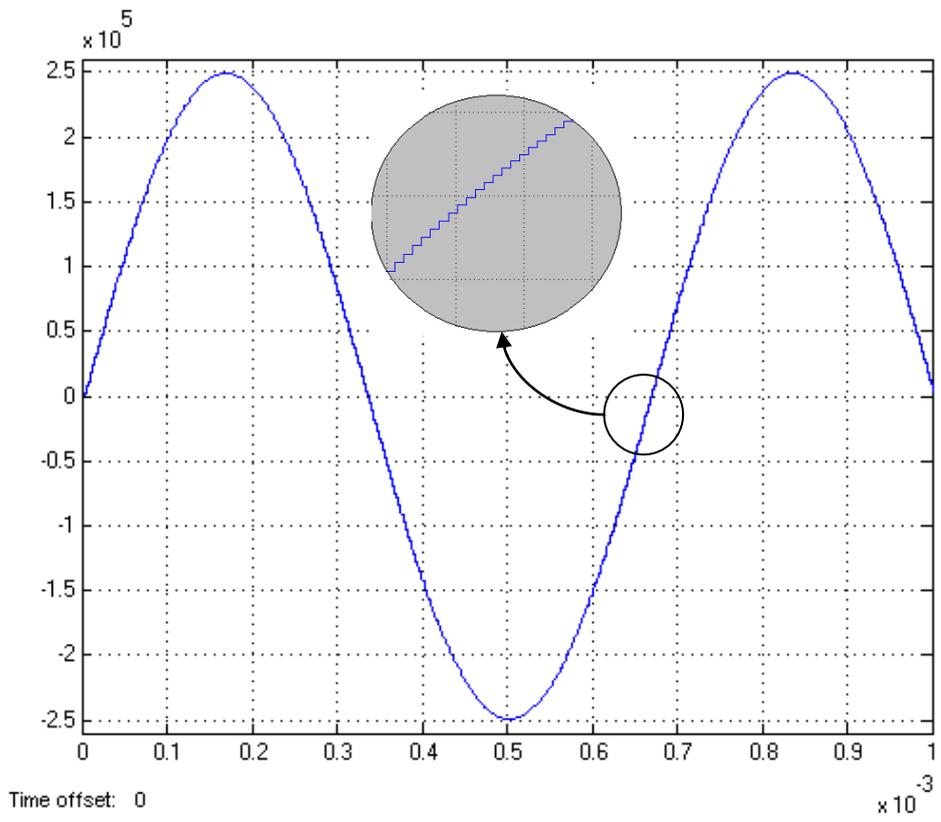

Figure 6.7  Fifth-order CIC filter quantized output

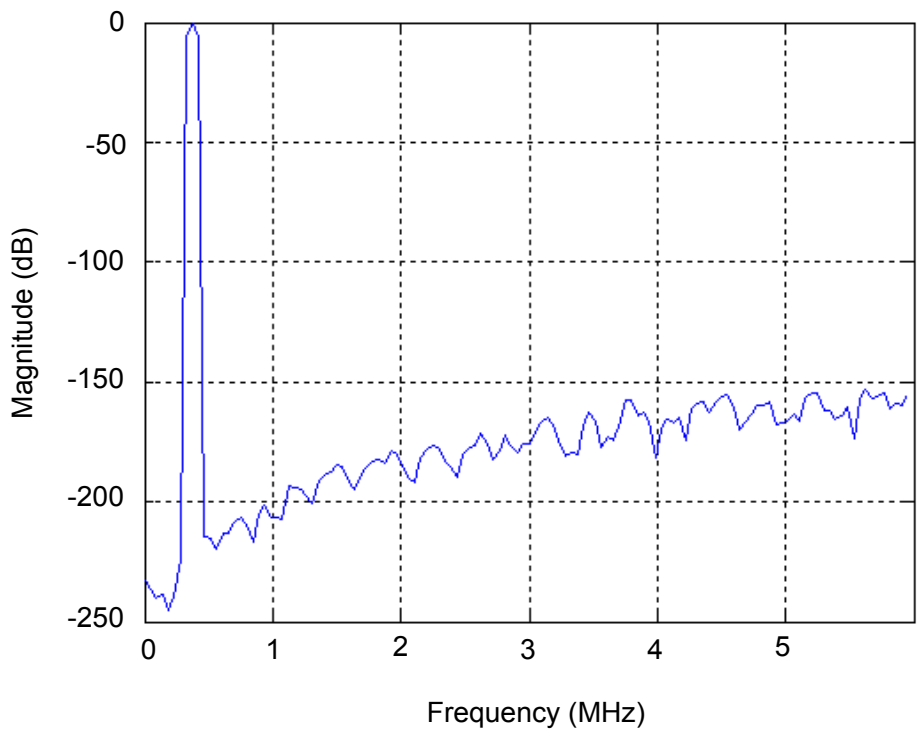

Figure 6.8  Signal to noise ratio of the sigma delta modulator

### 6.1.4 FIRST HALFBAND FILTER SIMULATION RESULTS

The first half band filter with the order of eight provides the decimation ratio of two. Figure 6.9 shows the frequency response of the 8-order halfband filter when passband frequency, stopband frequency and transition band are 32 kHz, 170 kHz and 138 kHz respectively.

The output digital signal of the second half band filter is given with the sampling frequency of 192 kHz, whilst the frequency components between 170 kHz to 192 kHz (Figure 6.11) are attenuated.

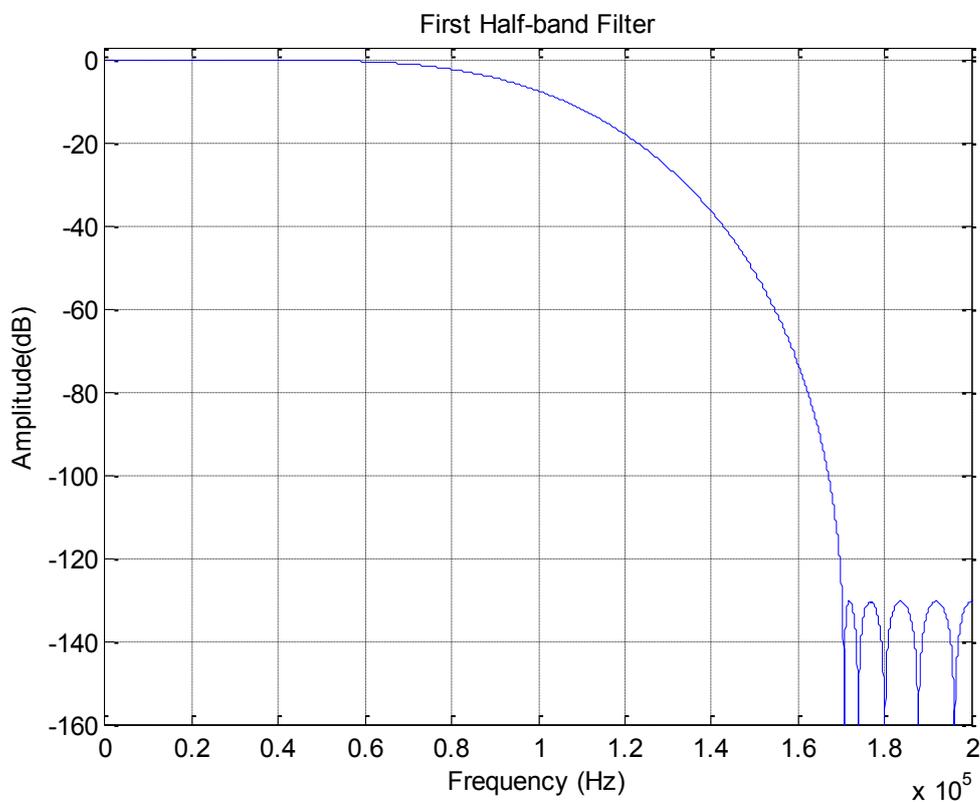

Figure 6.9  Frequency response of the 8-order halfband filter

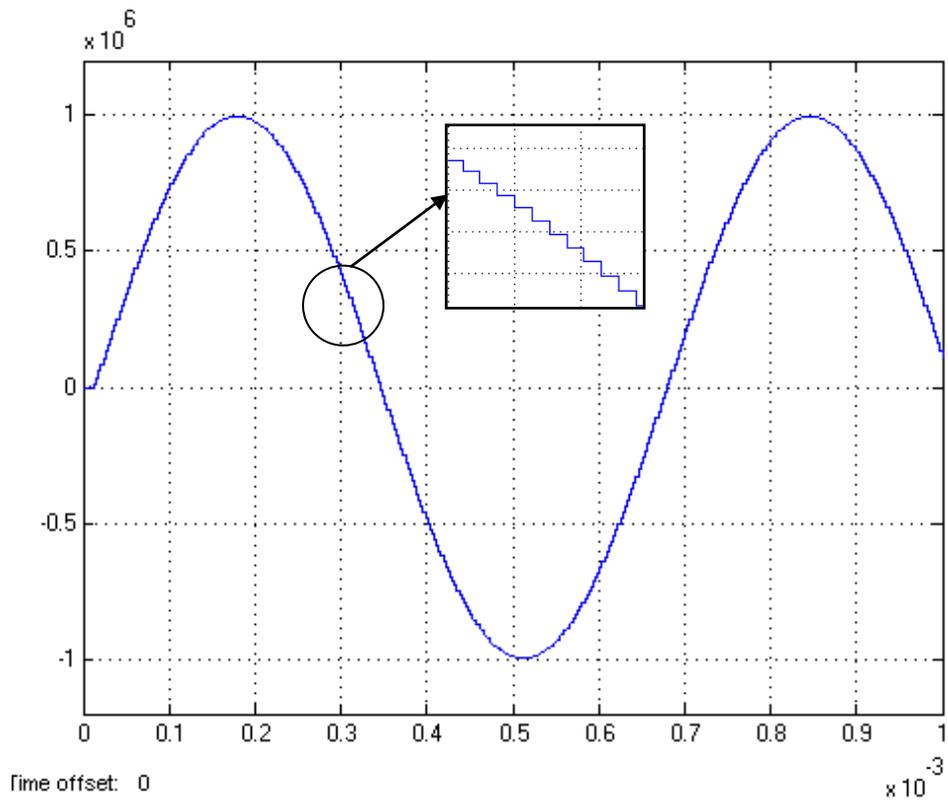

Figure 6.10 First halfband filter output with sampling frequency of 192 kHz

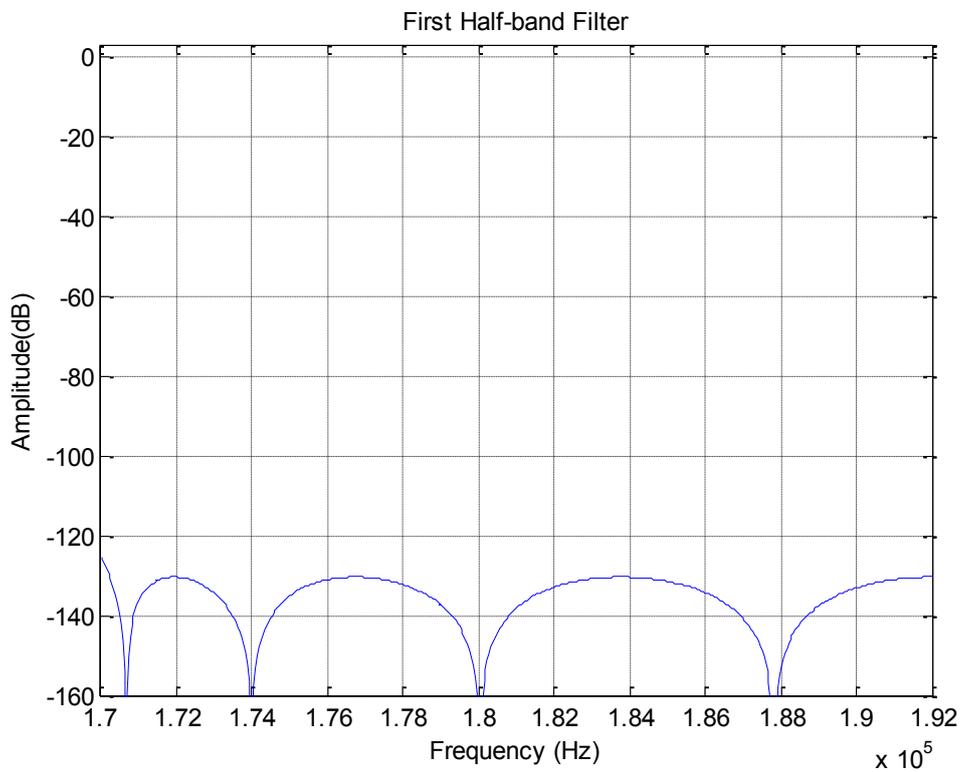

Figure 6.11 Frequency response of the halfband filter in frequency of 170 to 192 kHz

### 6.1.5 DROOP CORRECTION FILTER SIMULATION RESULTS

As described in chapter III, the droop correction filter is the 14-order FIR filter located after the half band filter. This filter also provides a decimation ratio of two. Thus the output signal appears with the sampling frequency of 96 kHz.

The frequency response and output signal of the droop correction filter are given in Figures 6.12 and 6.13. The result shows that droop correction filter compensate the drop produced by the CIC filter and creates the flat Pass-band in the frequency response. As shown in the Figure 6.12 the Pass-band, Stop-band and transition band are 32 kHz, 70 kHz and 38 kHz respectively.

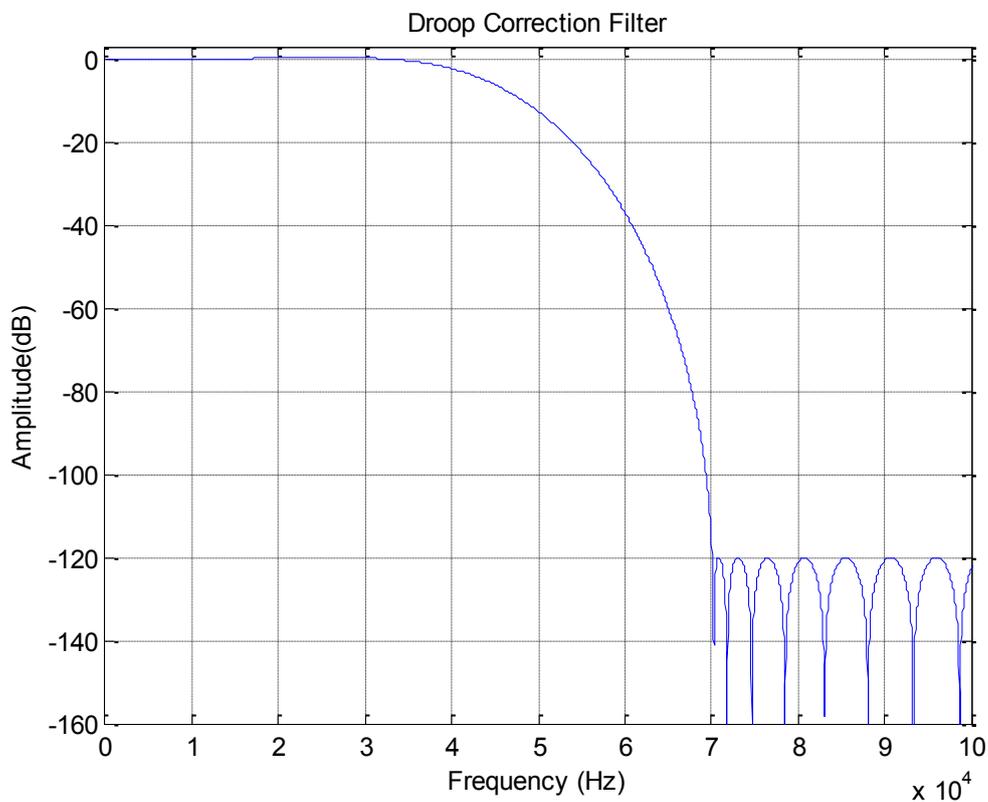

Figure 6.12  Frequency response of the 14-order droop correction filter

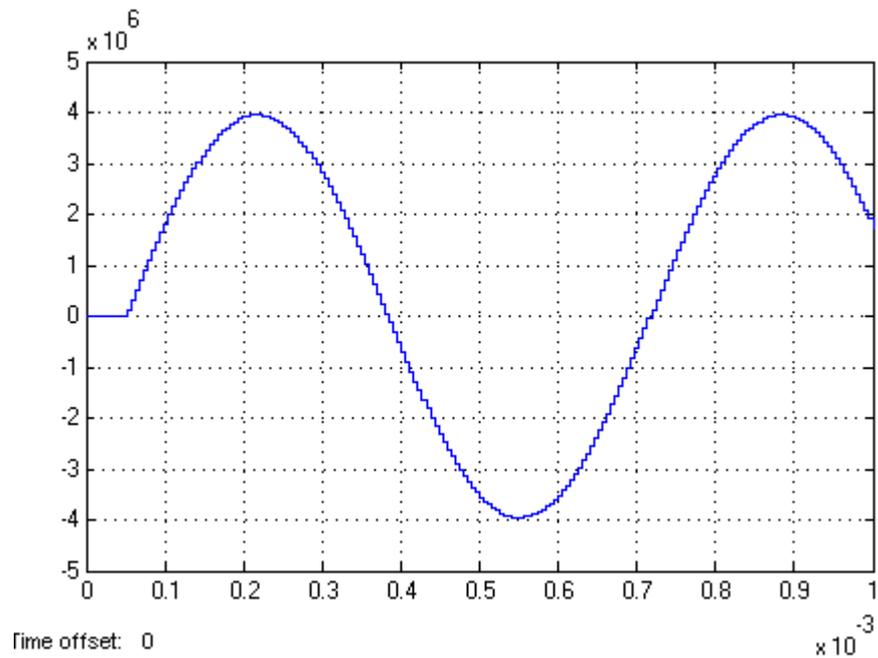

Figure 6.13 First halfband filter output with the sampling frequency of 96 kHz

### 6.1.6 SECOND HALFBAND FILTER SIMULATION RESULTS

The sharp filtering is done at the last stage and is more complex. The second Half-band filter with the order of 80 and decimation ratio of two provide the sharp Stop-band in lower sampling frequency. Its frequency response is shown in Figure 6.14 when the Pass-band, Stop-band frequency and transition band are 21.77 kHz, 26.53 kHz and 4.76 kHz respectively. The second Half-band filter operates with the sampling frequency of 48 kHz rate which is the Nyquist rate.

The digital output of the second Half-band filter with the Stop-band attenuation of 120 dB is given in Figure 6.15.

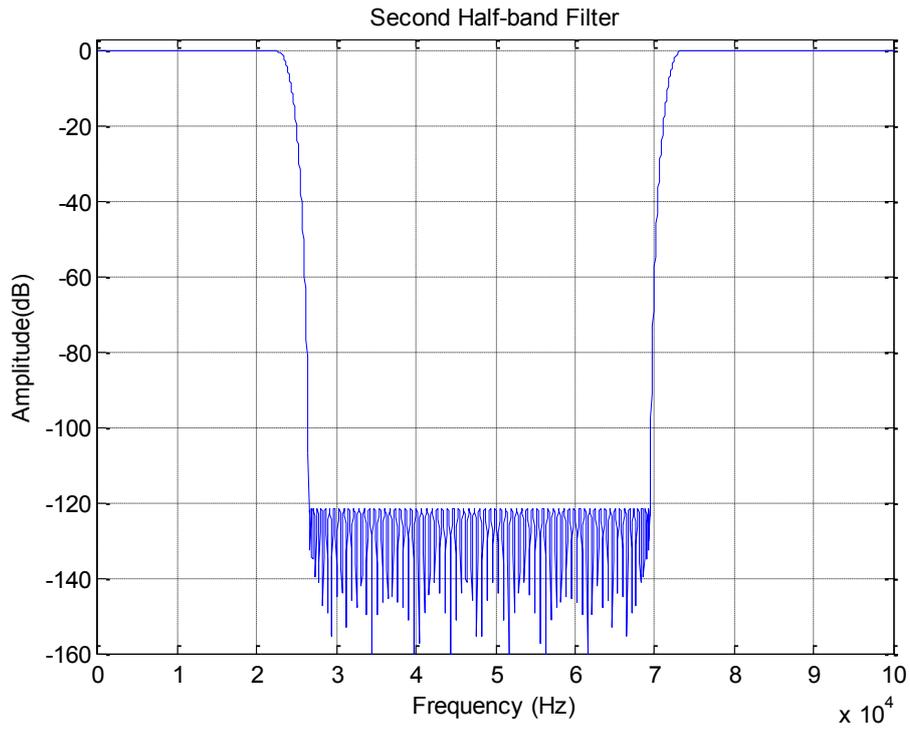

Figure 6.14  Frequency response of the second halfband filter

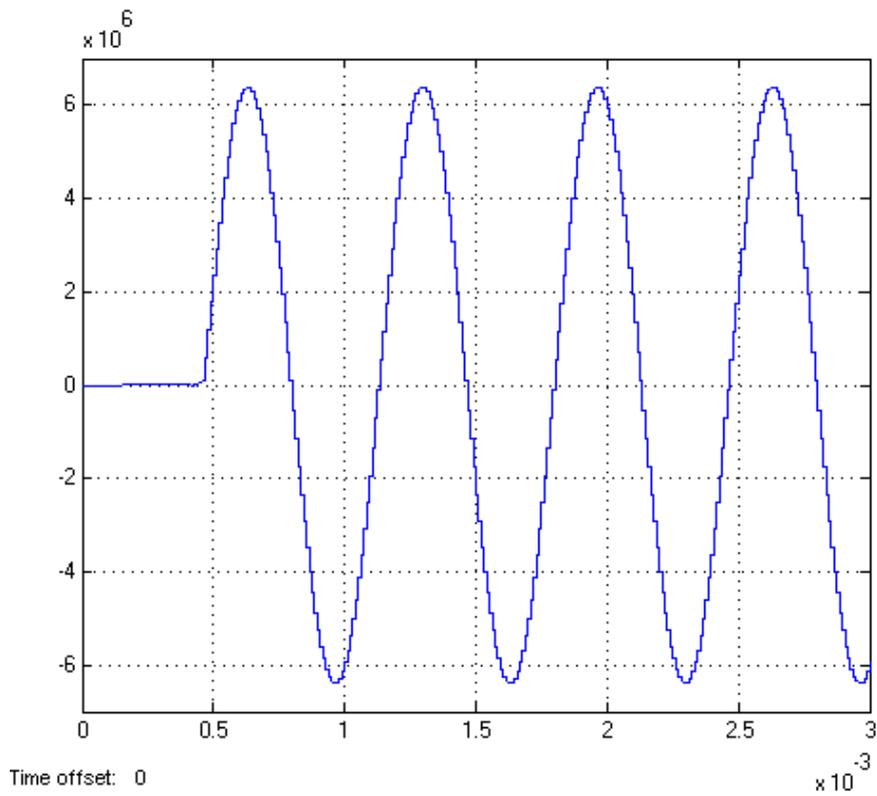

Figure 6.15  Second Half-band filter output with the sampling frequency of 48 kHz

## 6.1.7 OVERALL SYSTEM SIMULATION RESULTS FOR THE DECIMATION FILTERS

The combination of CIC filter, first and second Half-band filters and the droop correction filter creates overall decimation system with a flat and sharp frequency response. We carried out the MATLAB simulation and the frequency response of the decimation filters are shown in Figure 6.16 and 6.17. The frequency response of overall decimation filter has the same specifications with the frequency response of second half band filter.

This combination also reduces the filter ripple. The Pass-band ripple, after second Half-band filter is limited within $\pm$ 0.0001 dB. Figure 6.18 shows filter ripple in the Pass-band frequency which is negligible.

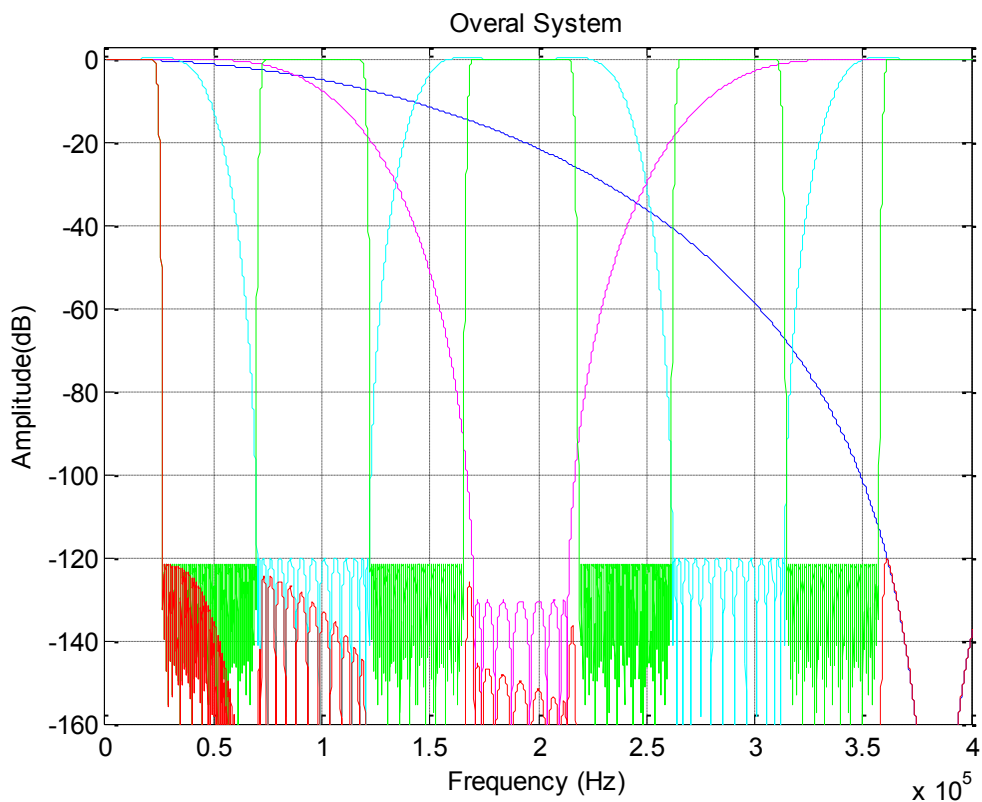

Figure 6.16  Frequency response of all filter stages

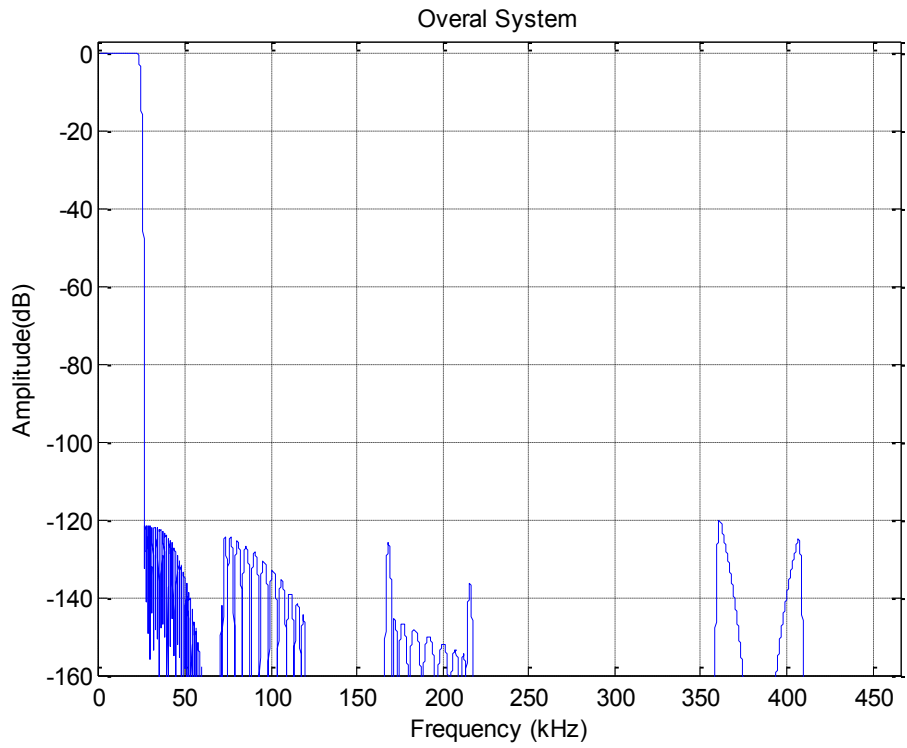

Figure 6.17  Frequency response of the overall decimation filters

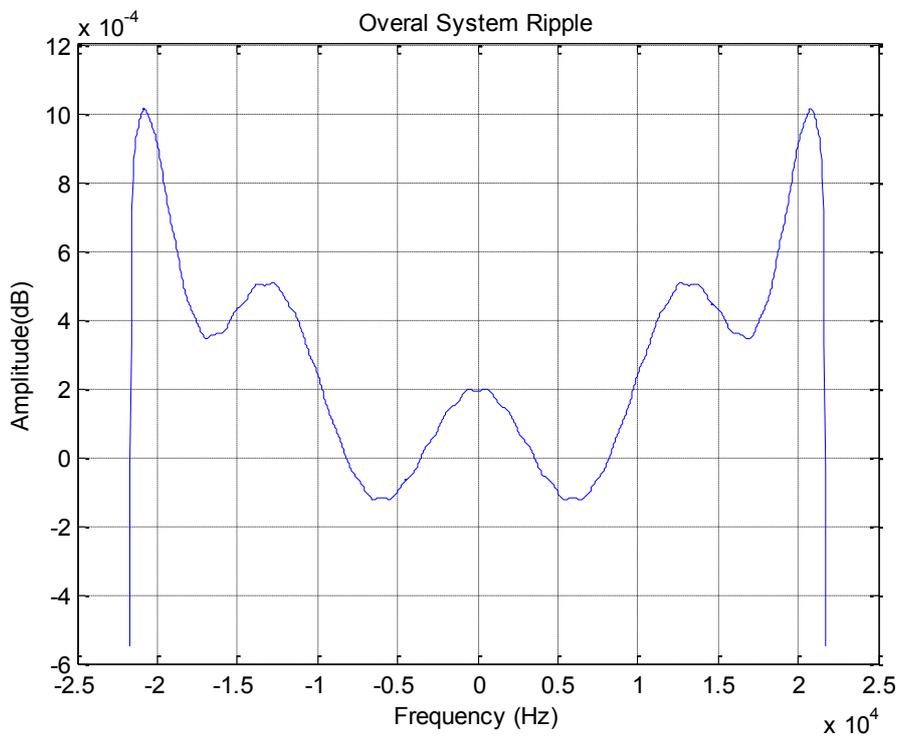

Figure 6.18  Passband ripple of overall decimation system

## 6.2 CIC FILTER IMPLEMENTATION

After the CIC filter function verification, then we proceed with the FPGA implementation of the CIC filter. In summary the FPGA downloading process are as follow:

1) Verilog HDL coding for the filter and simulating the design by Modelsim simulator.

The Modelsim simulation result of the CIC filter is shown in Figure 6.19 and 6.20. These Figures illustrate the output and the intermediate input signal of the top-module CIC filter in FPGA board. There are four sub-modules in the design which are clock divider, counter, memory and the CIC filter. This structure was explained in chapter V in detail. The sub-module of the CIC filter after feeding by memory as input data and clock divider as trigger clock frequency produces digital output data for top-module of CIC filter. The output data is shown as quantized digital sine wave (Figure 6.20) and normal digital output data (Figure 6.19). The result shows the output latency of 23 clock cycle in the CIC filter output.

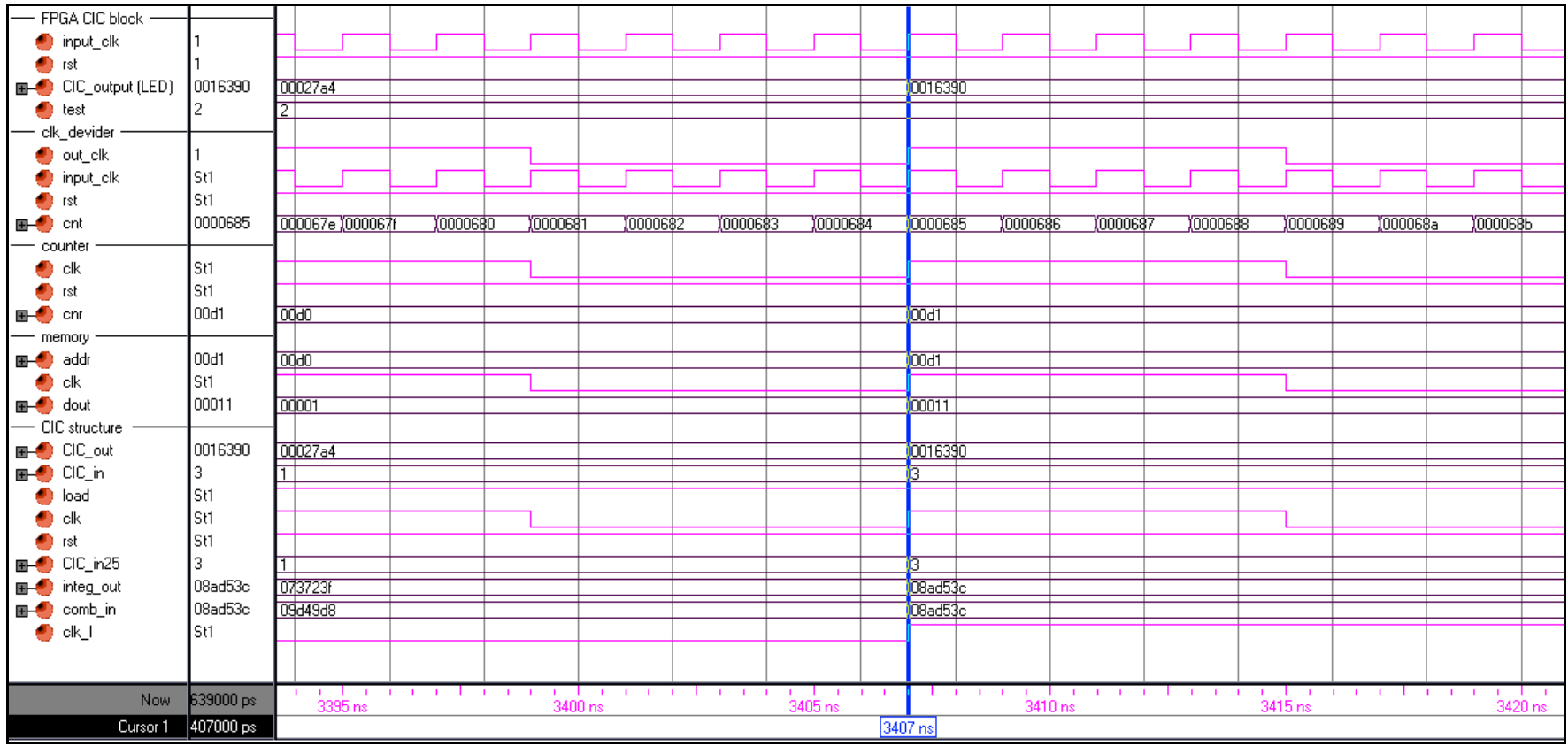

Figure 6.19  Simulation of the CIC filter as digital data by using Modelsim software

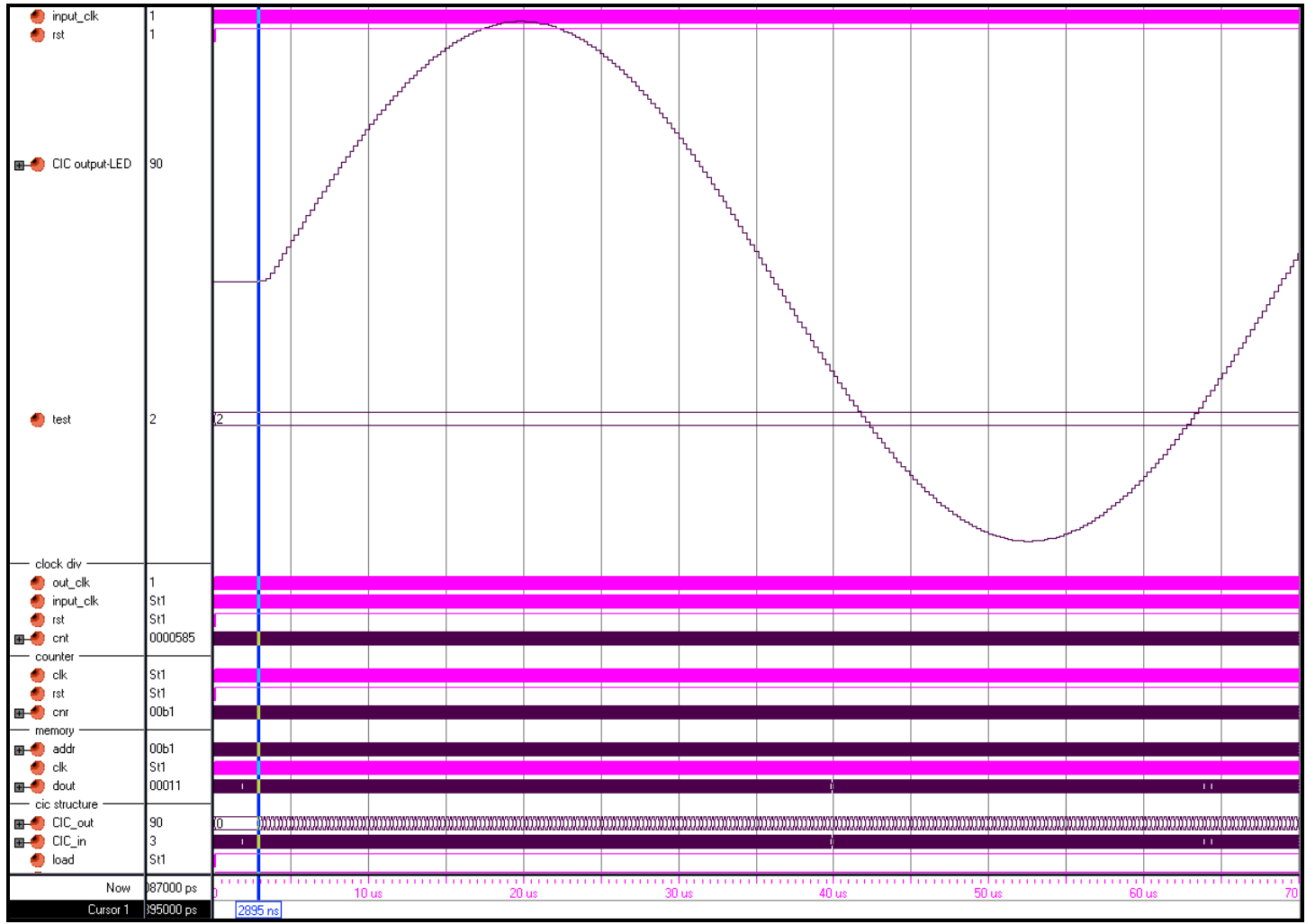

Figure 6.20  Modelsim simulation of the CIC filter as analog wave form

2) Past over from the PC (in the Modelsim) we optimized the filter in Synopsys tools to achieve the gate level. In this level the filter is re-simulated by Cadence CAD design tools for verification and the layout is produced. In order to refer the verification, the data loaded in FPGA board is compared with the simulation data obtain from NCLaunch in CAD tools.

Figure 6.21 shows layout of the CIC filter active core using Silterra 0.18 $\mu m$ technology. This layout was routed by Encounter tools. Figure 6.22 demonstrates gate level output result of the CIC filter produced by NCLaunch using CAD tools.

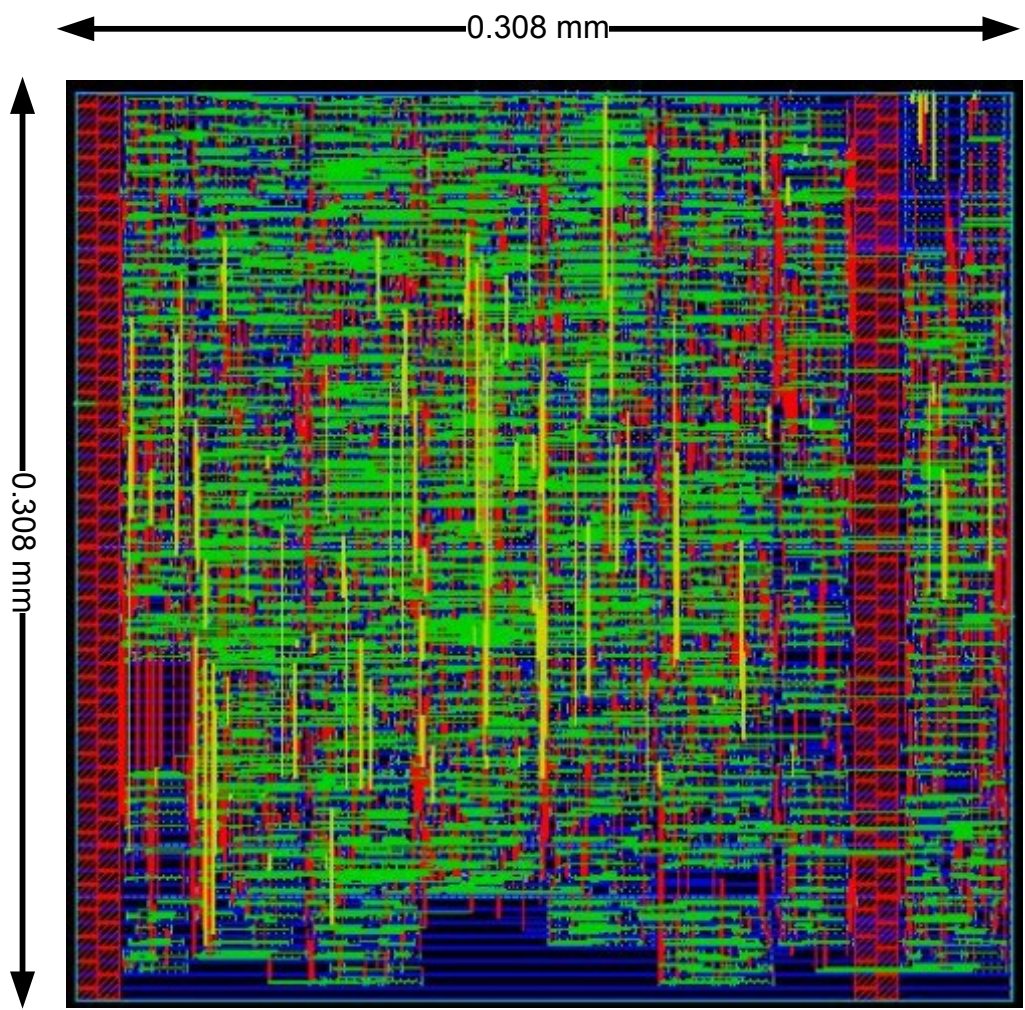

6.21 Layout of the CIC decimation filter

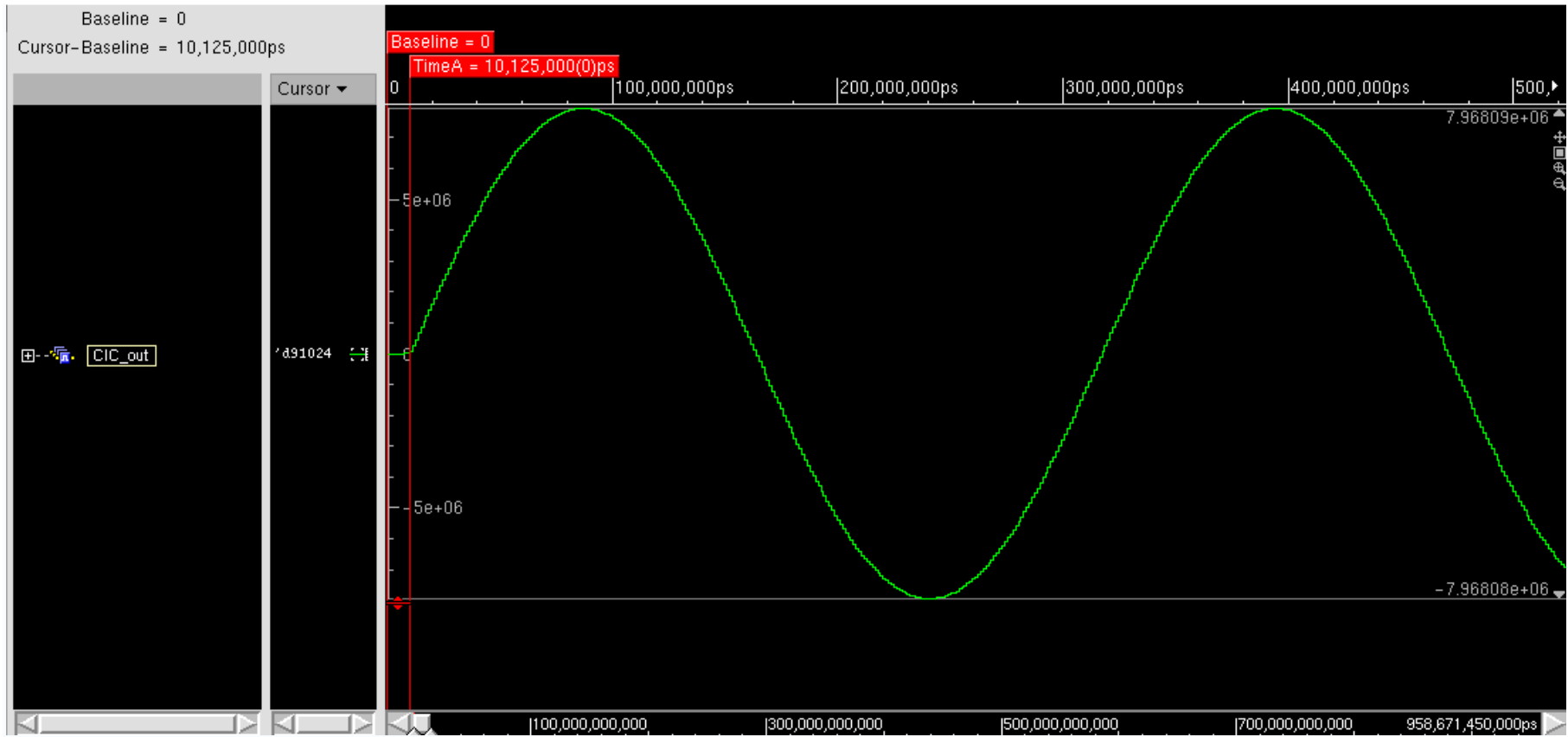

Figure 6.22 NCLaunch result of the CIC filter in gate level

3) The verified CIC Verilog codes were downloaded to the FPGA using Xilinx ISE soft tools. The downloading was successfully done and the routing for the CIC filter is shown in Figure 6.24

Figure 6.23 shows the FPGA board (Virtex II) that was used to implement the CIC filter.

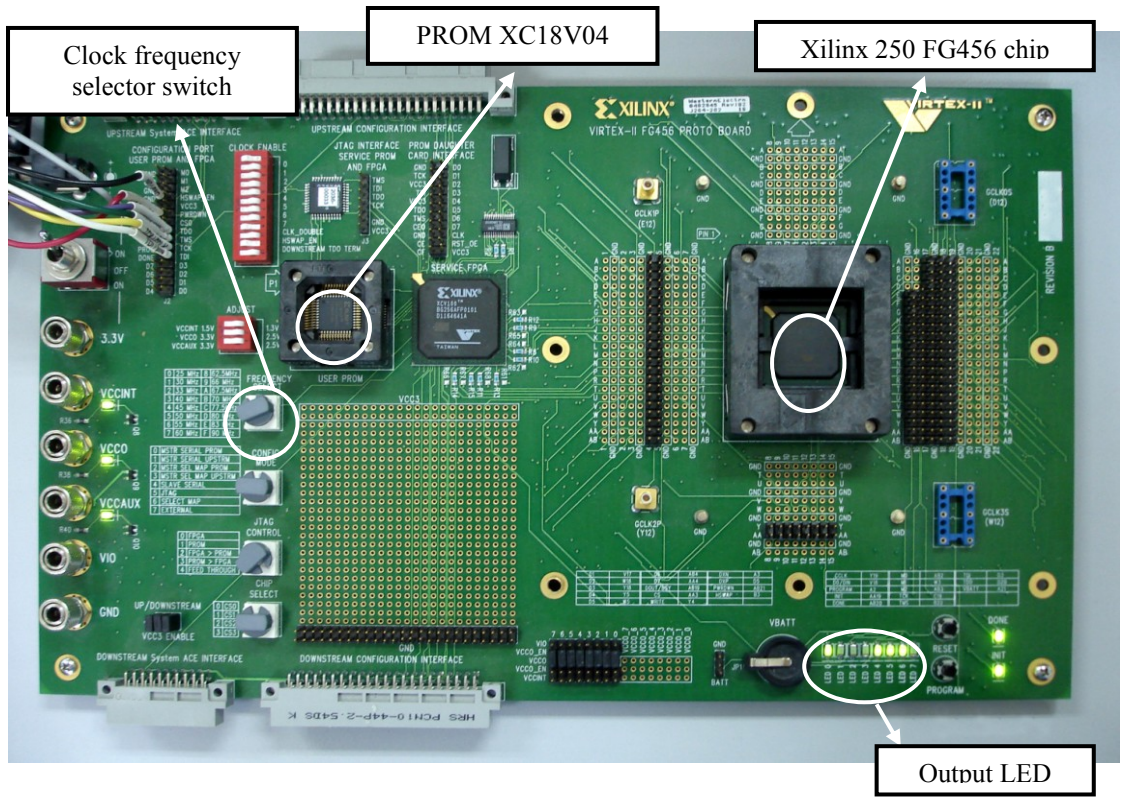

Figure 6.23  Implementation of the CIC filter on FPGA board

For area and power consumption comparisons, we implement the CIC filter using the CLA and the MCLA. Based on the report produced by Xilinx ISE, we found the MCLA implementation reduces the area and power dissipation by using less number of gates (Table 5.4).  High speed CIC filter is implemented on FPGA board (Virtex II). The Xilinx 250 FG456 chip routing is shown in Figure 6.24.

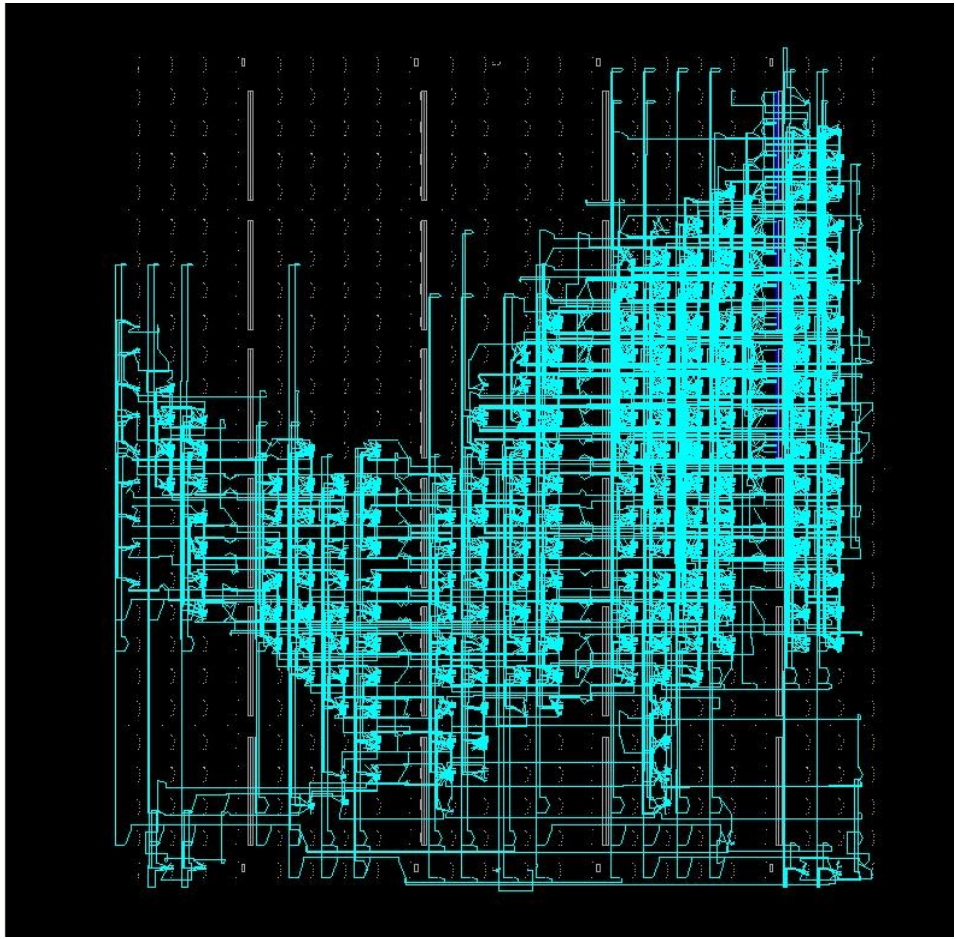

Figure 6.24  Xilinx Chip routing of the CIC filter

This CIC FPGA chip was tested using logic analyzer. The logic analyzer result shows that the CIC filter can function correctly up to 189 MHz as shown in 6.26, whilst 6.25 show the equipment setup for the Logic Analyzer.

Figure 6.26 shows the testing result of the FPGA board on logic analyser when the 25-bit CIC filter is downloaded. The clock frequency is set on the maximum rate (189 MHz).

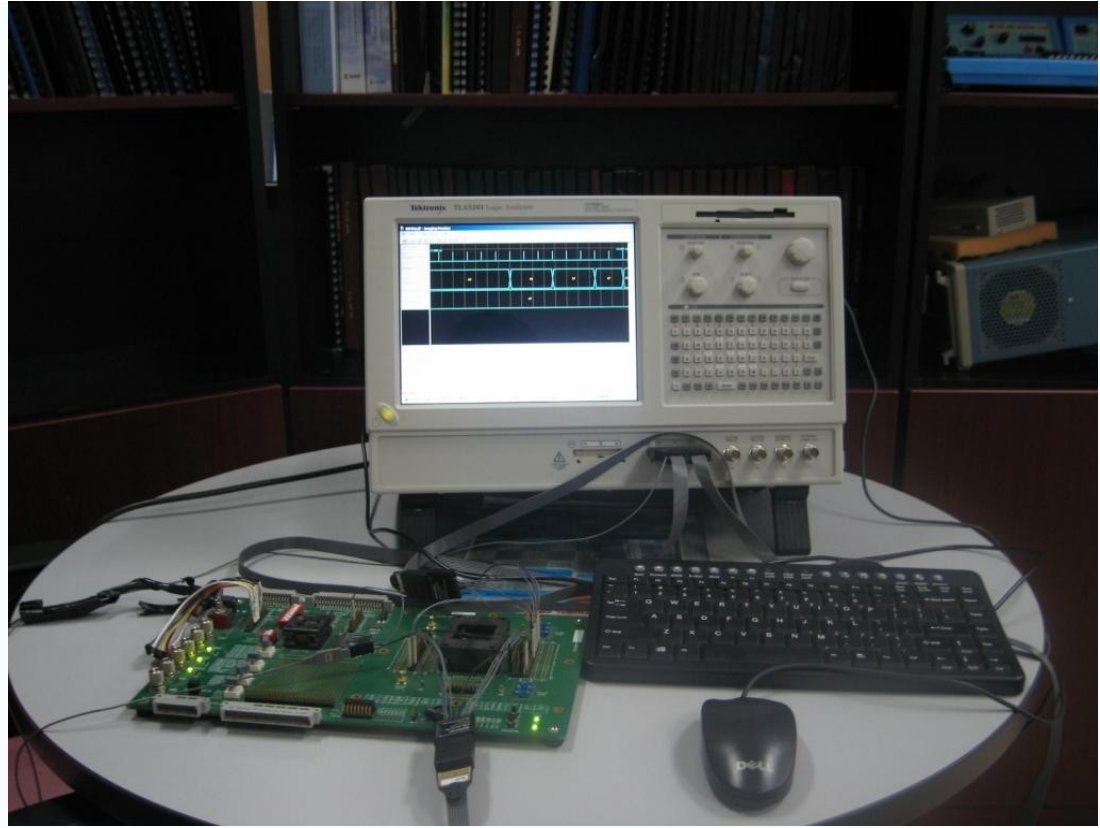

Figure 6.25 CIC filter test circuit using Logic Analyzer

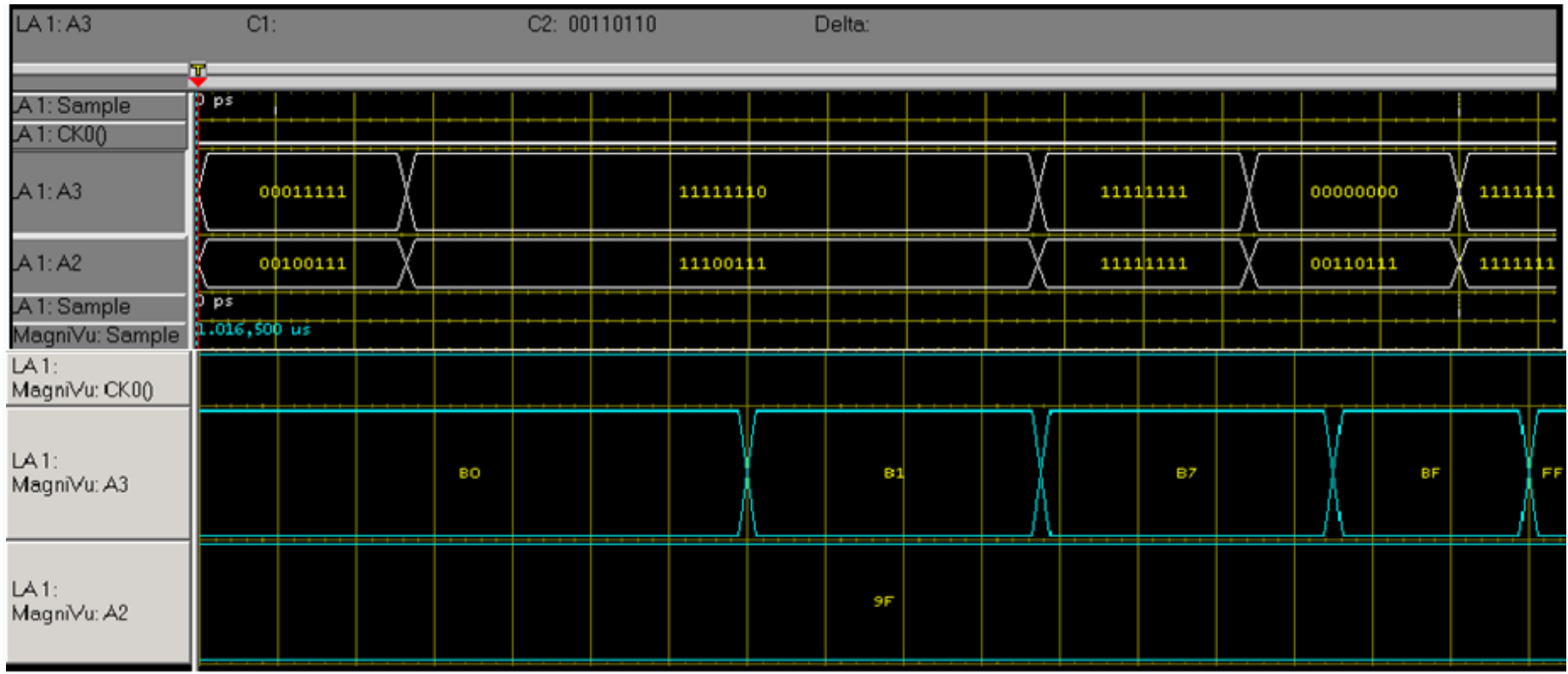

Figure 6.26 CIC filter test result using Logic Analyzer

For separate testing of the CIC filter blocks such as integrator, comb and downsampler a certain configuration of the CIC filter is used. In this configuration a number of multiplexers and de-multiplexers were used. Figure 6.27 shows the circuit configuration that was used to test the sub-blocks of the CIC filter.

As seen in the Figure 6.27, when signals $C_5$ and $C_6$ are high, only the integrator is tested, while other stages do not produce any output signal. If signal $C_1$, $C_2$ and $C_6$ are high and signal $C_5$ is low, down-sampler provides output signal. If $C_3$, $C_4$ and $C_6$ are low the comb stage is tested and when $C_1$, $C_2$ and $C_6$ are low and then $C_3$ and $C_4$ are high the system works under normal condition and the result can be evaluated. Figures 6.28, 6.29 and 6.30 show the result of testing different stages of the CIC filter.

Figure 6.28 shows the integrator stage testing when digital quantized sine wave feed to the integrator. The output is increased without bounded which is the integrator property.

Figure 6.29 illustrates the down-sampler stage testing when the input rate is changed to lower rate by decimation factor of 16. Every multiple data of 16 is kept to the register and the rest are discarded.

Figure 6.30 demonstrates the comb stage result test when the integrator and down-sampler do not produce the output signal. This block limits the input data by subtractor.

Figure 6.27  Testability of the CIC decimation filter

Figure 6.28  Integrator result testing

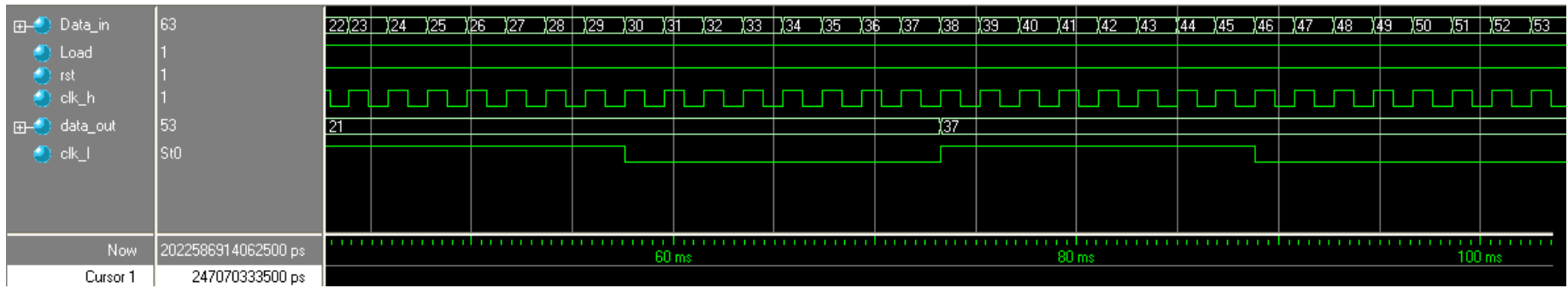

Figure 6.29 Down-sampler result test

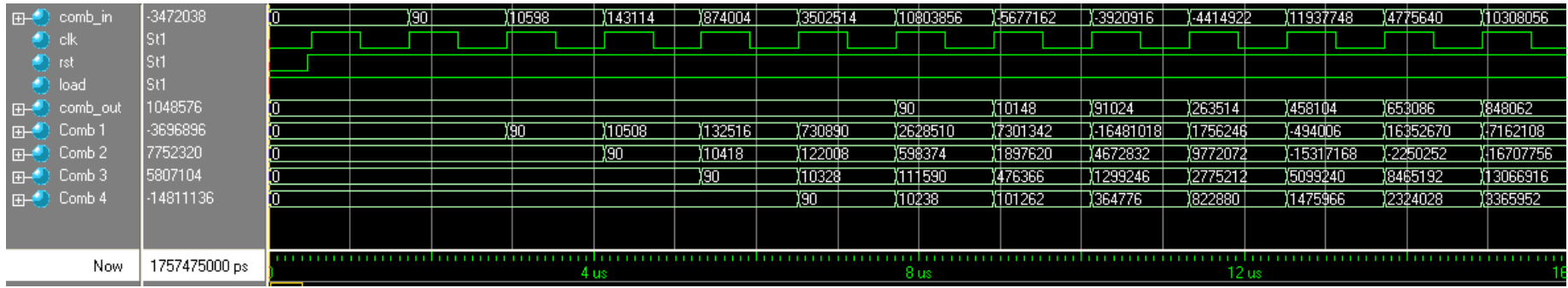

Figure 6.30 Comb result test

The CIC filter specifications are summarized in table 6.1.

Table 6.1  Summary performance of the CIC filter

| | | |
|---|---|---|
| Filter order | $N$ | 5 |
| Decimation Ratio | $R$ | 16 |
| Differential delay | $M$ | 1 |
| Max. register growth | $G_{max}$ | 120 dB |
| Input resolution | - | 5 bit |
| Output resolution | - | 25 bit |
| Sampling frequency | $f_s$ | 6.144 MHz |
| Max. clock frequency | $f_{s.max}$ | 189 MHz |
| Passband frequency | $f_{pass}$ | 7 kHz |
| Stopband frequency | $f_{stop}$ | 384 kHz |
| Transition band | $f_T$ | 377 kHz |
| Stop band attenuation | $A_s$ | 120 dB |
| Pass band attenuation | $A_p$ | 0.29 dB |
| Signal to noise ratio | SNR | 145.35 dB |
| Power consumption (Silrerra 0.18 $\mu m$ library) | $P_o$ | 3.14 mW |
| Active core Area (Silrerra 0.18 $\mu m$ library) | - | $0.308mm \times 0.308mm$ |

## 6.3  SUMMARY

In this chapter, simulation result of the oversampling and decimation system using MATLAB tools are presented. The frequency response and digital output signal of decimation filters are given separately. This was found to give the required results of the overall frequency response with the Pass-band, Stop-band frequency and transition band of 21.77 kHz, 26.53 kHz and 4.76 kHz respectively. The simulation result also shows the Pass-band ripple less than 0.0001 dB. This chapter also demonstrates the FPGA implementation result of the CIC filter with the maximum throughput of 189 MHz in Xilinx synthesis report. For area and power consumption, the CIC filter was optimized in ASIC under Silterra 0.18 $\mu m$ technology file. The Layout was shown in this chapter.

Additionally the output of the CIC filter is tested by using Logic Analyzer which shows the operation of the CIC filter in maximum frequency of 189 MHz. A certain configuration was implemented to test different part of the CIC filter structure and the results were given.

# CHAPTER VII

# CONCLUSIONS AND FUTURE WORK

## 7.1  CONCLUSIONS

The oversampling technique followed by the decimation process discussed was evaluated with the focused on the development CIC. The sigma delta modulator acts as an over-sampler and the decimation system acts as down-sampler. The third order sigma delta modulator with sampling frequency of 6.144 MHz converts analogue audio signal to digital sampled data and the high speed cascaded integrator comb filter (CIC), first and second half band filters and droop correction filter decimate the sampling rate from 6.144 MHz to 48 kHz in Nyquist rate. Among these filters, implementation of high speed CIC filter with the property of maximum decimation ratio, quantization noise rejection and anti-aliasing was the research topic of this project. The CIC filter has three main blocks which are the integrator, the down-sampler and the comb stages. The specifications of the filter in this research work are as follows: number of integrator is 5, number of comb filter is 5, decimation ratio is 16, and maximum number of bit is 25-bit that corresponds to maximum register growth of 120 dB. This filter was designed, investigated and implemented using MATLAB, Verilog HDL code, Xilinx and CAD synthesis tools. Enhanced high speed CIC filter was obtained by utilizing three methods. The pipeline structure which is applied for overall CIC filter blocks, using the modified carry look-ahead adder (MCLA) and truncation for integrator parts lead us to have high speed CIC filter with the maximum throuput of 189 MHz. The evaluation indicates that the pipelined CIC filter with MCLA adder is attractive due to high speed when both the decimation ratio and filter order are not high (the order and the decimation ratio of less than 10 and 64 respectively). Since the first stage of the CIC filter requires maximum word length and also because of the recursive loop in its structure, the power consumption is limited by the throughput. Reducing power consumption is one of the key issues in designing the CIC filter. It will be achieved by decreasing the number of calculation (achieved by truncation) and eliminating unnecessary elements (achieved by pipeline structure and using MCLA) in the filter.

The result of these efforts is the power consumption of 3.5 mW in Silterra 0.18 $\mu m$ and 6.03 mW in Mimos 0.35 $\mu m$ technology libraries at the maximum clock frequency.

## 7.2  FUTURE WORK

The layout was done. We proposed to have programmable CIC filter with the schematic of the chip Figure 7.1.

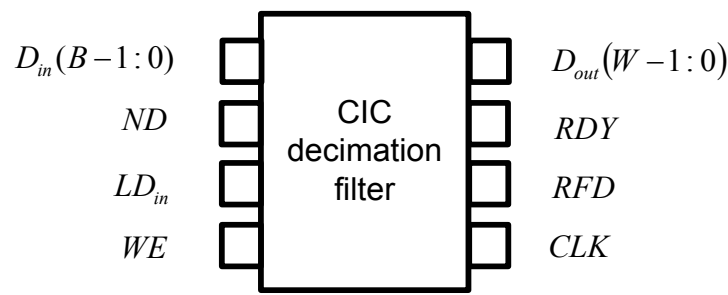

Figure 7.1  External control pin of the CIC fitter

where *CLK* is clock master and become active by rising edge. $D_{in}$ is input data with the number of *B* bit. *ND* is new data which is considered active high control pin. When this signal is asserted the data sample presented on the $D_{in}$ port is loaded in to the filter. $D_{out}$ is filter output with the number of *W*-bit. *W* must be greater than *B*. *RDY* is filter output sample ready pin. This control pin indicates a new filter output sample which is available on the $D_{out}$ port. *RFD* is ready for data pin which is active high. This control pin indicates when the filter can accept a new input sample. $LD_{in}$ is input bus which is used to supply the sample rate change value when the programmable rate change option for the CIC decimator is selected. $LD_{in}$ is an optional port and is only available with decimation filters. *WE* is active high write enable pin. This signal is associated with the $LD_{in}$ bus. The value on the $LD_{in}$ bus is latched by the core on the rising edge of the clock qualified with *WE* equal high. *WE* is an optional port and may only be selected for decimation filters (Logicore, 2002).

External control pins must operate together as parallel. Figure 7.2 shows the procedure of external control pins in the CIC decimation chip.

The CIC filter timing is also shown in Figure 7.3. In this Figure new data sample is considered to supply to the chip on every clock cycle. $D_{out}$ is available after passing some clock pulse interval depends on filter latency and *R*. fixed latency is related to internal pipeline registers in the CIC chip.

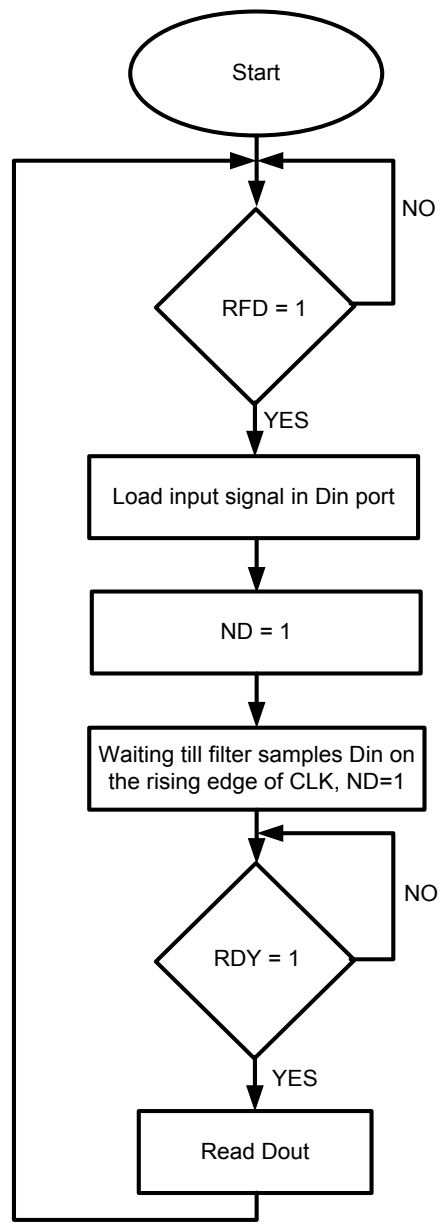

Figure 7.2   Performance of the CIC filter control pins

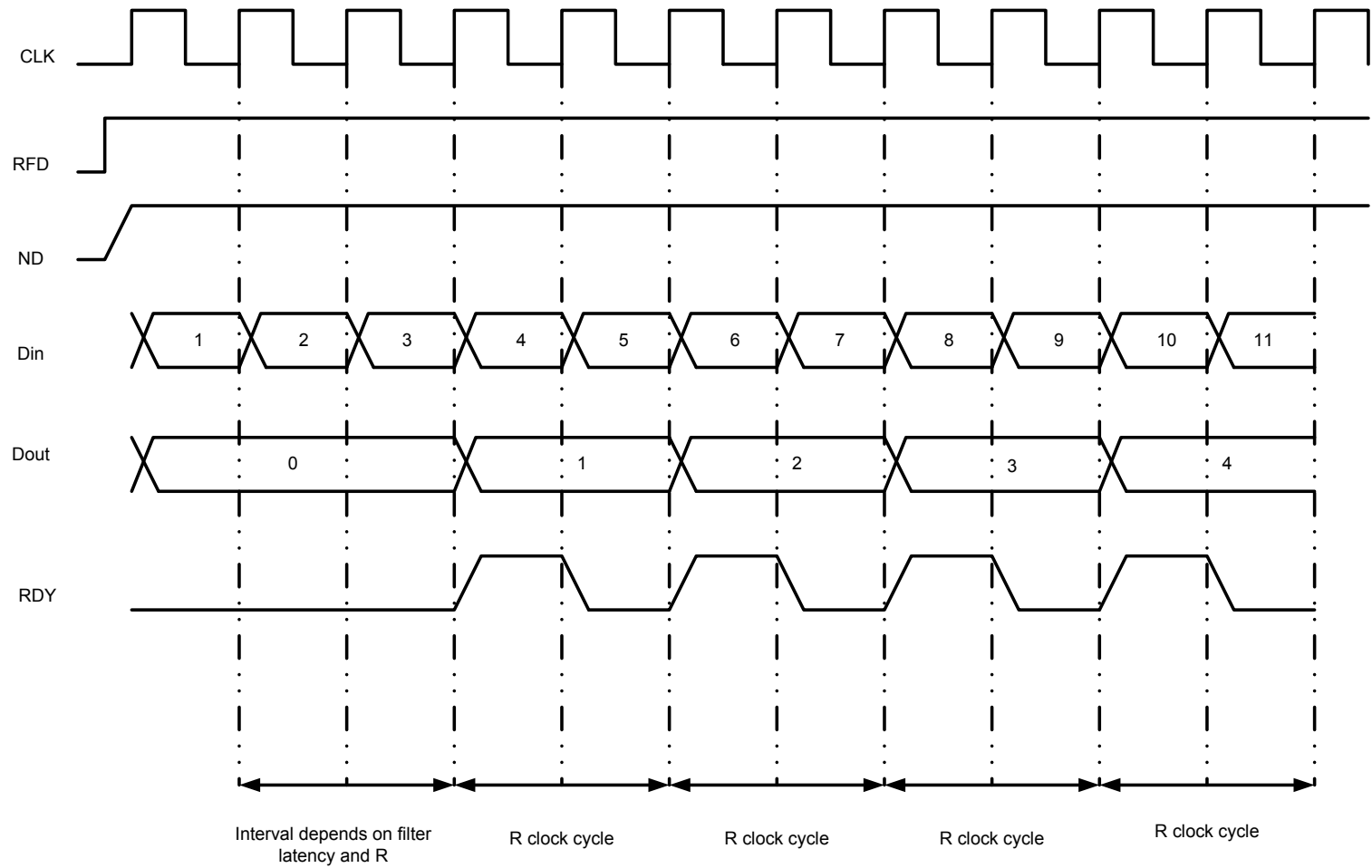

Figure 7.3 CIC decimation filter timing

# CURRICULUM VITAE

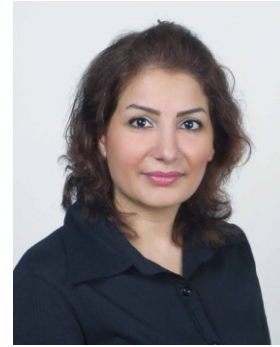

Asst. Prof. Dr. Rozita Teymourzadeh, CEng. has received her Bachelor of Science (B.Sc) in Electronic Engineering in 2004 and the M.Sc. degree from the Faculty of Electrical & Electronic Engineering at the National University of Malaysia (UKM) in 2007. Her PhD degree was awarded by the same university (UKM) in the year 2011. Furthermore, she is active member of Institution of Engineering and Technology (MIET) and the member of Institute of Electrical and Electronics Engineers (IEEE). In addition, she was entitled as Chartered Engineer (CEng) from Engineering council (UK) and the title of "**Young Women Engineer**" from IET in 2012. Her research area is in the field of Portable ASIC implementation, VLSI Design, Digital Signal Processing, FFT Processor & Digital Filter Design, Nano Technology , Front-end Chip implementation (SOC), embedded design, LAB view and GUI MATLAB design, system tool analysis and finally, Image processing. Moreover, Asst. Prof. Dr. Rozita has the vast of academic and industry experience in the Electrical field in MV design and has completed several industry projects and has published several valuable papers.